\documentclass[a4paper,11pt]{article}
\pdfoutput=1
\usepackage{geometry}
\usepackage{a4wide}
\usepackage{graphicx}
\usepackage{epsf}
\usepackage{amsmath}
\usepackage[normalem]{ulem}
\usepackage{amssymb}

\usepackage{cite}
\usepackage{multirow,tabularx}
\usepackage{appendix}
\usepackage{tikz}
\usepackage{amsmath,amsfonts,amssymb,amsthm,euscript,braket,xcolor}
\newcommand{\be}{\begin{equation}}
\newcommand{\ee}{\end{equation}}

\newcommand{\Rmnum}[1]{\expandafter\@slowromancap\romannumeral #1@}
\newcommand{\bea}{\begin{eqnarray}}
\newcommand{\eea}{\end{eqnarray}}

\numberwithin{equation}{section}

\usepackage{subfigure}
\begin{document}

\title{\bf Analytic topological hairy dyonic black holes and thermodynamics}

\author{\textbf{Supragyan Priyadarshinee}\thanks{518ph1009@nitrkl.ac.in}, \textbf{Subhash Mahapatra}\thanks{mahapatrasub@nitrkl.ac.in}, \textbf{Indrani Banerjee}\thanks{banerjeein@nitrkl.ac.in}
 \\\\
 \textit{{\small Department of Physics and Astronomy, National Institute of Technology Rourkela, Rourkela - 769008, India}}
}
\date{}


\maketitle
\abstract{}
We present and discuss a new family of topological hairy dyonic black hole solutions in asymptotically anti–de Sitter (AdS) space. The coupled Einstein-Maxwell-Scalar gravity system, that carries both the electric and magnetic charges is solved, and exact hairy dyonic black hole solutions are obtained analytically. The scalar field profiles that give rise to such black hole solutions are regular everywhere. The hairy solutions are obtained for planar, spherical, and hyperbolic horizon topologies. In addition, analytic expressions of regularized action, stress tensor, conserved charges and free energies are obtained. We further comment on different prescriptions for computing the black hole mass with hairy backgrounds. We analyze the thermodynamics of these hairy dyonic black holes in canonical and grand canonical ensembles, and we find that both electric and magnetic charges have a constructive effect on the stability of the hairy solution. For the case of planar and hyperbolic horizons, we find thermodynamically stable hairy black holes that are favoured at low temperatures compared to the non-hairy counterparts. We further find that, for a spherical hairy dyonic black hole, the thermodynamic phase diagram resembles to that of a Van der Waals fluid not only in canonical but also in the grand canonical ensemble.

\section{Introduction}
Black holes, which are one of the most intriguing and celebrated predictions of general relativity, and yet are still far from being fully understood, are conjectured to follow the no-hair theorem \cite{Ruffini}. The black hole no-hair theorem simply states that a black hole with a spherical horizon can be completely characterized by only three parameters: its mass, angular momentum, and charge in the asymptotically flat space. Put in another way, black holes do not support additional matter fields, such as scalar fields, outside their horizon in asymptotically flat space. The prime reason for belief in the no-hair theorem is somewhat based on the strong absorbing nature of the horizon, which tries to absorb and pull everything around it. Although the initial no-hair theorem has been advocated by several works \cite{Bekenstein:1971hc,Israel:1967wq,Israel:1967za,Wald:1971iw,Carter:1971zc,Robinson:1975bv,Mazur:1982db,Mazur:1984wz,Teitelboim:1972qx,Chase}, it is not a theorem in a rigorous mathematical sense. Indeed, by now several counterexamples to the no-hair theorem in a variety of contexts exist \cite{Volkov:1990sva,Bizon:1990sr,Kuenzle:1990is,Straumann:1990as,Zhou:1991nu,Bizon:1991hw,Bizon:1991nt,Volkov:1995np,Brodbeck:1994vu,Garfinkle:1990qj,Herdeiro:2014goa,Berti:2013gfa,Greene:1992fw,Lavrelashvili:1992ia,
Torii:1993vm}.

The investigation of the no-scalar hair theorem and interplay between gravity-scalar systems are not just of theoretical concern, and there are many reasons to examine them. Scalar fields play a central role in cosmology and particle physics \cite{Svrcek:2006yi}. Scalar fields appear naturally as a basic constituent of fundamental theories, such as string theory, and also arise naturally in the high energy unification theories \cite{Arvanitaki:2009fg,Visinelli:2018utg}. They are also arguably among the most economical and suitable candidates for describing dark matter, dark energy, and inflation physics \cite{McDonald:1993ex,Linde:1982uu,Peebles:1998qn}. From the astronomical observational perspective, the discovery of gravitational wave and black hole image opened a new window to test the no-hair theorem \cite{LIGOScientific:2016aoc,EventHorizonTelescope:2019dse}, as they are believed to be sensitive to the geometry around the black hole, especially in the vicinity of the horizon, and might contain valuable information of additional matter fields around the black hole. See \cite{Sadeghian:2011ub,Khodadi:2020jij,Khodadi:2021gbc,Bambi:2019tjh,Vagnozzi:2019apd} for a discussion in this direction, and \cite{Cardoso:2016ryw} for a review on observational tests of the no-hair theorem.

There have been many attempts to endow black holes with hair by means of additional scalar fields in the last few decades. In principle, there are two essential requirements for a physically acceptable scalar hairy black hole solution (i) scalar field should be regular in the near horizon region and should fall off sufficiently fast at the asymptotic boundary, (ii) the hairy geometry should be smooth and should not contain any additional singularity. The stability of hairy solutions under perturbations is also desirable. The initial black hole scalar hair solution in the asymptotic flat spaces, unfortunately, turned out to be un-physical as the scalar field diverged on the horizon \cite{Bocharova,Bekenstein:1974sf,Bekenstein:1975ts,Bronnikov:1978mx}. A direct proof and many compelling arguments in favour of the no-scalar hair theorem in the asymptotic flat spacetime was provided in \cite{Bekenstein:1971hc,Bekenstein:1995un,Sudarsky:1995zg,Heusler:1992ss}. For a review and discussion on the interplay of scalar field and gravity in asymptotic flat spaces, see \cite{Herdeiro:2015waa,Hertog:2006rr,Astefanesei:2019mds}. One way of ensuring regular scalar field at the horizon, thereby evading the no-hair theorem, is by introducing a new scale in the gravitational sector, \textit{i.e.}, via a cosmological constant. This amounts to studying black holes in asymptotically de Sitter (dS) or anti-de Sitter (AdS) spaces. The essential idea here is that the cosmological constant can create an effective potential and therefore may stabilize the scalar field outside the horizon. Some of these ideas led to stable hairy black holes with interesting physical properties. In recent years, many works discussing diverse physical scenarios of the hairy black holes in various asymptotic spaces have appeared, see \cite{Zloshchastiev:2004ny,Torii:1998ir,Torii:2001pg,Winstanley:2002jt,Martinez:2004nb,Martinez:2005di,Martinez:2006an,Hertog:2004dr,Henneaux:2004zi,Henneaux:2006hk,Amsel:2006uf,Kolyvaris:2009pc,
Dias:2011at,Dias:2011tj,Bhattacharyya:2010yg,Basu:2010uz,Anabalon:2012ih,Anabalon:2012tu,Kleihaus:2013tba,Kolyvaris:2011fk,Kolyvaris:2013zfa,Gonzalez:2013aca,Anabalon:2009qt,Charmousis:2009cm,Lemos:1995cm,
Ballon-Bayona:2020xls,Guo:2021ere,Erices:2021uyu,Fahim:2021nly,Karakasis:2021rpn,Rahmani:2020vvv}  for a necessarily biased selection.

Black holes in AdS spaces, in particular, have attracted much attention of late for many reasons. First, the black holes in AdS spaces, as opposed to their asymptotically flat counterparts, are thermodynamically stable with their surroundings and exhibit rich phase structure. In particular, AdS black holes often exhibit critical phenomena akin to phase transitions in familiar liquid gas systems \cite{Hawking:1982dh,Chamblin:1999tk,Chamblin:1999hg,Dey:2015ytd,Mahapatra:2016dae,Cvetic:1999ne,Sahay:2010tx,Sahay:2010wi}. Second, the horizon topology of AdS black holes can be a planar $\mathbb{R}^2$, sphere $\mathbb{S}^2$, or hyperbolic $\mathbb{H}^2$, as compared to asymptotically flat black holes where the horizon topology must be a round sphere $\mathbb{S}^2$ \cite{Friedman:1993ty,Birmingham:1998nr,Brill:1997mf,Mann:1996gj,Lemos:1994xp,Vanzo:1997gw,Cai:1996eg}. This again makes the associated thermodynamic properties more interesting to analyze in AdS spaces \cite{Cai:2004pz,Sheykhi:2007wg,Dehghani:2014caa}. Third, and perhaps the most important reason, is the discovery of the gauge/gravity duality \cite{Maldacena:1997re,Witten:1998qj,Gubser:1998bc}.  The gauge/gravity duality maps a theory of gravity in AdS space to quantum field theory living at the AdS boundary in one less dimension. The duality provides a unique and attractive tool to address important questions related to strong couplings in quantum field theories (which otherwise are difficult to solve) using the classical AdS gravity theory. Indeed, AdS black holes have been used to discuss important questions in field theories such as confinement physics \cite{Witten:1998zw}, quark-gluon plasma \cite{Kovtun:2004de},  non-equilibrium physics \cite{Buchel:2013lla} etc.

Recent progress in the understanding of the gauge/gravity duality advocate for a deeper and broad study of the behaviour of matter fields in the surroundings of charged AdS black holes, which might develop hair  \cite{Gubser:2008px}. In particular, a good understanding of hairy charged AdS black holes could shed new light on our understanding of a number of condensed matter systems from the gauge/gravity duality perspective. Prominent examples include superfluidity and superconductivity (where a charged scalar field condenses and develops a non-zero vacuum expectation value at low temperatures) \cite{Hartnoll:2008vx,Hartnoll:2008kx}, quantum liquids \cite{Iqbal:2011ae}, non-conformal plasmas \cite{Jain:2014vka,Casalderrey-Solana:2011dxg}.

In a similar context, dyonic black holes, that carry both electric and magnetic charges have also appeared abundantly in the literature. Because of electromagnetic duality, it is possible to construct a black hole that carries both electric and magnetic charges in four dimensions. In the dual boundary description, these dyonic black holes correspond to a field theory in ($2+1$) dimensions with a $U(1)$ conserved charge ($q_e$) and in a constant magnetic field background ($q_M$). The presence of magnetic charge not only enriches the thermodynamic phase diagram of AdS black holes in the gravity side but also allows us to introduce a background magnetic field in the dual boundary side, thereby providing an approach to probe physics related to the Hall effect \cite{Hartnoll:2007ai}, ferromagnetism \cite{Dutta:2013dca},  magnetohydrodynamics \cite{Caldarelli:2008ze}, the Nernst effect \cite{Hartnoll:2007ih} etc. See \cite{Goldstein:2010aw,Kundu:2012jn,Caldarelli:2016nni,Donos:2015bxe,Sadeghi:2016dvc,Kruglov:2020aqm,Kruglov:2019ybs,Panahiyan:2018fpb,Hajkhalili:2018thm,Hendi:2016uni,Chaturvedi:2014vpa,Kim:2015wba,
Amoretti:2020mkp,Bhatnagar:2017twr,Lindgren:2015lia,Zhou:2015dha,Khimphun:2017mqb,Bai:2020ezy,Li:2016nll,BravoGaete:2019rci} for other related discussions on dyonic black holes and their holographic applications to field theory.

However, despite their profound importance in the context of holography, the discussion of hairy dyonic black holes is rather limited \cite{Caldarelli:2016nni}. The main reason for this is the difficulty in constructing such solutions analytically, as it requires a simultaneous solution of the Einstein-Maxwell-Scalar system with a non-trivial profile for the scalar and gauge fields. Accordingly, most hairy dyonic black hole solutions have been obtained numerically \cite{Cadoni:2011kv}. Moreover, analytic expressions of conserved charges and thermodynamics observables are often difficult to obtain in these systems. Cases where analytic hairy dyonic solutions have been obtained, are also mostly restricted to the planar case. This, therefore, has hindered our understanding of hairy dyonic black holes. In this paper, we remedy some of these issues and discuss a number of exact analytic dyonic black hole solutions that are simultaneously supported by a scalar field.

More specifically, our aim in this work is to first construct static hairy dyonic black holes and then study their thermodynamic properties. To carry out this objective,  we consider the Einstein-Maxwell-Scalar gravity system and solve the coupled Einstein-Maxwell-Scalar equations of motion simultaneously in terms of a function $A(z)$ (see the next section for details) using the potential reconstruction technique \cite{Mahapatra:2020wym,Dudal:2017max,Dudal:2018ztm,Mahapatra:2018gig,Mahapatra:2019uql,Bohra:2019ebj,He:2013qq,Arefeva:2018hyo,Arefeva:2018cli,Alanen:2009xs,Arefeva:2020byn}. The different forms of $A(z)$ then allow us to construct a different family of hairy dyonic black hole solutions. To make the analysis and results more comprehensive, we choose two particular forms of $A(z)$. These forms of $A(z)$ allow us to introduce a parameter $a$, which controls the strength of the scalar hair. We find that for these forms the obtained gravity solution exhibits desirable features such as the scalar field being regular and the Kretschmann scalar being finite everywhere outside the horizon. Moreover, the hairy dyonic solutions are obtained not only for planar but for spherical and hyperbolic horizon topologies as well. We further obtain conserved charges of these black holes analytically. In particular, we employ two methods: (i) Ashtekar-Magnon-Das (AMD) prescription \cite{Ashtekar:1999jx} and (ii) holographic renormalization method \cite{Balasubramanian:1999re,Papadimitriou:2005ii,Skenderis:2002wp,deHaro:2000vlm} to find the mass of the hairy black hole. We find that the mass expression matches, albeit with some subtleties, from these two methods for the planar black hole. However, because of the presence of additional logarithmic terms in the near boundary expansion of the metric function, the AMD prescription is difficult to implement for the spherical and hyperbolic black holes; therefore, for these black holes, the mass expressions are obtained from the holographic renormalization method only.

We then discuss the thermodynamic properties of these hairy dyonic black holes in both canonical and grand-canonical ensembles. We obtain the Gibbs and Helmholtz free energies analytically and find that the specific heat is always positive for the planar and hyperbolic cases, thereby establishing the local stability of these hairy black holes, in both these ensembles. Moreover, the hairy black holes are not only thermodynamically stable but also thermodynamically favoured. In particular, the free energies of the hairy black holes are lower than the non-hairy black holes at low temperatures. We further analyse the influence of parameters $\{a, q_e, q_M \}$ on the temperature range for which the hairy black holes are thermodynamically favoured and find that they have a constructive effect on the thermodynamic stability of the hairy black hole. We find that $q_M$ can make the free energy of the uncharged hairy black hole smaller than the uncharged non-hairy black hole. This is an important result considering that the free energy of the RN-AdS black hole is generally found to be smaller than the hairy black hole for $q_e=0$. Similarly, for the spherical horizon, like their non-hairy counterpart, we find
Hawking/Page and small/large Van der Waals type phase transitions. Interestingly, with scalar hair, unlike their non-hairy counterpart, the small/large black hole phase transition can appear in the grand-canonical ensemble as well. In the constant charge ensemble, we get the same thermodynamic properties as in \cite{Chamblin:1999tk,Chamblin:1999hg}, with $q_e^2\rightarrow q_e^2 + q_M^2$.

The paper is organized as follows. In the next section, we introduce the gravity model and present its analytic solution corresponding to the topological hairy dyonic black holes. In sections 3 and 4, we study the thermodynamic stability of these hairy black holes for two different $A(z)$ forms. Finally, in Section 5, we summarize our main results and discuss the future directions.

\section{Hairy dyonic black hole solution}
In this section, our main motivation is to construct exact analytic solutions of hairy dyonic black holes. For this purpose, we start with the Einstein-Maxwell-Scalar action,
\begin{eqnarray}
S_{EMS} = \frac{1}{16 \pi G_4} \int_{\mathcal{M}} \mathrm{d^4}x \ \sqrt{-g}  \ \left[R - \frac{f(\phi)}{4}F_{MN}F^{MN} -\frac{1}{2}\partial_{M}\phi \partial^{M}\phi -V(\phi)\right]\,,
\label{actionEF}
\end{eqnarray}
where $R$ is the Ricci scalar of the manifold $\mathcal{M}$, $G_4$ is the four-dimensional Newton constant, $F_{MN}$ is the electromagnetic field strength tensor of $U(1)$ gauge field $B_M$,  $\phi$ is the scalar field, and $V(\phi)$ is the potential of the scalar field $\phi$. The information about the electric and magnetic charges, hence the dyonic properties of the gravity system, lie within the structure of the electromagnetic field strength tensor. The function $f(\phi)$ represents the coupling between scalar and $U(1)$ gauge fields.  \\

The variation of the action (\ref{actionEF}) leads to the following Einstein, Maxwell, and scalar equations of motion,
\begin{eqnarray}
& & R_{MN}- \frac{1}{2}g_{MN} R + \frac{f(\phi)}{4}\biggl(\frac{g_{MN}}{2} F^2 - 2 F_{MP}F_{N}^{\ P}\biggr)    \nonumber \\
& &  + \frac{1}{2} \biggl(\frac{g_{MN}}{2} \partial_{P}\phi \partial^{P}\phi -\partial_{M}\phi \partial_{N}\phi  + g_{MN} V(\phi)  \biggr)  =0 \,,
\label{EinsteinEE}
\end{eqnarray}
\begin{eqnarray}
& & \nabla_{M} \biggl[ f(\phi) F^{MN}  \biggr] = 0\,,
\label{MaxwellEE}
\end{eqnarray}
\begin{eqnarray}
& & \frac{1}{\sqrt{-g}}\partial_{M} \biggl[ \sqrt{-g}  \partial^{M}\phi \biggr] - \frac{F^2}{4} \frac{\partial f(\phi)}{\partial \phi} - \frac{\partial V(\phi)}{\partial \phi} = 0 \,.
\label{dilatonEE}
\end{eqnarray}
Since we want to construct static hairy dyonic black hole solution for various horizon topologies, we consider the following Ans\"atze for the metric $g_{MN}$, gauge field $B_M$, and scalar field $\phi$:
\begin{eqnarray}
& & ds^2=\frac{L^2}{z^2}\biggl[-g(z)dt^2 + \frac{ e^{2A(z)} dz^2}{g(z)} + d\Omega_{\kappa, 2}^2 \biggr]\,, \nonumber \\
& & \phi=\phi(z), \ \ B_{M}=B_{t}(z)\delta_{M}^{t} + q_M \mathcal{X} \,,
\label{ansatz}
\end{eqnarray}
where $L$ is the AdS length scale, and the parameter $\kappa$ indicates the curvature of the two-dimensional metric $d\Omega_{\kappa, 2}^2$. In particular, $\kappa$ can take three different values, $\{-1,0,+1 \}$, corresponding to hyperbolic, planar, and spherical horizon topologies, respectively.

\[
    d\Omega_{\kappa, 2}^2=
\begin{cases}
    dx_1^2 +  \sin^2x_1 dx_2^2, &  \kappa=1 \\
    dx_1^2 +  dx_2^2, &  \kappa=0 \\
   dx_1^2 + \sinh^2x_1 dx_2^2, &  \kappa=-1 \,,
\end{cases}
\]
$B_t$ and $q_M$ contain the information about the electric and magnetic charges. In this work, we keep $q_M$ (or the background magnetic field) fixed, \textit{i.e.}, $q_M$ is considered as a parameter rather than a thermodynamic variable. We will see in the next section that this is a consistent treatment. And $\mathcal{X}$ is
\[
    \mathcal{X} =
\begin{cases}
   \cos x_1, &  \kappa=1 \\
    x_1, &  \kappa=0 \\
   \cosh x_1, &  \kappa=-1 \,.
\end{cases}
\]
As usual, the radial coordinate $z$ runs from $z=0$ (asymptotic boundary) to $z=z_h$ (horizon radius), or to $z=\infty$ for thermal-AdS (without horizon). Sometimes, we will also use the coordinate $r=1/z$.

Substituting the above Ans\"atze into Eq.~(\ref{EinsteinEE}), we get the following three Einstein equations of motion,
\begin{eqnarray}
& & tt: \  \frac{A'(z)}{z} - \frac{g'(z)}{2zg(z)} + \frac{\phi'(z)^2}{8}  + \frac{3}{2z^2}  + \frac{z^2f(z)B_{t}'(z)^2}{8L^2g(z)}  \nonumber \\
& & +  \frac{q_{M}^{2} e^{2A(z)} z^2 f(z)}{8L^2g(z)}  +\frac{e^{2A(z)}L^2V(z)}{4z^2g(z)} - \frac{e^{2A(z)}\kappa}{2g(z)} =0\,,
\label{Einsteintt}
\end{eqnarray}
\begin{eqnarray}
& & zz: \ - \frac{g'(z)}{z} + g(z) \left(\frac{3}{z^2} - \frac{\phi'(z)^2}{4} \right)  + \frac{z^2f(z)B_{t}'(z)^2}{4L^2}   \nonumber \\
& & +  \frac{q_{M}^{2} e^{2A(z)} z^2 f(z)}{4L^2}  +\frac{e^{2A(z)}L^2V(z)}{2z^2} - e^{2A(z)}\kappa=0 \,,
\label{Einsteinzz}
\end{eqnarray}
\begin{eqnarray}
& &  x_{1}x_{1}: \ g''(z) -  g'(z) \left(A'(z)+ \frac{4}{z}\right) + g(z)\left(\frac{6}{z^2} + \frac{4A'(z)}{z} + \frac{\phi'(z)^2}{2}\right)  \nonumber \\
& &  - \frac{z^2f(z)B_{t}'(z)^2}{2L^2} - \frac{q_{M}^{2} e^{2A(z)} z^2 f(z)}{2L^2} + \frac{e^{2A(z)}L^2V(z)}{z^2} =0 \,.
\label{Einsteinxixi}
\end{eqnarray}
Importantly, one can rearrange these Einstein equations and put them into the following simpler forms, which are then easier to solve
\begin{eqnarray}
g''(z) - g'(z) \left(A'(z)+\frac{2}{z}\right) - \frac{z^2f(z)B_{t}'(z)^2}{L^2}  - \frac{q_{M}^{2} e^{2A(z)} z^2 f(z)}{L^2} + 2 e^{2A(z)} \kappa = 0\,,
\label{EOM11}
\end{eqnarray}
\begin{eqnarray}
\phi'(z)^2 + \frac{4 A'(z)}{z} = 0 \,,
\label{EOM22}
\end{eqnarray}
\begin{eqnarray}
\frac{g''(z)}{4g(z)} + A'(z)\left(\frac{1}{z}- \frac{g'(z)}{4g(z)} \right) - \frac{3 g'(z)}{2 z g(z)}
  +\frac{e^{2A(z)}L^2V(z)}{2z^2g(z)} +\frac{3}{z^2} -\frac{ e^{2A(z)} \kappa}{2g(z)}   = 0 \,.
\label{EOM33}
\end{eqnarray}
Similarly, one gets the following equation of motion for the scalar field,
\begin{eqnarray}
 \phi ''(z) +\phi '(z) \left(\frac{g'(z)}{g(z)}- A'(z)-\frac{2}{z}\right) + \frac{z^2 B_{t}'(z)^2}{2 L^2 g(z)}\frac{\partial f(\phi)}{\partial \phi}  \nonumber \\
  - \frac{q_{M}^{2} e^{2A(z)} z^2}{2 L^2 g(z)}\frac{\partial f(\phi)}{\partial \phi}  - \frac{L^2 e^{2A(z)}}{z^2 g(z)} \frac{\partial V(\phi)}{\partial \phi} =0\,,
\label{dilatonEOM}
\end{eqnarray}
and for the gauge field,
\begin{eqnarray}
B_{t}''(z)+ B_{t}'(z) \left(\frac{f'(z)}{f(z)} - A'(z)\right) =0\,.
\label{MaxwellAtEOM}
\end{eqnarray}
Accordingly, there are a total of five equations. However, only four of them are independent. Here, we consider the scalar equation (\ref{dilatonEOM}) as a constrained equation and take the remaining equations as independent. To solve these equations, we impose the following boundary conditions:
\begin{eqnarray}
&& g(0)=1 \ \ \text{and} \ \ g(z_h)=0, \nonumber \\
&& B_{t}(0)= \mu_e \ \ \text{and} \ \  B_{t}(z_h)=0, \nonumber \\
&& A(0) = 0 \,.
\label{boundaryconditions}
\end{eqnarray}
These boundary conditions are chosen to ensure that the spacetime asymptotes to AdS at the boundary $z \rightarrow 0$. The parameter $\mu_e$ is the leading term of the near boundary expansion of the gauge field $B_t(z)$, and corresponds to the chemical potential of the theory. Using Gauss's theorem, one can also find a relation between $\mu_e$ and the electric charge of the black hole (see the discussion below). Apart from these boundary conditions, we further demand that the scalar field $\phi$ remains real everywhere in the bulk and goes to zero at the asymptotic boundary  $\phi(0)=0$.\\

Interestingly, one can find a complete closed form solution of the Einstein-Maxwell-Scalar equations (\ref{EOM11})-(\ref{MaxwellAtEOM}) in terms of two unknown functions $A(z)$ and $f(z)$ by the following approach:

\begin{itemize}
\item Solve Eq.~(\ref{EOM22}) and find $\phi'(z)$ in terms of $A(z)$.

\item Solve Eq.~(\ref{MaxwellAtEOM}) and find $B_t(z)$ in terms of $A(z)$ and $f(z)$.

\item From the obtained $B_t(z)$ solution, solve Eq.~(\ref{EOM11}) and find $g(z)$ in terms of $A(z)$ and $f(z)$.

\item Lastly, we solve Eq.~(\ref{EOM33}) and obtain $V$ in terms of $A(z)$ and $g(z)$.
\end{itemize}
Using this approach, we get the solution for $B_t(z)$ as
\begin{eqnarray}
B_{t} (z) = C_{1} \int_0^z \, d\xi \frac{e^{A(\xi)}}{f(\xi)} + C_2 \,,
\label{Atsol}
\end{eqnarray}
where the integration constants $C_1$ and $C_2$ can be found from the boundary conditions [Eq.(\ref{boundaryconditions})] as
\begin{eqnarray}
C_2 = \mu_e,  \ \ \ \ \ C_1 = -\frac{\mu_e}{ \int_0^{z_h} \, d\xi \frac{e^{A(\xi)}}{f(\xi)} }  \,.
\end{eqnarray}
The solution for $B_{t}(z)$ then becomes
\begin{eqnarray}
B_{t} (z) = \mu_e \biggl[ 1 - \frac{\int_0^z \, d\xi \frac{e^{A(\xi)}}{f(\xi)}}{\int_{0}^{z_h} \, d\xi \frac{e^{A(\xi)}}{f(\xi)}} \biggr] = \tilde{\mu}_e \int_z^{z_h} \, d\xi \frac{e^{A(\xi)}}{f(\xi)}  \,.
 \label{Atsol}
\end{eqnarray}
Now, using Eq.~(\ref{Atsol}) into Eq.~(\ref{EOM11}), we get the following solution for $g(z)$,
\begin{eqnarray}
& & g(z) =  C_4 + \int_0^z \, d\xi \ e^{A(\xi)} \xi^{2} \biggl[ C_{3} + \mathcal{K}(\xi) \biggr] \,,
\label{gsol}
\end{eqnarray}
where,
\begin{eqnarray}
& & \mathcal{K}(\xi)= \int \, d\xi \ \biggl[ \frac{\tilde{\mu}_e^2 e^{A(\xi)}}{L^2 f(\xi)} + \frac{q_{M}^{2} e^{A(\xi)} f(\xi) }{L^2} -2\kappa \frac{e^{A(\xi)}}{\xi^2} \biggr] \,,
\label{gsol}
\end{eqnarray}
$C_3$ and $C_4$ being the integration constants that can be again obtained from Eq.~(\ref{boundaryconditions}),
\begin{eqnarray}
C_4 = 1,  \ \ \ \ \ C_3 =- \frac{1+ \int_0^{z_h} \, d\xi e^{A(\xi)} \xi^{2} \mathcal{K}(\xi) }{ \int_0^{z_h} \, d\xi e^{A(\xi)} \xi^{2} }  \,.
\end{eqnarray}
Similarly, the scalar field $\phi$ can be solved in terms of $A(z)$ from Eq.~(\ref{EOM22})
\begin{eqnarray}
\phi(z) = \int \, dz \ 2 \sqrt{\frac{ - A'(z)}{z}} + C_{5} \,,
\label{phisol}
\end{eqnarray}
where $C_{5}$ can be obtained by demanding $\phi$ to vanishes near the asymptotic boundary, \textit{i.e.}, $\phi |_{z=0}\rightarrow 0$. Lastly, the potential $V$ can be found from Eq. (\ref{EOM33}),
\begin{eqnarray}
V(z) =  - \frac{z^2 e^{-2 A(z)} g''(z)}{2 L^2} + \frac{g'(z) e^{-2 A(z)}}{L^2} \left(\frac{z^2 A'(z)}{2}+ 3 z \right)  \nonumber \\
- \frac{g(z) e^{-2 A(z)}}{L^2} \left(2 z  A'(z) + 6 \right) + \kappa \frac{z^2}{L^2} \,.
\label{Vsol}
\end{eqnarray}
It is therefore clear that we can obtain a closed form analytic solution of the Einstein-Maxwell-Scalar gravity system of Eq.~(\ref{actionEF}) in terms of two arbitrary functions, \textit{i.e.}, $A(z)$ and $f(z)$, and construct a dyonic hairy black hole solution with various horizon topologies. The different forms of these functions $A(z)$ and $f(z)$  will however correspond to different $V(z)$, \textit{i.e.}, different $A(z)$ and $f(z)$ will attribute to different dyonic hairy black hole solutions.
Therefore, we can construct a large family of physically allowed dyonic hairy black hole solutions for the Einstein-Maxwell-Scalar gravity system of Eq.~(\ref{actionEF}) by choosing different forms of $A(z)$ and $f(z)$.

Nonetheless, in the context of AdS/CFT correspondence, these functions are usually fixed by taking inputs from the dual boundary theory. In particular, depending upon what kind of boundary physics one is interested in one usually consider different forms of these functions. For example, in holographic QCD, the expressions of $f(z)$ and $A(z)$ are usually determined by demanding the dual QCD theory to exhibit physical QCD properties such as the linear Regge trajectory for meson mass spectrum, confinement/deconfinement phase transition etc. \cite{Dudal:2017max,Dudal:2018ztm} \footnote{In holographic QCD model context, unlike in our case, the function $A(z)$ generally appears as an overall conformal factor in the spacetime metric \cite{Dudal:2017max,Dudal:2018ztm}.}.

One can also take a more pragmatic approach and consider various different forms of $A(z)$ and $f(z)$ to thoroughly investigate the effects of scalar hair and make a comprehensive argument of the stability and thermodynamics of the hairy dyonic black holes, without worrying too much about the dual boundary theory. Here, we take such an approach. Particularly, we consider two different forms of $A(z)$: (i) $A(z)=-\log(1+a z)$ and (ii) $A(z)=-a z$.   These forms of $A(z)$ are chosen not just for their simplicity but also to have better control over the integrals that appear in Eqs.~(\ref{Atsol})-(\ref{Vsol}), again without bothering greatly about their dual boundary theory \footnote{In principle, one can consider other forms, such as $A(z)=-az^n$ with $n>1$, as well. However, we will not dwell into such forms here.}. With these forms of $A(z)$, one can see from Eq.~(\ref{phisol}) that the strength of the scalar hair is characterised by the parameter $a$. Therefore, when $a$ vanishes so does the scalar field. Hence, as desired, in the limit $a\rightarrow 0$, we get back to the RN-AdS solution. Similarly, we can consider different forms of $f(z)$. Here, we mostly concentrate on the case $f(z)=1$, corresponding to no direct coupling between the scalar and gauge field. It is also possible to take other useful couplings, such as the non-minimal linear $f(\phi)\propto \phi$ and exponential $f(\phi)\propto e^{-\phi}$ couplings, which have been considered in the literature. A detailed discussion with these non-minimal coupling functions will appear in a companion paper.

Another reason for considering the above-mentioned $f(z)$ and $A(z)$ forms is that it makes sure our constructed hairy spacetime asymptotes to AdS at the boundary $z\rightarrow 0$. In particular, near the  boundary, we have
\begin{eqnarray}
& & V(z)|_{z\rightarrow 0} = -\frac{6}{L^2} + \frac{m^2\phi^2}{2}+\dots \nonumber\\
 & &  V(z)|_{z\rightarrow 0} =  2\Lambda + \frac{m^2\phi^2}{2}+\dots   \,.
\label{Vsolexp}
\end{eqnarray}
where $\Lambda=-\frac{3}{L^2}$ is the negative cosmological constant in four dimensions. Furthermore, $m^2=-5/4$ is the mass of the scalar field, satisfying the Breitenlohner-Freedman bound for stability in AdS space \textit{i.e}, $m^2\geq-9/4$ \cite{Breitenlohner:1982jf}. Together with the fact that $g(z)|_{z\rightarrow 0}=1$, it indeed makes sure that the constructed geometry asymptotes to AdS at the boundary.\\

Apart from the above obtained black hole solution, there also exists a second solution having no horizon. This corresponds to a thermal-AdS solution and it can be obtained by taking the limit $z_h\rightarrow \infty$ in the black hole solution given above \footnote{Here, we are referring this without horizon solution as thermal-AdS for simplicity even though the curvature is not constant throughout the spacetime.}. Depending upon the form of $A(z)$, the thermal-AdS solution can have a non-trivial structure in the bulk spacetime, however, it will always go to AdS asymptotically. Interestingly, as we will see shortly, there can be a Hawking-Page type thermal-AdS/black hole phase transition between these two solutions. \\

It is imperative to emphasize once again that Eqs.~(\ref{Atsol})-(\ref{Vsol}) are a consistent solution to the Einstein-Maxwell-Scalar action for any $A(z)$ and $f(z)$. Correspondingly, depending upon the forms of $A(z)$ and $f(z)$, a hairy dyonic black hole solution with various horizon topologies can be constructed analytically. However, it is important to note that different forms of $A(z)$ and $f(z)$ will correspond to  different potentials $V(z)$. Therefore, by choosing different forms of $A(z)$ and $f(z)$, one is actually constructing a different family of hairy dyonic black holes, as different forms of $A(z)$ and $f(z)$ will give different potentials $V(z)$. However, once the forms of $A(z)$ and $f(z)$ are fixed/chosen then the form of $V(z)$ also is, and in return Eqs.~(\ref{Atsol})-(\ref{Vsol}) correspond to a self-consistent solution to a particular action with predetermined $A(z)$, $f(z)$, $V(z)$, and then there is no ambiguity in the solution itself \footnote{Although the functions $A(z)$ and $f(z)$ seem to be arbitrary in our construction, however, we need to be careful in choosing their forms. In particular, we have to make sure that all the fields in the gravity systems such as scalar field, gauge field, metric function remain real throughout the bulk. This puts a lot of constraint on the forms of $A(z)$ and $f(z)$ we can choose. For example, the scalar field would be complex if $A(z)=a z$, with $a$ being positive, is considered. Similarly, we can not choose forms like $A(z)=a/z^n$, with $n$ being positive, as this will make the asymptotic boundary different from AdS.}. \\

To further establish the consistency of the gravity model, we moreover check validity of the null energy condition (NEC) in our model. The NEC can be expressed as
\begin{eqnarray}
T_{MN}\mathcal{N}^M \mathcal{N}^N \geqslant 0 \,,
\label{NEC}
\end{eqnarray}
where $T_{MN}$ is the energy-momentum tensor of the matter fields. The null vector $\mathcal{N}^{M}$ satisfies the condition $g_{MN}\mathcal{N}^M \mathcal{N}^N=0$, and can be chosen as
\begin{eqnarray}
\mathcal{N}^M= \frac{1}{\sqrt{g(z)}}\mathcal{N}^{t} + \frac{\cos{\alpha}\sqrt{g(z)}}{e^{A(z)}} + \mathcal{N}^{z} + \frac{\sin{\alpha}}{\sqrt{2}}\mathcal{N}^{x_1}+\frac{\sin{\alpha}}{\sqrt{2 g_{x_2 x_2}}}\mathcal{N}^{x_2}  \,,
\label{nullvector}
\end{eqnarray}
for arbitrary parameter $\alpha$. The NEC then becomes
\begin{eqnarray}
e^{-2A(z)} \left[z^2 f(z) \sin^2{\alpha} \left(B_{t}'(z) + q_{M}^2 e^{2A(z)}  \right)   \right] + e^{-2A(z)} \cos^2{\alpha} g(z) \phi'(z)^2 \geq 0\,,
\label{nullvector}
\end{eqnarray}
which is always satisfied in our model for the positive and real gauge kinetic function $f(z)$ and scalar field $\phi(z)$. \\

We can also write down the expressions of various thermodynamic observables associated with the constructed hairy black holes. This will be useful in the discussion of black hole thermodynamics. The temperature and entropy of the black hole are given by
\begin{eqnarray}
& & T= \frac{z_{h}^{2}}{ 4 \pi } \biggl[-\mathcal{K}(z_h) +  \frac{1+ \int_0^{z_h} \, d\xi e^{A(\xi)} \xi^{2} \mathcal{K}(\xi) }{ \int_0^{z_h} \, d\xi e^{A(\xi)} \xi^{2} }  \biggr] \, , \nonumber\\
& &  S_{BH}=\frac{L^{2}\Omega_{2,\kappa}}{4 G_4 z_{h}^{2}} \,,
\label{STexp}
\end{eqnarray}
where $\Omega_{2,\kappa}$ is the unit volume of the boundary space constant hypersurface. To find the charge of the black hole, notice from Eq.~(\ref{MaxwellEE}) that
\begin{eqnarray}
 \left(\sqrt{-g}f(\phi)F^{z t}\right) = q_e \,,
 \label{qeqn}
\end{eqnarray}
where $q_e$ is a $z$-independent constant related to the electric charge. By measuring the flux of the electric field at the boundary, the electric charge can be computed as
\begin{eqnarray}
Q_e=\frac{1}{16 \pi G_4} \int  f(\phi)F_{\alpha\beta}u^\alpha n^\beta \ d\Omega_{2,\kappa}\,
\end{eqnarray}
where $u^\alpha$ and $n^\beta$ are the unit spacelike and timelike normals, respectively, to the constant radial surface
\begin{eqnarray}
& & u^\alpha = \frac{1}{\sqrt{-g_{tt}}} dt =\frac{z}{\sqrt{g(z)}} dt \,, \nonumber\\
& & n^\beta = \frac{1}{\sqrt{g_{zz}}}dz=\frac{z \sqrt{g(z)}}{e^{A(z)}} dz \,,
\end{eqnarray}
this leads to the charge as
\begin{eqnarray}
Q_e=\frac{q_e \Omega_{2,\kappa}}{16 \pi G_4} \,.
\end{eqnarray}
We can further find an explicit relation between $Q_e$ and the corresponding conjugate chemical potential $\mu_e$. Substituting the $B_{t}(z)$ solution from Eq.~(\ref{Atsol}) into Eq.~(\ref{qeqn}), we get
\begin{eqnarray}
& & B_{t}'(z)= - \frac{q_e e^{A(z)}}{f(z)} \,, \nonumber\\
& & \tilde{\mu}_e = \frac{\mu_e}{\int_0^{z_h} \, d\xi \frac{e^{A(\xi)}}{f(\xi)}} = q_e \,.
\end{eqnarray}
The explicit relation between $\mu_e$ and $Q_e$ depends on the forms of $A(z)$ and $f(z)$, and therefore is model dependent.

To discuss the thermodynamic stability of the constructed hairy dyonic black hole solutions, we also need to examine the notion of an energy function, or ``mass'' for the black hole solutions. However, there are various alternative definitions for calculating conserved charges in AdS spaces. This includes the conformal Weyl procedure of Ashtekar, Magnon and Das (AMD) \cite{Ashtekar:1999jx}, and the holographic renormalize stress tensor procedure \cite{Balasubramanian:1999re,Papadimitriou:2005ii,Skenderis:2002wp,deHaro:2000vlm} \footnote{One can also evaluate the mass using the Hamiltonian formulism \cite{Henneaux:2006hk}. However, we will not concentrate on this method here.}. Moreover, these definitions do not always agree with each other especially for gravity systems containing additional matter fields, unless desirable boundary conditions are met for the matter fields at the asymptotic boundary \cite{Henneaux:2006hk} (see the next section for more details). In this work, we will compute the black hole mass using these two definitions and explicitly evaluate the effect of scalar hair on the black hole mass. This computation will not only help us to make a better comparison between these methods but also help us to further understand the differences between them. We find that even for the complicated Einstein-Maxwell-Scalar like gravity system one can obtain an analytic expression of mass from both these definitions. However, since these definitions explicitly depend on the near asymptotic structure of gravity and matter fields, it is difficult to write down the mass expression for different horizon topologies in full generality. For this reason, we have to investigate the mass, and therefore the thermodynamics, on a case-by-case basis.

\section{Black hole thermodynamics with $A(z)=-\log(1+az)$}
In this section, we first examine the stability and thermodynamics of the dyonic hairy gravity solution for the case $A(z)=-\log(1+az)$. The discussion for $A(z)=-az$ is postponed to the next section.

\subsection{Planar horizon: $\kappa=0$}
Let us first discuss the results with planar horizon,  corresponding to $\kappa=0$. With $A(z)=-\log(1+az)$, the solutions of $\phi(z)$ and $B_t(z)$ reduce to
\begin{eqnarray}
& & \phi(z) = 4 \sinh(\sqrt{az}) \,, \nonumber\\
& & B_{t}(z)= \mu_e \left(1- \frac{\log(1+ a z)}{\log(1+ a z_h)}    \right) \,.
\label{Atsolk0case1}
\end{eqnarray}
This also gives us the relation
\begin{eqnarray}
& & \tilde{\mu}_e = \frac{a \mu_e}{\log (1+ a z_h)} = q_e \,.
\end{eqnarray}
Similarly, we get the following solution for $g(z)$,
\begin{eqnarray}
 g(z) =1 + \frac{1}{4 a^4} \left(q_M^2 + \frac{\mu_e^2 a^2}{ \log^2(1+ a z_h)} \right) \Bigl(\frac{a z (2-a z)-2 \log (a z+1)}{a z_h (a z_h-2)+2 \log (az_h+1)} \Bigr) \times \nonumber\\
\Bigl( a z_h (6-a z_h)+2 \log (a z_h+1) ((a z_h-3) + (a z_h+1)+\log (a z_h+1)) \Bigr) \nonumber\\
+ \frac{-a z (a z-2)-2 \log (a z+1)}{a z_h (az_h-2)+2 \log (a z_h+1)} + \frac{1}{4 a^4}\left(q_M^2 + \frac{\mu_e^2 a^2}{\log^2(1+ a z_h)} \right) \times \nonumber\\
\Bigl(a z (6-az)+2 \log (a z+1) ((a z-3) (a z+1)+\log (a z+1)) \Bigr)
   \label{gsolk0case1}
\end{eqnarray}

We can similarly write down the analytic expression of $V(z)$. However, it is too complicated and lengthy, and at the same time not very illuminating; therefore, we skip to reproduce it here
for brevity. Notice that in the limit $a\rightarrow0$, the scalar field goes to zero and all other expressions reduce to the standard non-hairy dyonic expressions.

\begin{figure}[ht]
	\subfigure[]{
		\includegraphics[scale=0.4]{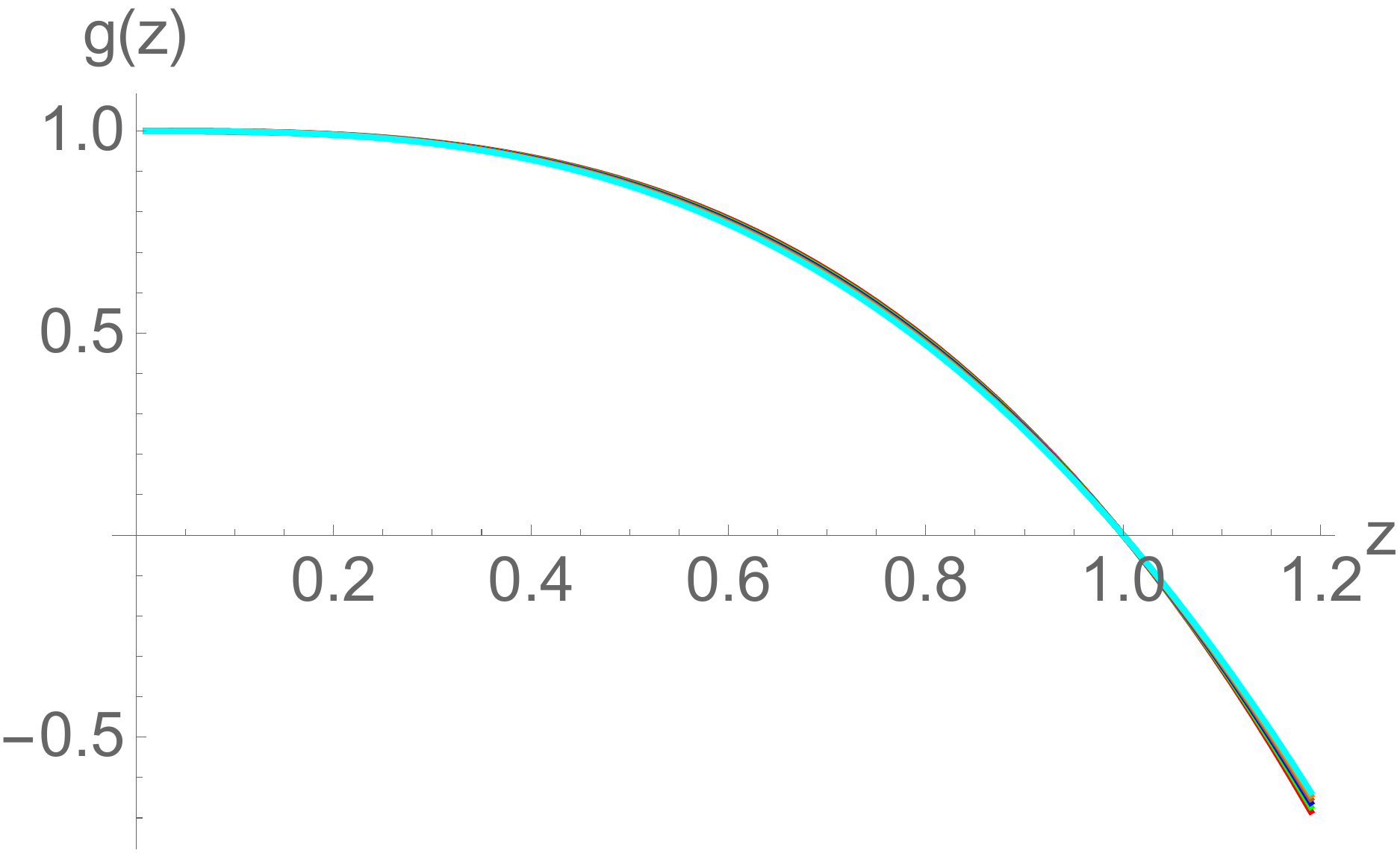}
	}
	\subfigure[]{
		\includegraphics[scale=0.4]{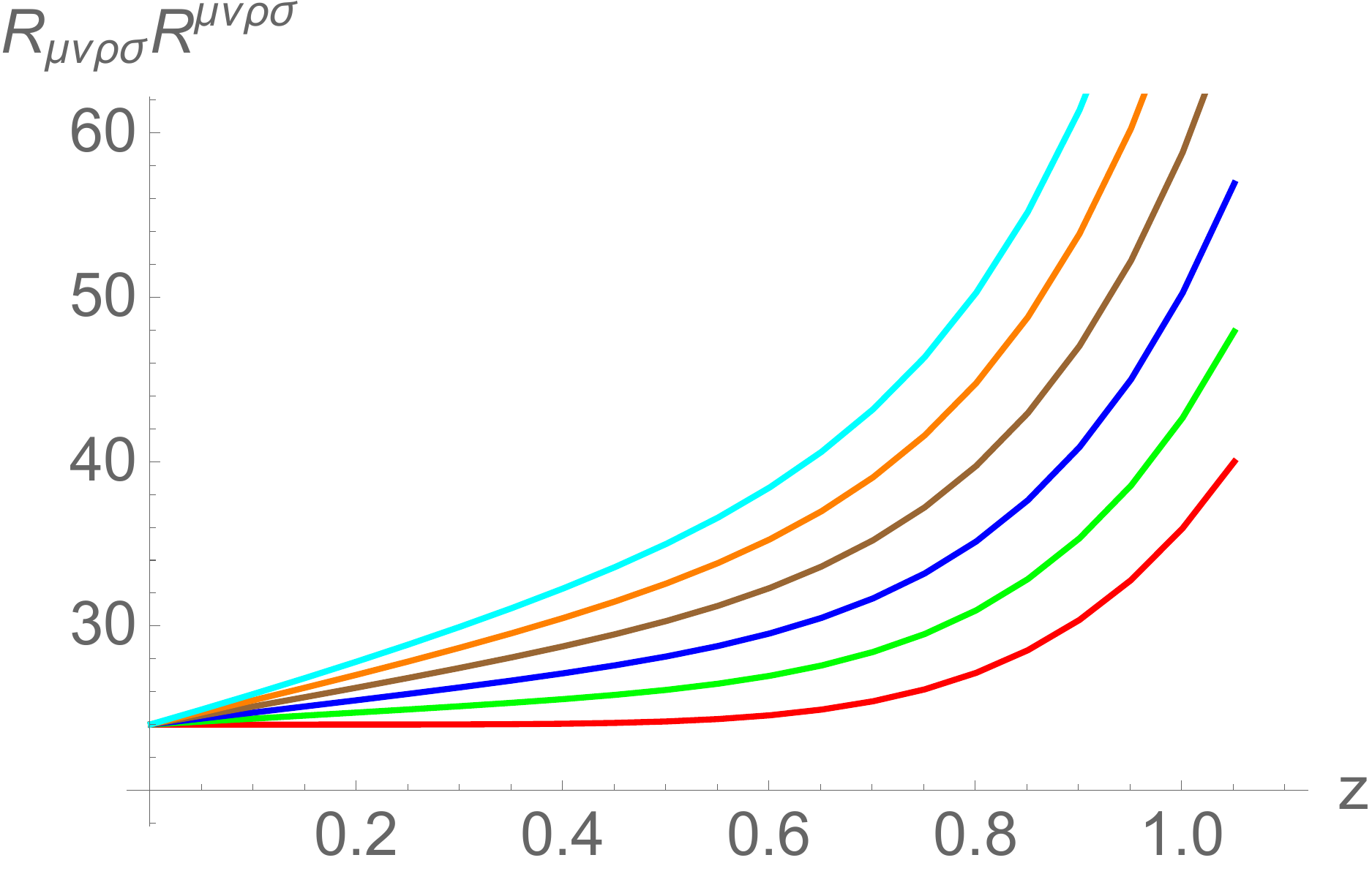}
	}
	\subfigure[]{
		\includegraphics[scale=0.4]{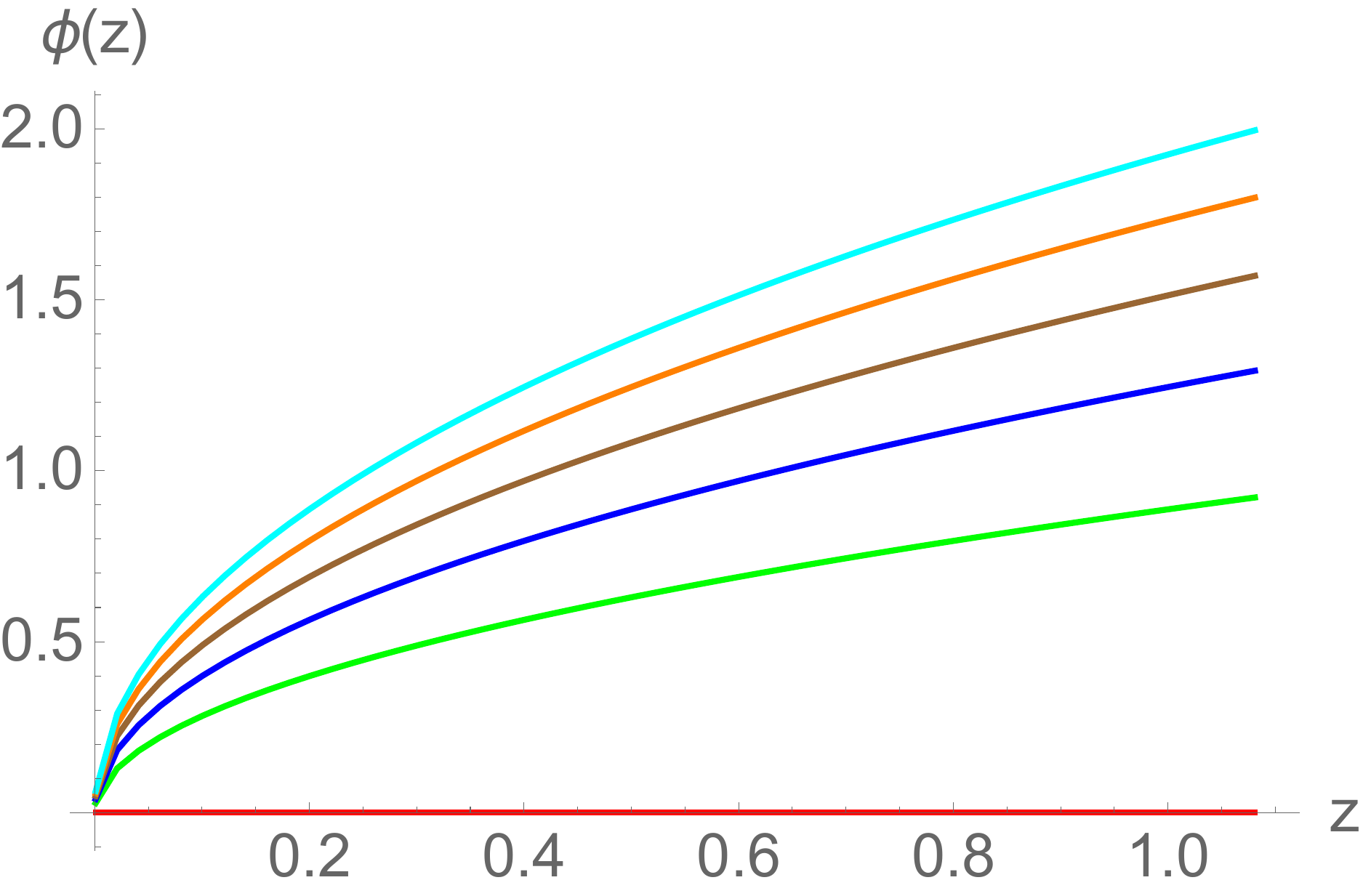}
	}
\subfigure[]{
		\includegraphics[scale=0.4]{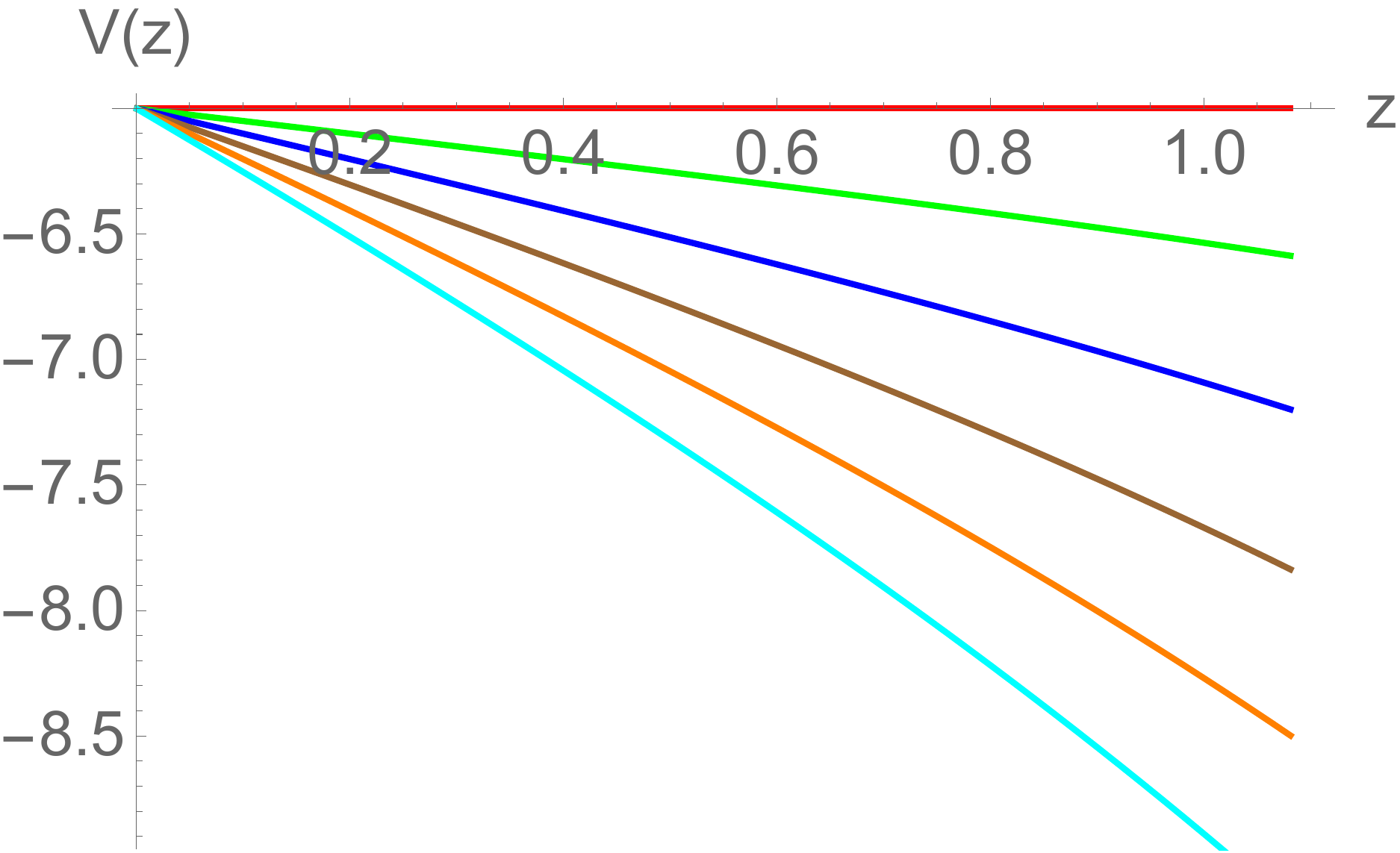}
	}
	\caption{\small The behavior of $g(z)$, $R_{MNPQ}R^{MNPQ}$, $\phi(z)$ and $V(z)$ for different values of hair parameter $a$. Here $z_h=1$, $\mu_e=0.1$, $\kappa=0$, and $q_M=0.1$ are used. Red, green, blue, brown, orange, and cyan curves correspond to $a=0$, $0.05$, $0.10$, $0.15$, $0.20$, and $0.25$ respectively.}
	\label{zvsgvsamuPt1qMPt1zh1f1planarcase1}
\end{figure}

In Fig.~\ref{zvsgvsamuPt1qMPt1zh1f1planarcase1}, the behaviour of these functions for different values of hair parameter $a$ is illustrated. The results here are illustrated for a particular value of $z_h=1$, $\mu_e=1$, and $q_M=1$; however, analogous results occur for their other values as well. The blackening function $g(z)$ changes sign at $z=z_h$ for all values of $a$, indicating the presence of a horizon. The finiteness of the Kretschmann scalar $R_{MNPQ}R^{RNPQ}$ outside the horizon further indicates the non-singular nature of the bulk spacetime. Similarly, the scalar field is regular and real everywhere in the exterior horizon region. Notice from Eq.~(\ref{Atsolk0case1}) that $\phi$ goes to zero only at the asymptotic boundary, implying the existence of a well-behaved planar dyonic hairy black hole solution in our model. We will discuss the thermodynamic stability of this hairy dyonic black hole against the non-hairy dyonic black hole shortly when we will examine their free energies. Similarly, the potential is also regular everywhere in the exterior horizon region, and it asymptotes to $V(z=0)=-6$ at the AdS boundary for all $a$. Additionally, provided that $\mu_e$ and $q_M$ are not too large, the potential is also bounded from above by its boundary value,\textit{ i.e.}, $V(0)\geq V(z)$, thereby satisfying the Gubser criterion to have a well-defined dual boundary theory \cite{Gubser:2000nd}. However, for higher values of $\mu_e \gtrsim 2$ and $q_M \gtrsim 2$, the Gubser criterion can be violated.

\begin{figure}[h!]
\begin{minipage}[b]{0.5\linewidth}
\centering
\includegraphics[width=2.8in,height=2.3in]{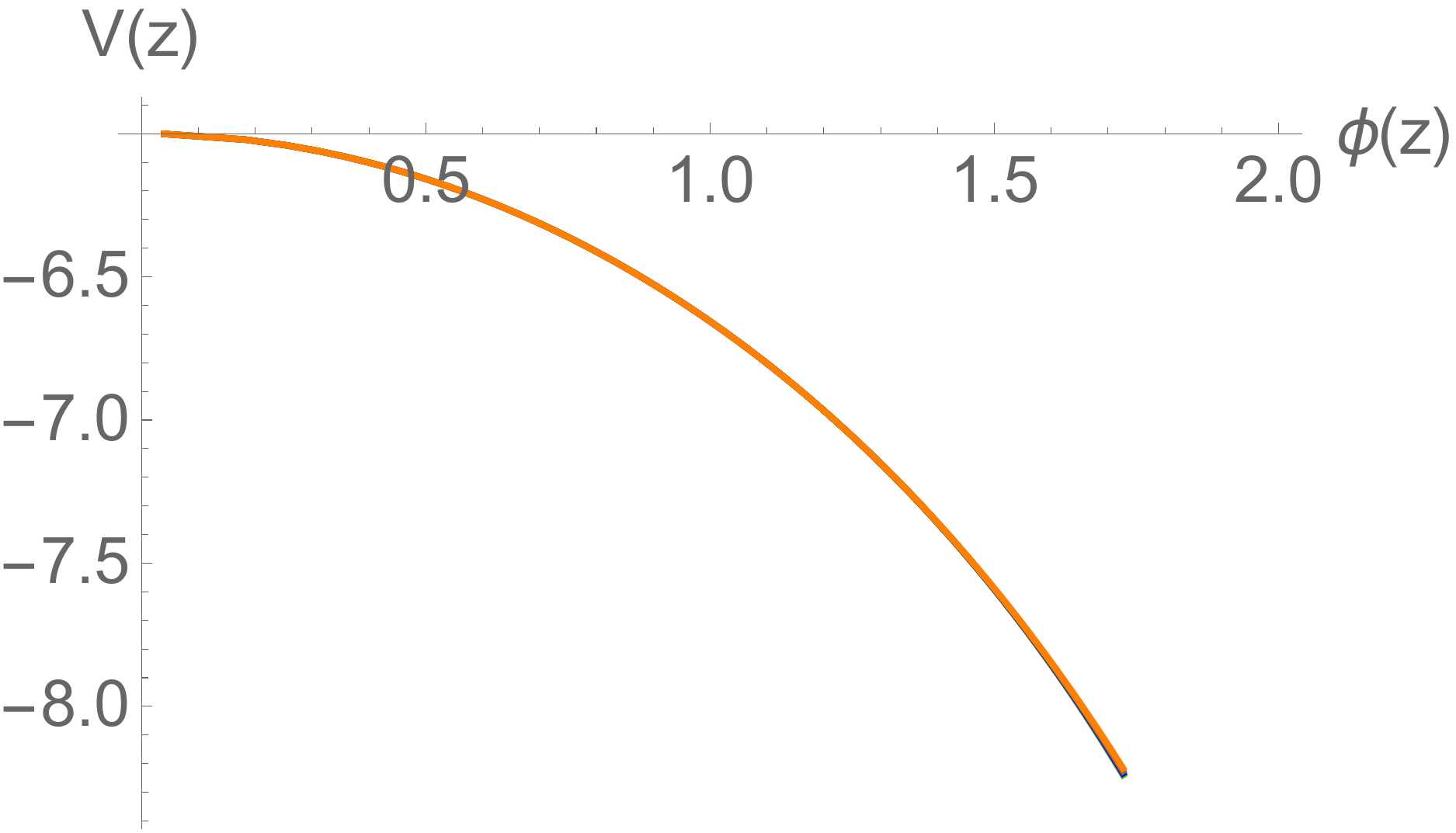}
\caption{ \small  Scalar potential as a function of scalar field for various values of $\mu_e$.  Here $a=0.1$, $z_h=2.0$, and $q_M=0.1$ are used. Red, green, blue, brown, and orange curves correspond to $\mu_e=0$, $0.1$, $0.2$, $0.3$, and $0.4$ respectively. }
\label{PhivsVvsmuaPt1qMPt1zh2f1case1}
\end{minipage}
\hspace{0.4cm}
\begin{minipage}[b]{0.5\linewidth}
\centering
\includegraphics[width=2.8in,height=2.3in]{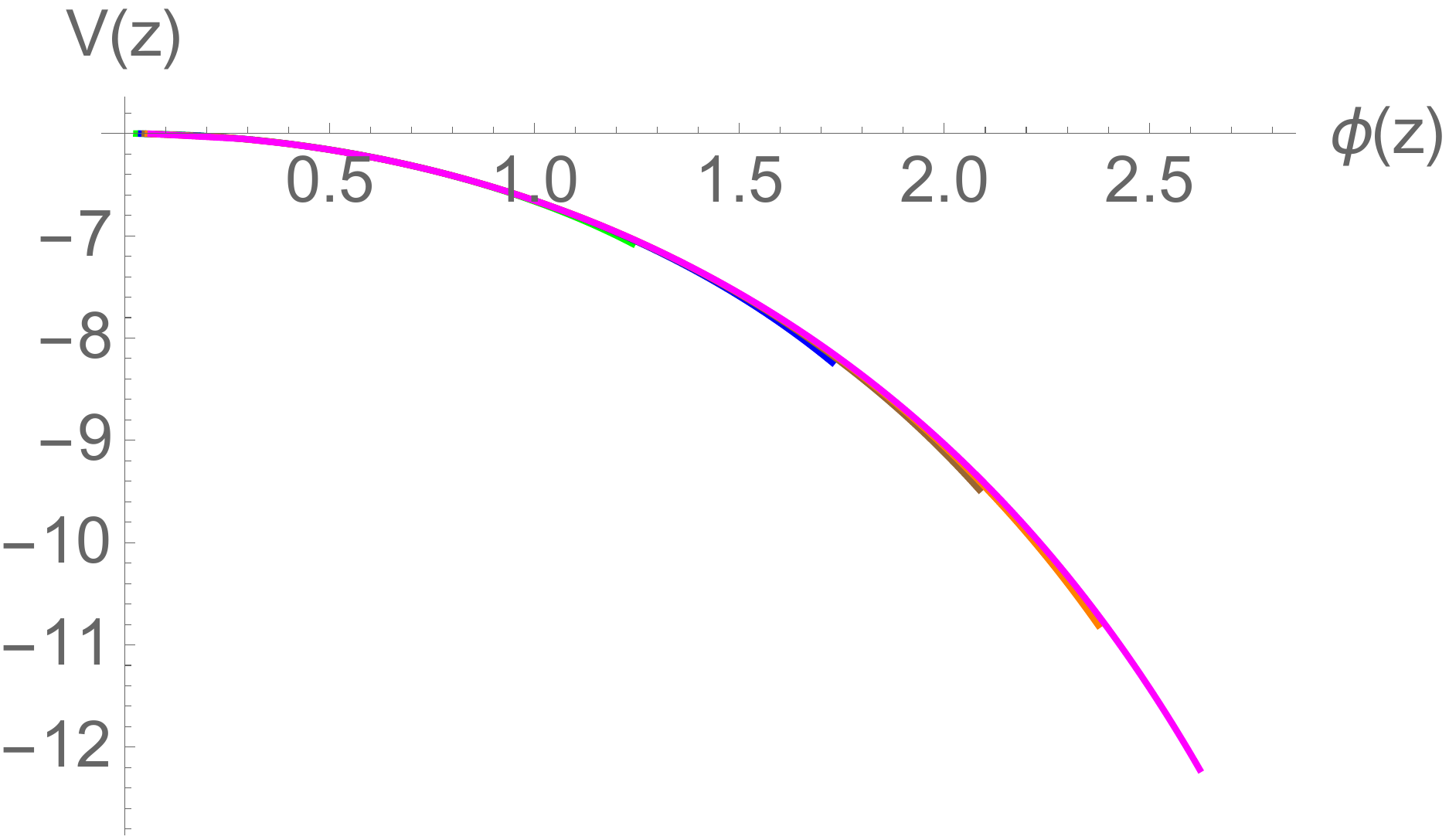}
\caption{\small  Scalar potential as a function of scalar field for various values of $a$.  Here $\mu=0.1$, $q_M=0.1$, and $z_h=2.0$ are used. Green, blue, brown, orange, and magenta curves correspond to $a=0.05$, $0.10$, $0.15$, $0.20$ and $0.25$ respectively. }
\label{PhivsVvsamuPt1qMPt1zh2f1case1}
\end{minipage}
\end{figure}

We can further analyse the behaviour of $V(z)$ with respect to $\phi(z)$. We find that $\phi$ vs $V$ profile for different values of $a$ and $z_h$ almost overlap with each other, demonstrating the independence of the potential on these parameters (which it should be). This is shown in Figures~\ref{PhivsVvsmuaPt1qMPt1zh2f1case1} and \ref{PhivsVvsamuPt1qMPt1zh2f1case1}. The $V(z)$ profile for different values of $\mu_e$ and $q_M$ also overall with each other, provided again that they are not too large. This is actually a consequence of the above-mentioned violation of the Gubser criterion. In particular, the parameter values for which the Gubser criterion is violated also lead to different and un-physical $\phi$ vs $V$ behavior. In the rest of the work, we will concentrate on only those parameter values for which the Gubser criterion is respected.

\subsubsection{Mass from AMD prescription}
In this subsection, we calculate the mass of the hairy black hole using the AMD prescription \cite{Ashtekar:1999jx}. This prescription is particularly useful as it can be straightforwardly applied. The essential idea here is to evaluate the electric part of the Weyl tensor of conformally recalled metric $\tilde{ds^2}=\omega^2 ds^2$, with $\omega$ having a zero of order one at infinity. The conserved quantity $C[K]$ associated
with a Killing field $K$ is then given by
\begin{equation}
\mathcal{C}[K]=\frac{1}{8 \pi G_{4}} \oint \tilde{\epsilon}^{\mu}_{ \ \nu}  K^{\nu} d\tilde{\varSigma}_{\mu} \,,
\label{charge}
\end{equation}
where $\tilde{\epsilon}^{\mu}_{ \ \nu}=\omega^{-1} \tilde{n}^\rho \tilde{n}^\sigma \tilde{C}^{\mu}_{\ \ \rho \nu \sigma}$, $\omega=z$ and $\tilde{n}^\rho$ is the unit normal vector to the constant $\omega$ surface. $\tilde{C}^{\mu}_{\ \ \rho \nu \sigma}$ is the Weyl tensor constructed from $\tilde{ds^2}$ and $d\tilde{\Sigma}^{\mu}$ is the two-dimensional area element of the space section of the AdS boundary. For a
timelike killing vector $K$, we have the mass expression
\begin{equation}
\mathcal{C}[K]=M_{AMD}=\frac{\Omega_{2,\kappa}}{8 \pi G_{4}} \omega^{-1} (\tilde{n}^{\omega})^{2}\tilde{C}^{t}_{\ \ \omega t \omega} \,,
\end{equation}
here all the functions are meant to be calculated with the metric $\tilde{ds^2}$. Using $\tilde{n}^{\mu}=\sqrt{g(\omega)e^{-2A(\omega)}}$, substituting the expression of $\tilde{C}^{t}_{\ \ \omega t \omega}$, switching back to $r=1/\omega$ coordinate and simplifying, we get
\begin{eqnarray}
M_{AMD} &=& \frac{\Omega_{2,\kappa}}{8 \pi G_{4}} \left( \frac{e^{-2 A(r)} r^4}{3} \left[-g'(r) + \frac{r A'(r)g'(r)}{2} -  \frac{r g''(r)}{2}\right] + \kappa\frac{r}{3} \right)\,,
\end{eqnarray}
Substituting $A(r)$ and $g(r)$ expressions of the planar hairy black hole, we finally get
\begin{eqnarray}
M_{AMD}= \frac{\Omega_{2,0}}{12 \pi G_{4}}\frac{a^3 \left(\frac{\left(q_e^2+q_M^2\right) \left(2 \left(a^2 z_h^2-2 a z_h-3\right) \log \left(a
   z_h+1\right)+a z_h \left(6-a z_h\right)+2 \log ^2\left(a z_h+1\right)\right)}{4 a^4}+1\right)}{a
   z_h \left(a z_h-2\right)+2 \log \left(a z_h+1\right)}
\end{eqnarray}
Notice that, in the limit $a\rightarrow0$, the above expression reduces to standard non-hairy dyonic expression.

\subsubsection{Mass from holographic renormalization}
We now compute the mass using the holographic renormalization method \cite{Balasubramanian:1999re,Papadimitriou:2005ii,Skenderis:2002wp,deHaro:2000vlm}. This is one of the most elegant methods that is used not just to compute the mass but also other thermodynamic quantities of the black hole. In this method, the thermodynamic quantities are computed from the regularized action using boundary counterterms. For the Einstein-Maxwell-Scalar system in Eq.~(\ref{actionEF}), the renormalized action is given by
subtracting the boundary terms from the bulk action
\begin{eqnarray}
S_{ren}^{G} = S_{EMS}^{on-shell} + \frac{1}{8 \pi G_4} \int_{\partial \mathcal{M}} \mathrm{d^3}x \ \sqrt{-\gamma} \Theta -\frac{1}{16 \pi G_4} \int_{\partial \mathcal{M}} \mathrm{d^3}x \ \sqrt{-\gamma} \left(4 + R^{(3)}\right) \nonumber \\
+ \frac{2}{16 \pi G_4} \int_{\partial \mathcal{M}} \mathrm{d^3}x \ \sqrt{-\gamma} \left( b_1 \phi^2 + b_2 \phi^4 + b_3 \phi^6 \right) \,.
\label{actionregGibbs}
\end{eqnarray}
where the first term is the on-shell action, the second term is the usual Gibbons-Hawking surface term, the third term is the Balasubramanian-Kraus counterterms, and the fourth term is the scalar counterterms. Notice that no new counterterms are needed due to the $U(1)$ gauge field, as $F^2$ term falls off sufficiently quickly near the boundary. $\gamma$ is the induced metric on the boundary $\partial \mathcal{M}$, $R^{(3)}$ is the Ricci scalar constructed from the boundary metric $\gamma$, and $\Theta$ is the trace of the extrinsic curvature $\Theta_{\mu\nu}$. The constants $b_1$, $b_2$ and $b_3$ are fixed by demanding complete cancellation of the IR divergences coming from the scalar part of the action. Their values are computed as
\begin{eqnarray}
b_1 = -\frac{1}{8} , \ \ \ \  b_2 = -\frac{1}{384},  \ \ \  b_3 = -\frac{1}{46080} \,.
 \end{eqnarray}
A few important points about the above regularized action are in order:
\begin{itemize}
\item The variation of scalar part of the Einstein-Maxwell-Scalar action also gives a boundary term $(-1/16 \pi G_4)\int_{\partial \mathcal{M}} \mathrm{d^3}x \ \sqrt{-\gamma} n^r \partial_r \phi (\delta \phi)$. Therefore, in principle one could add a boundary term $(1/16 \pi G_4)\int_{\partial \mathcal{M}} \mathrm{d^3}x \ \sqrt{-\gamma} n^r \phi \partial_r \phi$ in the renormalized action in Eq.~(\ref{actionregGibbs}). However, since the latter term is not explicitly constructed from the boundary metric, we therefore do not include this term here.

\item Similarly, the variation of gauge part of the action also gives a boundary term  $\int_{\partial \mathcal{M}} \mathrm{d^3}x \ \sqrt{-\gamma} n_r f(\phi) F^{r\mu} (\delta B_\mu)$. This term goes to zero for the Dirichlet boundary condition $\delta B_\mu=0$. Since the Dirichlet boundary condition corresponds to the case of fixing the constant part of the gauge field, \textit{i.e.}, the chemical potential; therefore, in the grand canonical ensemble, no additional terms are required to be added in the above action.

\item However, if we want to study the system in the canonical ensemble, corresponding to fixed charge $q_e$, \textit{i.e.}, $\delta F^{r\mu}=0$, then the boundary term $\int_{\partial \mathcal{M}} \mathrm{d^3}x \ \sqrt{-\gamma} n_r F^{r\mu} (\delta B_\mu)$ is not zero and we have to add a boundary term $\int_{\partial \mathcal{M}} \mathrm{d^3}x \ \sqrt{-\gamma} n_r f(\phi) F^{r\mu} B_\mu$ in the action such that we get a thermodynamic function in terms of the variable ($q_e$) we wish to control. Therefore, the renormalized action in the case of canonical ensemble would be
\begin{eqnarray}
 S_{ren}^{F} = S_{EMS}^{on-shell} + \frac{1}{8 \pi G_4} \int_{\partial \mathcal{M}} \mathrm{d^3}x \ \sqrt{-\gamma} \Theta -\frac{1}{16 \pi G_4} \int_{\partial \mathcal{M}} \mathrm{d^3}x \ \sqrt{-\gamma} \left(4 + R^{(3)}\right) + \nonumber \\
 \frac{2}{16 \pi G_4} \int_{\partial \mathcal{M}} \mathrm{d^3}x \ \sqrt{-\gamma} \left( b_1 \phi^2 + b_2 \phi^4 + b_3 \phi^6 \right) +\frac{1}{16 \pi G_4}\int_{\partial \mathcal{M}} \mathrm{d^3}x \ \sqrt{-\gamma} n_r f(\phi) F^{r\mu} B_\mu \nonumber \\
\label{actionregHelmholtz}
\end{eqnarray}
As we will see, the last term would not only give an additional contribution to the Helmholtz free energy but is also necessary to get a standard thermodynamic relation between canonical and grand-canonical free energies.

 \item Generally, the canonical boundary condition in Einstein-Maxwell-Scalar theory mixes the gauge field variation with that of the scalar field variation,  \textit{i.e.}, the sum of $\delta F^{r\mu}$ and $\delta\phi$ variations \cite{Kim:2016dik}. The coupled scalar and gauge variations lead to a possibility of a generalized mixed boundary condition on both fields, corresponding to the \textit{semi canonical ensemble}. However, for the case when $\delta\phi$ vanishes, this semi canonical ensemble reduces to the standard canonical ensemble. Since in this work we are mostly interested in physical situations where the scalar field (or the parameter $a$) is fixed, we will not dwell much into the semi canonical case here.

\end{itemize}
From the renormalized action, and using the Arnowitt-Deser-Misner (ADM) decomposition, we can compute the corresponding stress energy tensor
\begin{eqnarray}
 T^{\mu\nu} =\frac{1}{8 \pi G_4} \left[ \Theta \gamma^{\mu\nu} - \Theta^{\mu\nu} + \frac{2}{\sqrt{-\gamma}}\frac{\delta\mathcal{L}_{ct}}{\delta \gamma_{\mu\nu}} \right] \,,
\label{stresstensordefplanar}
\end{eqnarray}
where $\mathcal{L}_{ct}$ is the Lagrangian of the counterterms only. Explicitly, for the grand canonical case  Eq.~(\ref{actionregGibbs}), we have
\begin{eqnarray}
T_{\mu\nu}^{G} =\frac{1}{8 \pi G_4} \left[ \Theta \gamma_{\mu\nu} - \Theta_{\mu\nu}- 2 \gamma_{\mu\nu} + G^{(3)}_{\mu\nu} + \gamma_{\mu\nu} \left( b_1 \phi^2 + b_2 \phi^4 + b_3 \phi^6 \right) \right]  \,,
\label{stresstensorgibbsplanar}
\end{eqnarray}
where $G^{(3)}_{\mu\nu}$ is the Einstein tensor of the boundary metric. Similarly, for the canonical case
\begin{eqnarray}
T_{\mu\nu}^{F} =\frac{1}{8 \pi G_4} \left[ \Theta \gamma_{\mu\nu} - \Theta_{\mu\nu}- 2 \gamma_{\mu\nu} + G^{(3)}_{\mu\nu} + \gamma_{\mu\nu} \left( b_1 \phi^2 + b_2 \phi^4 + b_3 \phi^6 \right) \right] \nonumber \\
 + \frac{1}{8 \pi G_4} \left[ \gamma_{\mu\nu} \frac{f(\phi)}{2}n_r F^{r\rho} B_\rho - f(\phi)n^{\rho}F_{\rho j}B_{i} \right]  \,.
\label{stresstensorHelmholtzplanar1}
\end{eqnarray}
In the above equations, the superscript $G$ and $F$ are used to denote grand canonical and canonical ensemble, respectively. The mass of the black hole is then related to the $tt$ component of $T_{\mu\nu}$. In particular, if $K^\mu$ is a Killing vector generating an isometry of the boundary space, then the associated conserved charge is
\begin{eqnarray}
C[K] = M_{HR} = \int_\Sigma \ d^2 x \sqrt{\sigma} u^{\mu} T_{\mu\nu} K^{\nu}  \,,
\label{massgibbsplanar}
\end{eqnarray}
where $\Sigma$ is a spacelike surface in $\partial\mathcal{M}$, with induced metric $\sigma$, and $u_{\mu}=-\sqrt{g(z)}\delta^{t}_{\mu}$ is the timelike unit normal to $\Sigma$.

\subsubsection{Thermodynamics with constant potential}
Let us first discuss the black hole thermodynamics in the grand canonical ensemble. Substituting the planar black hole expressions in Eqs.~(\ref{stresstensorgibbsplanar}) and (\ref{massgibbsplanar}), we get the mass of the black hole
\begin{eqnarray}
M_{HR}^{G}= \frac{\Omega_{2,0}}{12 \pi G_{4}}\frac{a^3 \left(\frac{\left(q_e^2+q_M^2\right) \left(2 \left(a^2 z_h^2-2 a z_h-3\right) \log \left(a
   z_h+1\right)+a z_h \left(6-a z_h\right)+2 \log ^2\left(a z_h+1\right)\right)}{4 a^4}+1\right)}{a
   z_h \left(a z_h-2\right)+2 \log \left(a z_h+1\right)} \,.
\end{eqnarray}
Notice that this black hole mass expression matches exactly with the AMD prescription. Moreover, this expression also matches with the $z^3$ coefficient of $g(z)$. In particular,
\begin{eqnarray}
M_{HR}^{G}= M_{AMD} = \frac{\Omega_{2,0}}{8 \pi G_4} \times \left[ \text{$z^3$ coefficient of $g(z)$}  \right]\,.
\end{eqnarray}
At this point, we like to emphasize that the scalar IR divergences could have been removed from the action [Eq.~(\ref{actionregGibbs})] by adding only $b_1$ and $b_2$ terms. The $b_3$ term is needed to match the holographic renormalized mass with the AMD mass, otherwise there would be a disagreement between them. This further highlights the important difference between the AMD and holographic renormalized masses in the presence of a scalar field. In particular, the holographic renormalized mass is sensitive to boundary terms whereas the AMD mass does not depend on boundary counterterms. For further discussion on this matter, see \cite{Anabalon:2014fla}. This freedom to redefine the holographic mass expression is actually like the Legendre transformations that one makes between different forms of energy in standard thermodynamics.  Note that since no additional counterterms are needed for the gauge field, the AMD and holographic renormalized mass agree with each other for the standard non-hairy dyonic black holes.

From the renormalized action, we can further calculate the Gibbs free energy $G=-S_{ren}^G/\beta$
\begin{eqnarray}
& & G = \frac{\mu _e^2 \left(a^3 z_h \left(a z_h-6\right)-2 a^2 \log ^2\left(a z_h+1\right)+2 a^2 \left(-a^2
   z_h^2+2 a z_h+3\right) \log \left(a z_h+1\right)\right)}{96 \pi  a G_4 \log ^2\left(a z_h+1\right)
   \left(a z_h \left(a z_h-2\right)+2 \log \left(a z_h+1\right)\right)} \nonumber \\
& &    +\frac{q_M^2 \left(2 \left(2 a^2
   z_h^2-4 a z_h+3\right) \log ^3\left(a z_h+1\right)+a^2 z_h^2 \log ^2\left(a z_h+1\right)+10 \log
   ^4\left(a z_h+1\right)\right)}{96 \pi  a G_4 \log ^2\left(a
   z_h+1\right) \left(a z_h \left(a z_h-2\right)+2 \log \left(a z_h+1\right)\right)} \nonumber \\
&  & -\frac{2 a^3+3 z_h q_M^2}{48 \pi  G_4 \left(a z_h \left(a z_h-2\right)+2 \log \left(a z_h+1\right)\right)} \,.
\label{Gibbspanar}
\end{eqnarray}
Importantly, this expression of Gibbs free energy matches with the expected thermodynamic relation $G=M_{HR}^{G} - T S_{BH} - Q_e \mu_e$. This is a consistency check for the thermodynamic formulae found here for the dyonic hairy black holes. It is indeed surprising that even for these complicated dyonic hairy black holes analytic expressions of various thermodynamic observables could be obtained. Further, holding the relation $G=M_{HR}^{G} - T S_{BH} - Q_e \mu_e$, without requiring to add a term for the magnetic charge, further supports the fact that treatment of $q_M$ as a constant external parameter is consistent. We further calculated the pressure in the grand canonical ensemble \footnote{The pressure can be computed from the $x_1 x_1$ component of $T_{\mu\nu}$.}, and find that the Einstein-Maxwell-Scalar gravity system satisfies the standard thermodynamic relation \footnote{Note that with finite $q_M$, $P$ differs from $T_{xx}$ by a term proportional to the magnetization.},
\begin{eqnarray}
G = - P^G\,.
\label{pressureplanar}
\end{eqnarray}
As expected the above expression for $G$ reduces to the standard non-hairy dyonic black hole expression in the limit $a\rightarrow0$, again providing a consistency check of the expression. However, it turns out that the Gibbs free energy does not satisfy the differential first law like relation
\begin{eqnarray}
dG = -S_{BH}dT - Q_e d\mu_e\,.
\label{pressureplanar}
\end{eqnarray}
This unfortunate result might be related to the fact that with the scalar field this form of first law needs to be modified by scalar contributions. Indeed, by now many works have strongly advocated for modification of the differential first law in the presence of a scalar field \cite{Liu:2013gja,Lu:2014maa}. It is of course important to explicitly establish the first law in our Einstein-Maxwell-Scalar model as well; however, since our primary focus here is on the construction and thermodynamic stability of the hairy black holes (and on the corresponding hairy/nonhairy phase transitions). We therefore postpone this curious exercise for future work.

\begin{figure}[h!]
\begin{minipage}[b]{0.5\linewidth}
\centering
\includegraphics[width=2.8in,height=2.3in]{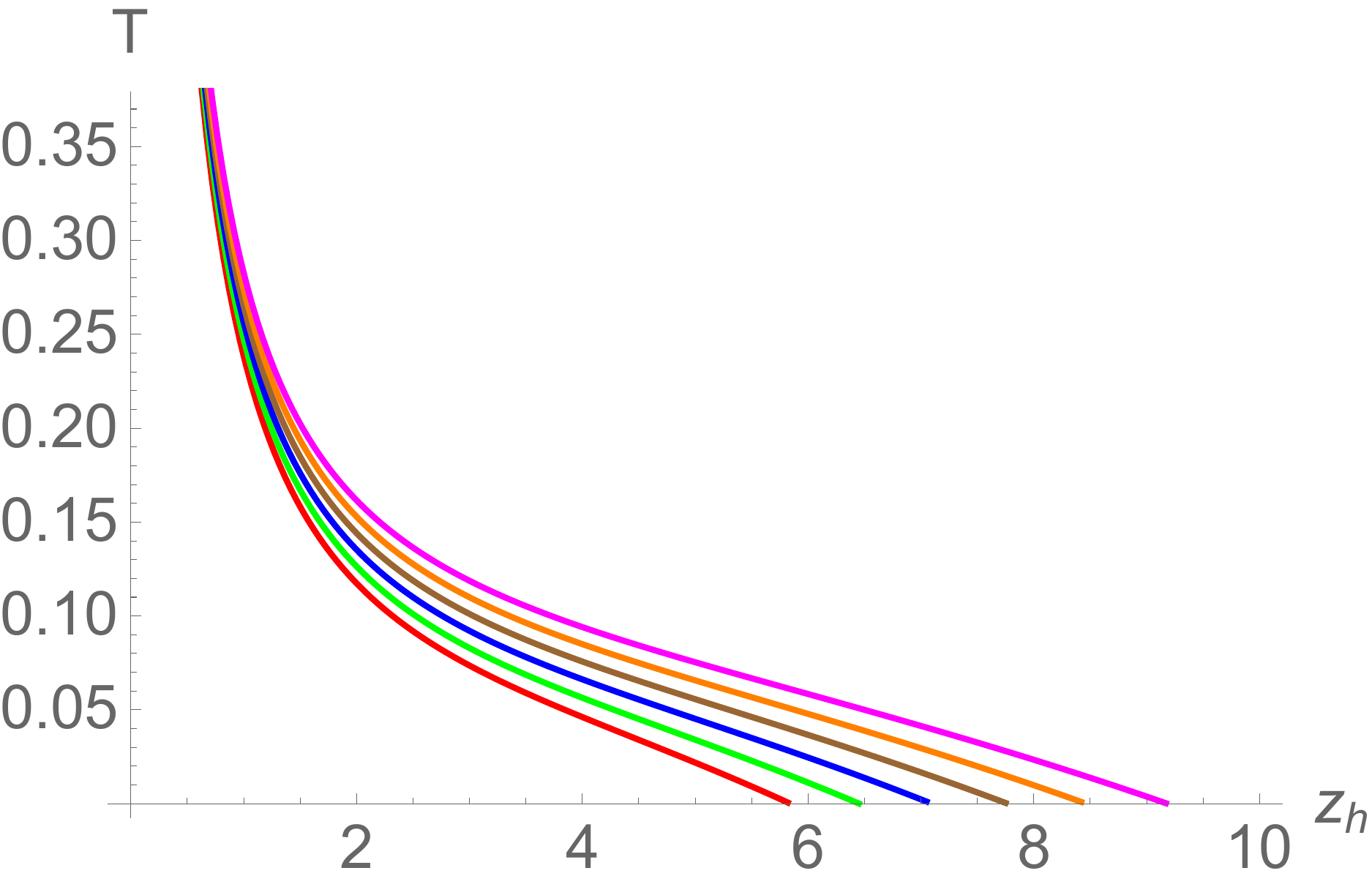}
\caption{ \small Hawking temperature $T$ as a function of horizon radius $z_h$ for various values of $a$.  Here $\mu_e=0.1$ and $q_M=0.1$ are used. Red, green, blue, brown, orange, and magenta curves correspond to $a=0$, $0.05$, $0.10$, $0.15$, $0.20$, and $0.25$ respectively.}
\label{zhvsTvsaMuPt1qMPt1}
\end{minipage}
\hspace{0.4cm}
\begin{minipage}[b]{0.5\linewidth}
\centering
\includegraphics[width=2.8in,height=2.3in]{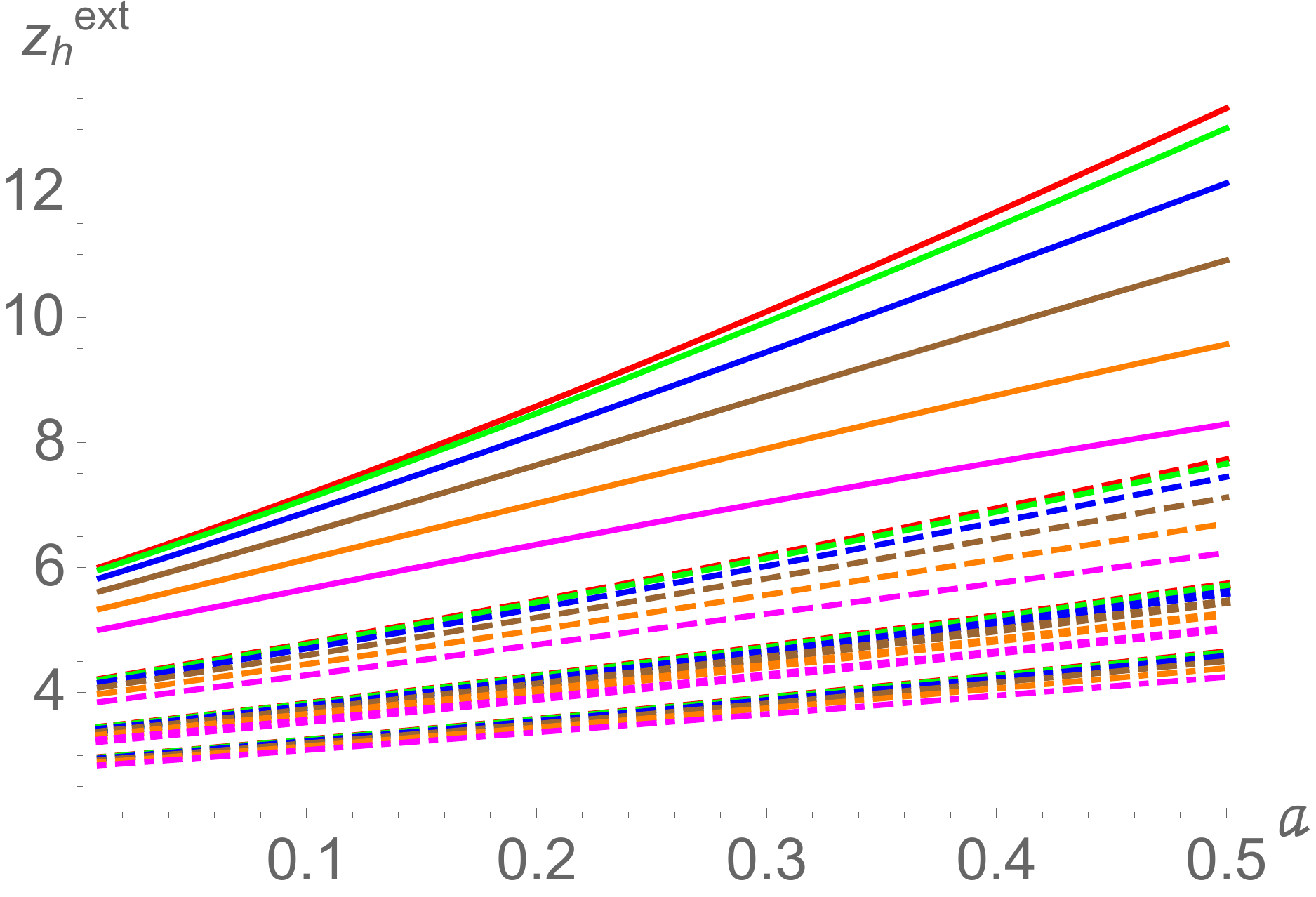}
\caption{\small The variation of extremal black hole horizon radius $z_{h}^{\text{ext}}$ as function of $a$. Red, green, blue, brown, orange, and magenta curves correspond to $\mu_e=0$, $0.1$, $0.2$, $0.3$, $0.4$, and $0.5$ respectively. Solid, dashed, dotted, and dot-dashed curves correspond to $q_M=0.1$, $0.2$, $0.3$, and $0.4$ respectively.}
\label{avszhextvsMuvsqMplanarcase1}
\end{minipage}
\end{figure}

With various analytic expressions in hand, we are now ready to discuss the thermodynamic properties of the planar hairy dyonic black holes. In Fig.~\ref{zhvsTvsaMuPt1qMPt1}, we have shown the behaviour of Hawking temperature with respect to inverse horizon radius $z_h$ for different values of $a$. Here, we have kept $\mu_e=0.1$ and $q_M=0.1$ fixed but similar results hold for other values of $\mu_e$ and $q_M$ as well. We find a one to one relation between the horizon radius and Hawking temperature. Remember that the usual non-hairy dyonic black hole becomes extremal only when $\mu_e$ and $q_M$ are finite. This property continues to hold for the hairy dyonic case as well. With finite $a$, the difference arises in the magnitude of $z_h^{ext}$ at which the hairy black hole becomes extremal. We find that the magnitude of $z_h^{ext}$ increases with $a$ for all $\mu_e$ and $q_M$. The overall dependence of extremal horizon radius on $a$, $\mu_e$ and $q_M$ is shown in Fig.~\ref{avszhextvsMuvsqMplanarcase1}.

It is also important to analyze the local stability, which measures the response of the equilibrium system under a small fluctuation in thermodynamical variables, of these hairy black holes.  In the constant potential ensemble, the local stability is quantified by the positivity of the specific heat $C_\mu=T(\partial S_{BH}/\partial T)$ at constant potential. Since $S_{BH} \propto z_h^{-2}$, it is easy to see that the slope of the $S_{BH}-T$ plane in these hairy black holes is always positive. This in turn implies that $C_\mu \geq 0$, indicating the local stability of hairy dyonic black holes.

\begin{figure}[ht]
	\subfigure[]{
		\includegraphics[scale=0.4]{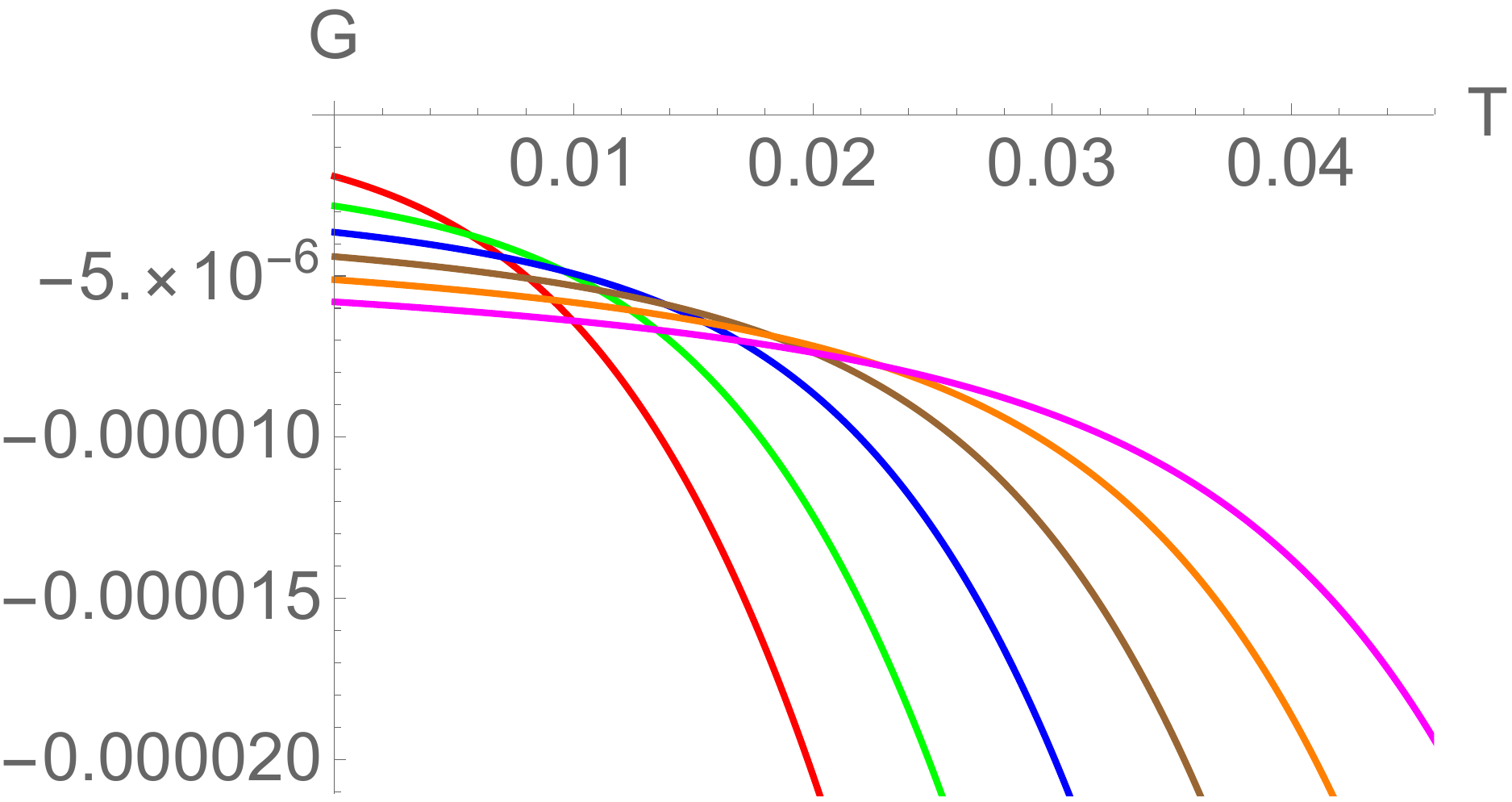}
	}
	\subfigure[]{
		\includegraphics[scale=0.4]{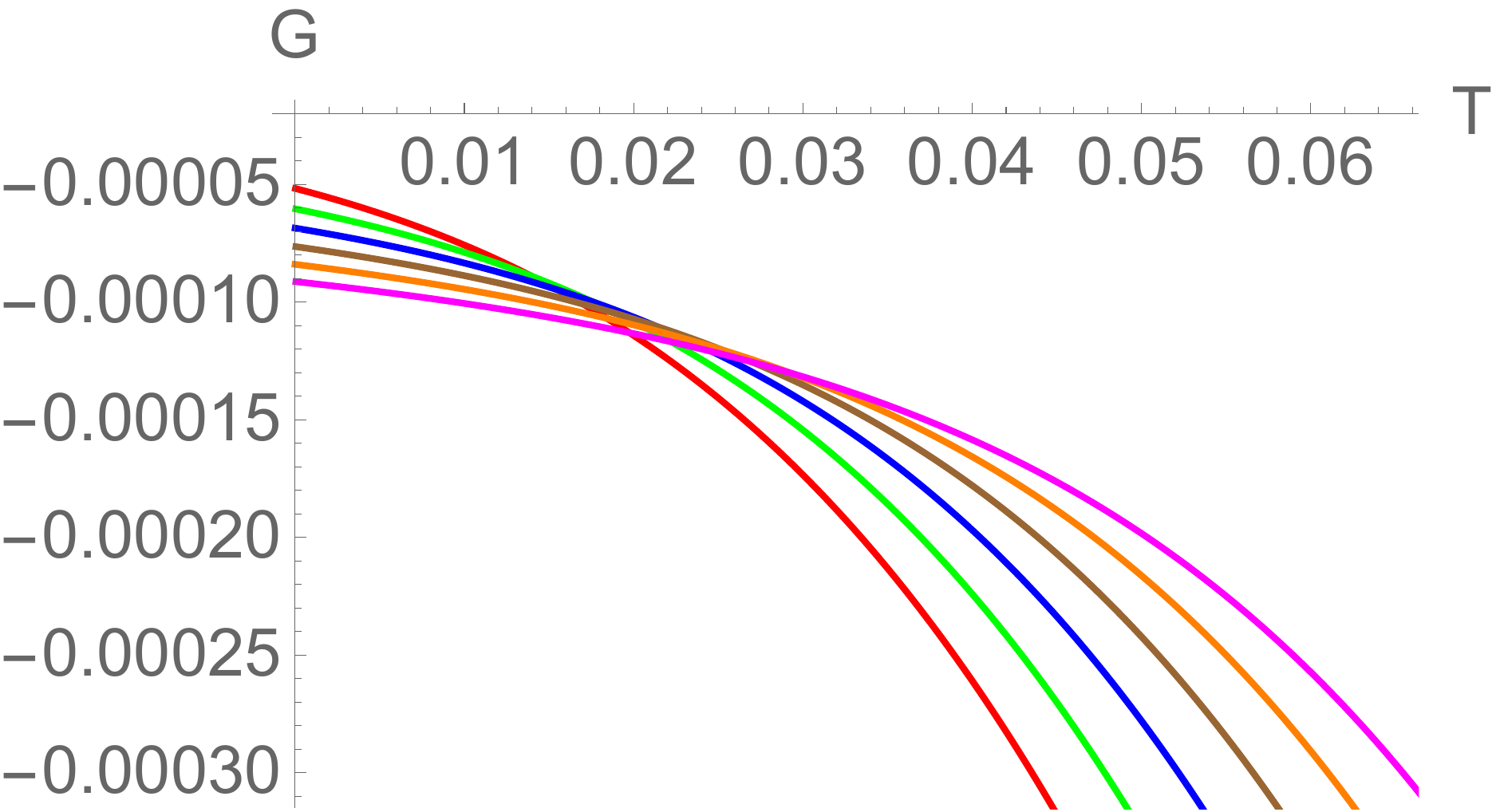}
	}
	\subfigure[]{
		\includegraphics[scale=0.4]{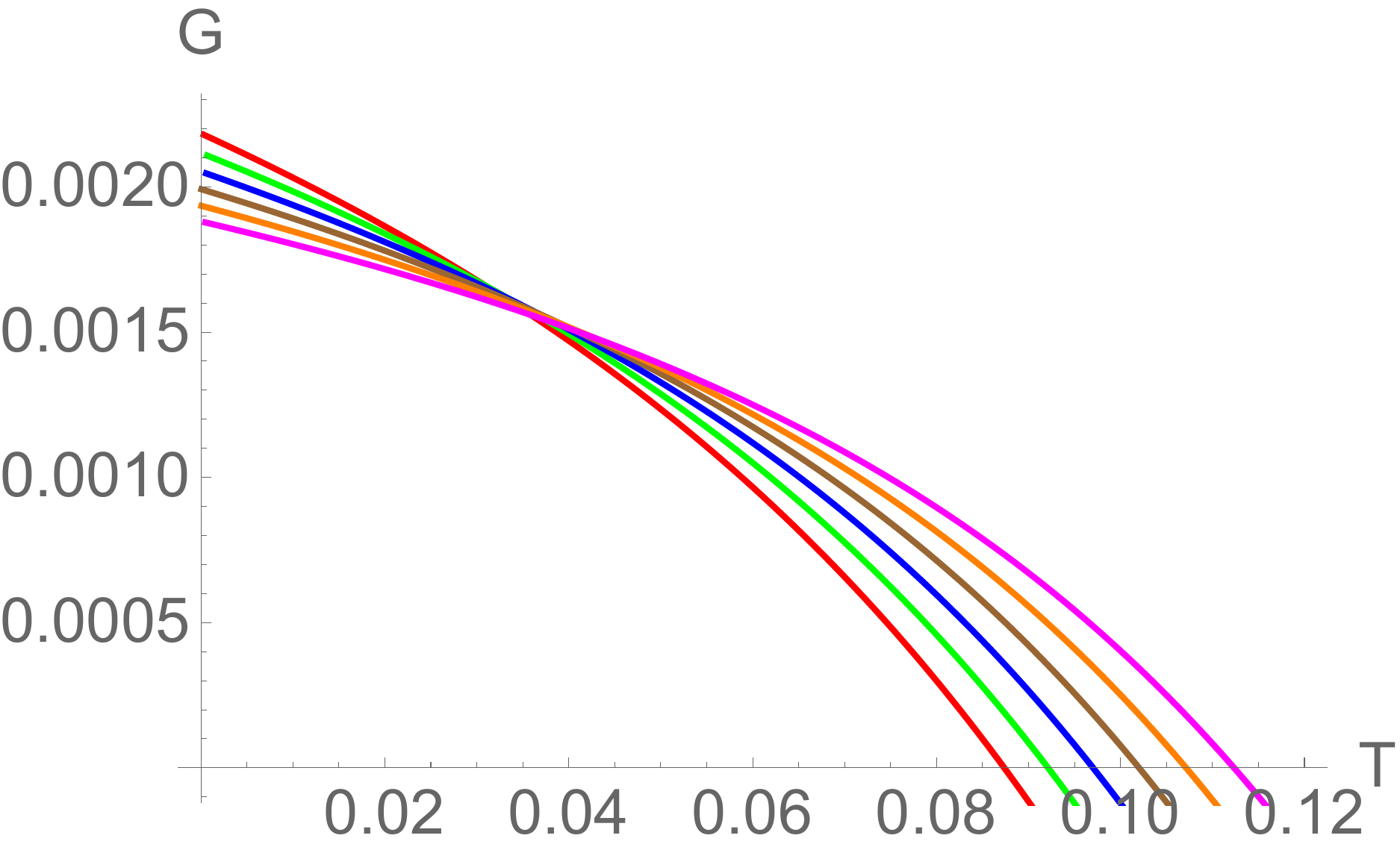}
	}
\subfigure[]{
		\includegraphics[scale=0.4]{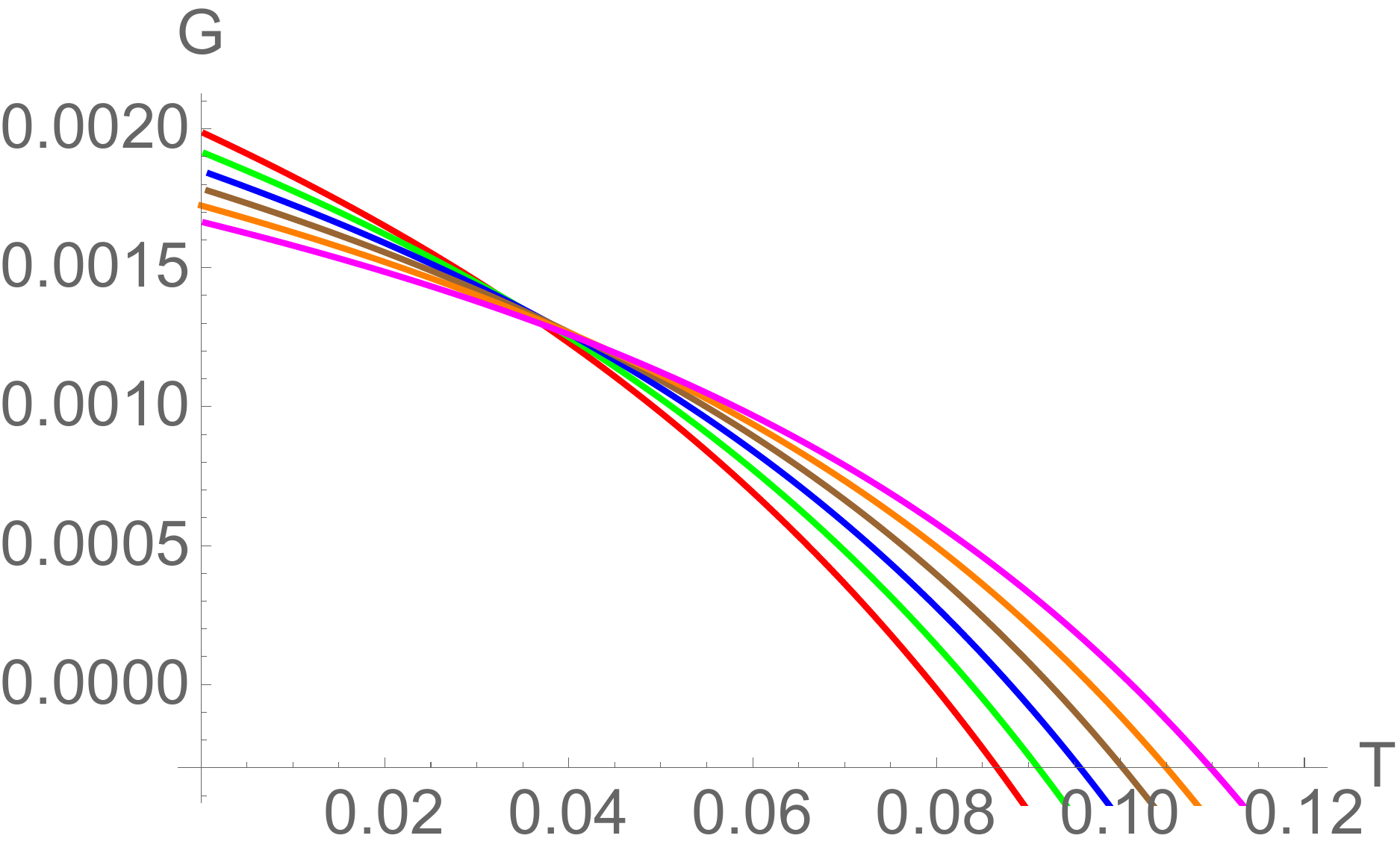}
	}
	\caption{\small Gibbs free energy $G$ as a function of Hawking temperature $T$ for various values of $a$. The upper-left graph (a) corresponds to $q_M=0$ and $\mu_e=0.1$; the upper-right graph (b) corresponds to $q_M=0$ and $\mu_e=0.3$; the lower-left graph (c) corresponds to $q_M=0.2$ and $\mu_e=0.1$; and the lower-right graph (d) corresponds to $q_M=0.2$ and $\mu_e=0.3$. Red, green, blue, brown, orange, and magenta curves correspond to $a=0$, $0.05$, $0.10$, $0.15$, $0.20$, and $0.25$ respectively.}
	\label{zvsgvsamu1qM1zh1f1case1}
\end{figure}

\begin{figure}[h!]
\centering
\includegraphics[width=2.8in,height=2.3in]{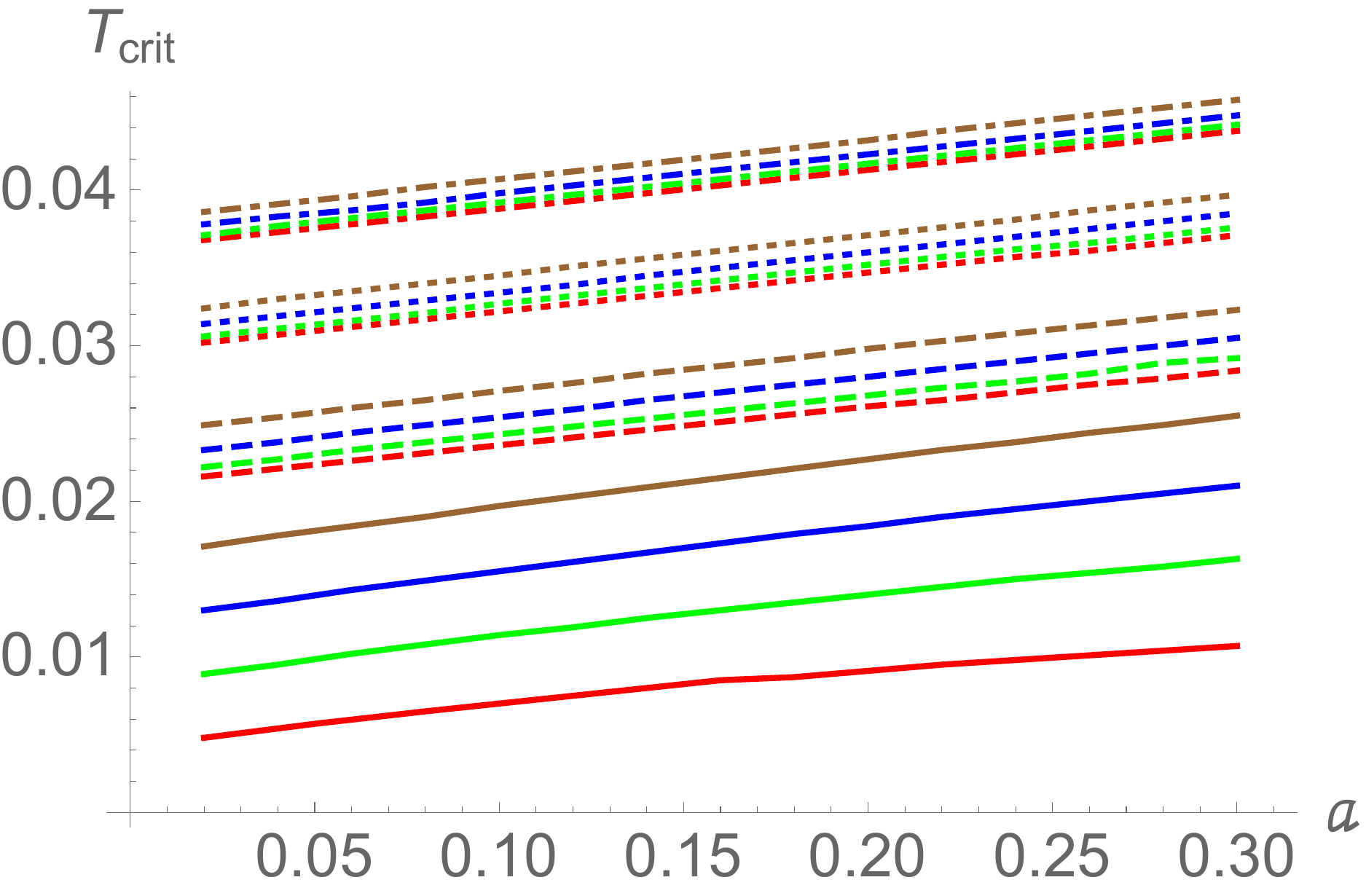}
\caption{\small The variation of $T_{crit}$ as function of $a$. Red, green, blue, and brown curves correspond to $\mu_e=0.1$, $0.2$, $0.3$, and $0.4$ respectively. Solid, dashed, dotted, and dot-dashed curves correspond to $q_M=0$, $0.1$, $0.2$, and $0.3$ respectively.}
\label{avsTcritvsMuvsqMplanatcase1}
\end{figure}

To further analyze the thermodynamic stability of the hairy dyonic black holes, we examine its free energy. The Gibbs free energy as a function of temperature for different values of $a$ is shown in Fig.~\ref{zvsgvsamu1qM1zh1f1case1}. One can immediately notice that Gibbs free energy of the hairy black hole ($a\neq0$) is smaller than the non-hairy black hole ($a=0$) at lower temperatures, whereas it is greater than the non-hairy black hole at higher temperatures. This suggests that although the non-hairy black hole is thermodynamically preferable at higher temperatures, however, it
is the hairy black hole structure that is thermodynamically more preferable at low temperatures. Moreover, the temperature range for which the hairy black hole is more preferable increases with $a$. In particular, the temperature $T_{crit}$ at which the free energy of hairy black hole becomes lower than the non-hairy black hole increases with $a$. Similarly, $T_{crit}$ increases with $\mu_e$ and $q_M$ as well. This indicates that the thermodynamic stability of the hairy dyonic black hole phase strengthens at higher temperatures as both $\mu_e$ and $q_M$ increase.  The complete phase diagram displaying the dependence of $T_{crit}$ on $a$, $\mu_e$, and $q_M$ is shown in Fig.~\ref{avsTcritvsMuvsqMplanatcase1}.

\begin{figure}[h!]
\begin{minipage}[b]{0.5\linewidth}
\centering
\includegraphics[width=2.8in,height=2.3in]{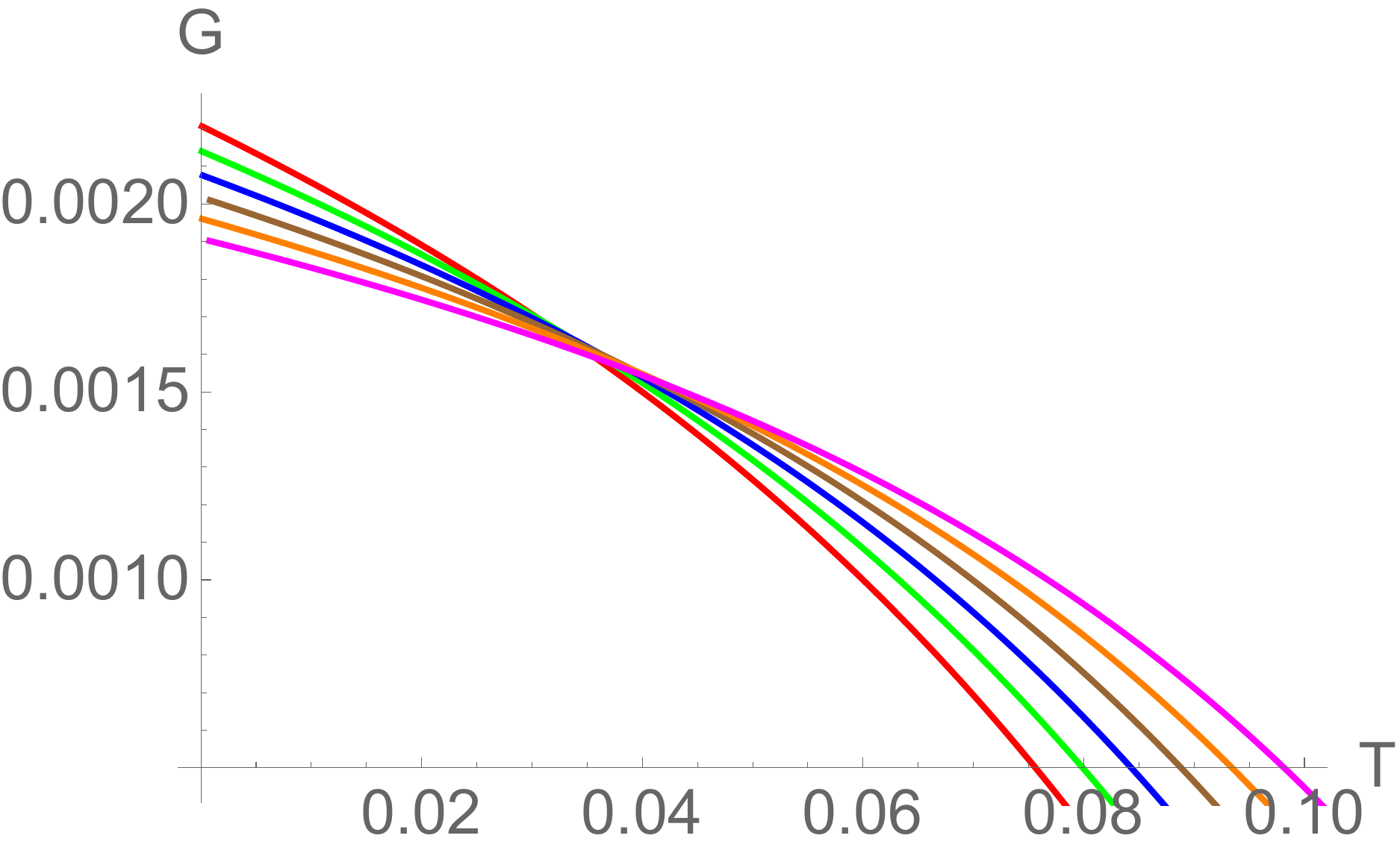}
\caption{ \small Gibbs free energy $G$ as a function of Hawking temperature $T$ for various values of $a$.  Here $\mu_e=0$ is used. Red, green, blue, brown, orange, and magenta curves correspond to $a=0$, $0.05$, $0.10$, $0.15$, $0.20$, and $0.25$ respectively.}
\label{TvsGvsaMu0qMPt2planarcase1}
\end{minipage}
\hspace{0.4cm}
\begin{minipage}[b]{0.5\linewidth}
\centering
\includegraphics[width=2.8in,height=2.3in]{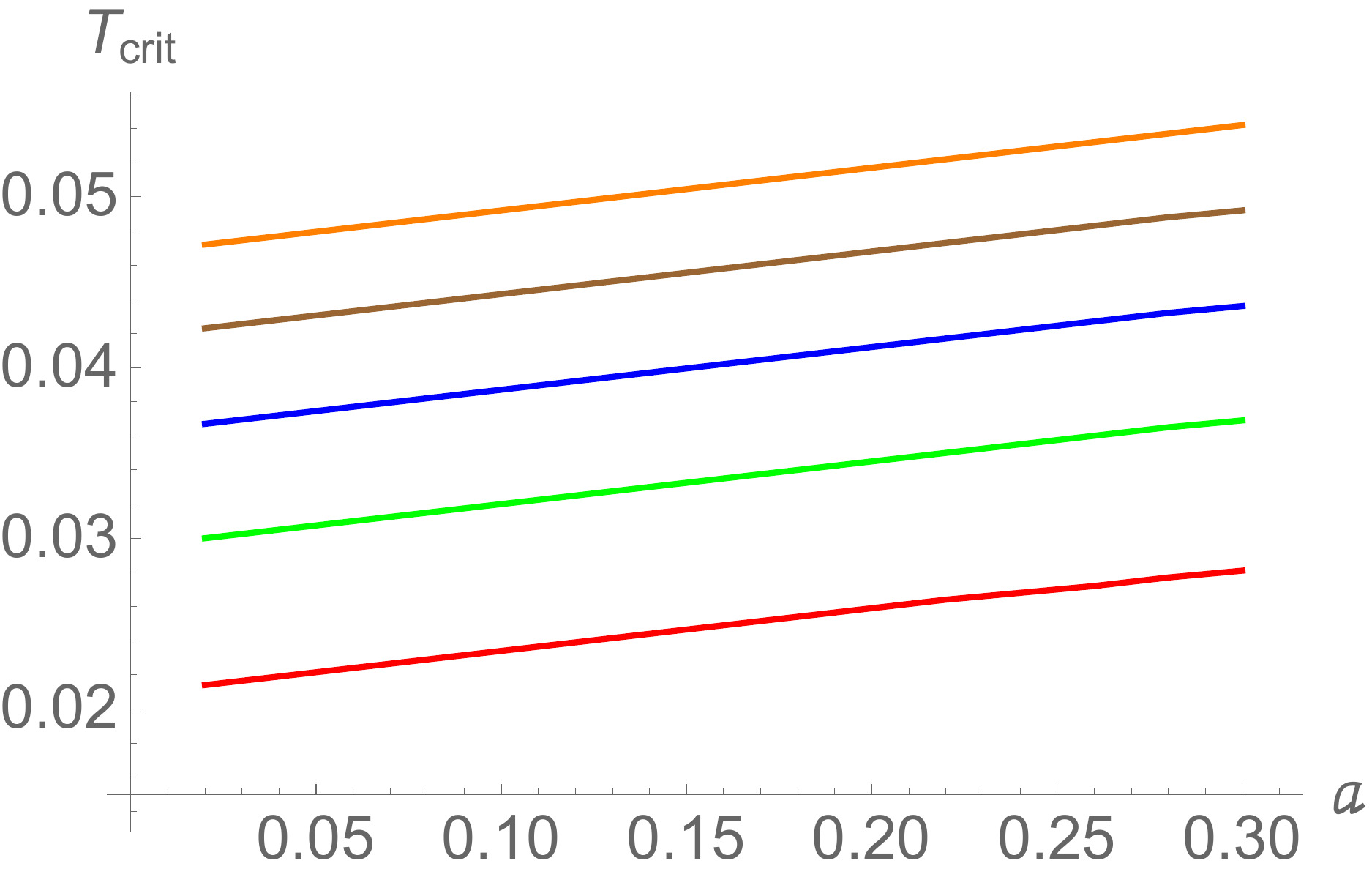}
\caption{\small The variation of $T_{crit}$ as function of $a$. Here $\mu_e=0$ is used. Red, green, blue, brown, and orange curves correspond to $q_M=0.1$, $0.2$, $0.3$, $0.4$, and $0.5$ respectively.}
\label{avsTcritvsqMMu0planatcase1}
\end{minipage}
\end{figure}

Importantly, the planar hairy black hole can become thermodynamically favourable even when $\mu_e = 0$. This is an interesting result considering that the free energy of planar RN-AdS black hole is generally found to be smaller than the planar hairy black holes for $\mu_e=0$ \cite{Mahapatra:2020wym}. We find that this situation can be circumvented for the dyonic case having finite $q_M$. In particular,  for $q_M\neq 0$, the free energy of an uncharged hairy black hole can be smaller than the uncharged non-hairy black hole at low temperatures. This is shown in Fig.~\ref{TvsGvsaMu0qMPt2planarcase1} for a particular value of $q_M=0.2$. Moreover, the temperature $T_{crit}$ at which the free energy of hairy uncharged black hole becomes lower than the non-hairy uncharged black hole increases with $q_M$. This further suggests that the possibility of thermodynamically stable hairy uncharged black hole increases with the dyonic parameter $q_M$. The dependence of $T_{crit}$ on $q_M$ and $a$ for the uncharged cases ($\mu_e=0$) is shown in Fig.~\ref{avsTcritvsqMMu0planatcase1}. Overall, our whole free energy analysis indicates the existence of a thermodynamically stable and well-behaved hairy dyonic planar black hole solution in asymptotically AdS space in our model.

\subsubsection{Thermodynamics with constant charge}
To discuss the thermodynamic in the canonical ensemble, we use the renormalized action Eq.~(\ref{actionregHelmholtz}) to compute the Helmholtz free energy $F=-S_{ren}^F/\beta$. We find that,
\begin{eqnarray}
& & F = \frac{\left(q_e^2+q_M^2\right) \left(a^2 z_h^2+2 \left(2 a^2 z_h^2-4 a z_h+3\right) \log
   \left[a z_h+1\right]-6 a z_h+10 \log ^2\left[a z_h+1\right]\right)}{96 \pi  a G_4 \left(a z_h \left(a
   z_h-2\right)+2 \log \left[a z_h+1\right]\right)} \nonumber \\
& &    -\frac{a^3}{24 \pi  G_4 \left(a z_h \left(a
   z_h-2\right)+2 \log \left[a z_h+1\right]\right)} \,.
\label{Helmholtzpanar}
\end{eqnarray}
This is again consistent with the standard thermodynamic relation that the Helmholtz free energy is a Legendre transformation of the Gibbs free energy
\begin{eqnarray}
& & F =  G + Q_e \mu_e = M_{HR}^{G} - T S_{BH}  \,.
\label{Helmholtzpanar}
\end{eqnarray}
Similarly, we can compute mass and pressure in the canonical ensemble from ($tt$) and ($xx$) component of the stress energy tensor Eq.~(\ref{stresstensorHelmholtzplanar1}). The pressure again turns out to be negative of the Helmholtz free energy \footnote{Again, for finite $q_M$, $P$ differs from $T_{xx}$ by a term proportional to the magnetization.},
\begin{eqnarray}
& & P^F = - F \,.
\label{PvsFrelationpanar}
\end{eqnarray}
Whereas, the mass is given by
\begin{eqnarray}
& & M_{HR}^{F}= \frac{\Omega_{2,0}}{12 \pi G_{4}} \frac{a^3 \left(\frac{ \left(q_e^2+q_M^2\right) \left(2 \log \left[ 1 + a z_h \right] \left(\left(a-\frac{3}{z_h}\right)
   \left(a+\frac{1}{z_h}\right)+\frac{\log \left[ 1+ a z_h \right]}{z_h^2}\right)-a
   \left(a-\frac{6}{z_h}\right)\right)}{4 a^4 z_h^{-2}}+1\right)}{a z_h^2 \left(a-\frac{2}{z_h}\right)+2 \log
   \left[1+ a z_h \right]}  \nonumber \\
& &    - \frac{\Omega_{2,0} }{16 \pi G_{4}} \frac{q_e^2 \log\left[ 1+ a z_h \right]}{a} \,.
\end{eqnarray}
Here the last term is nothing but $Q_e \mu_e$. Notice that this mass $M_{HR}^{F}$ differs from the mass in Helmholtz free energy in  Eq.~(\ref{Helmholtzpanar}) by a factor of $Q_e \mu_e$. This difference arises precisely from the additional gauge field terms introduced in the action [Eq.~(\ref{actionregHelmholtz})] and stress energy tensor [Eq.~(\ref{stresstensorHelmholtzplanar})] to have a well defined fixed charge ensemble. For the usual RN-AdS black hole in the canonical ensemble, the mass $M_{HR}^{F}$ is usually identified as the energy  above the ground state \textit{i.e.}, the extremal black hole \cite{Chamblin:1999tk}. However, we find that for the hairy case the mass of the fixed charge differ by $Q_e \mu_e$, and not by the energy of the extremal black hole. It would certainly be interesting to find out the exact meaning of this holographic renormalized mass $M_{HR}^{F}$ in the canonical ensemble context, however, at this moment we are unsure about its correct physical interpretation. 

\begin{figure}[h!]
\begin{minipage}[b]{0.5\linewidth}
\centering
\includegraphics[width=2.8in,height=2.3in]{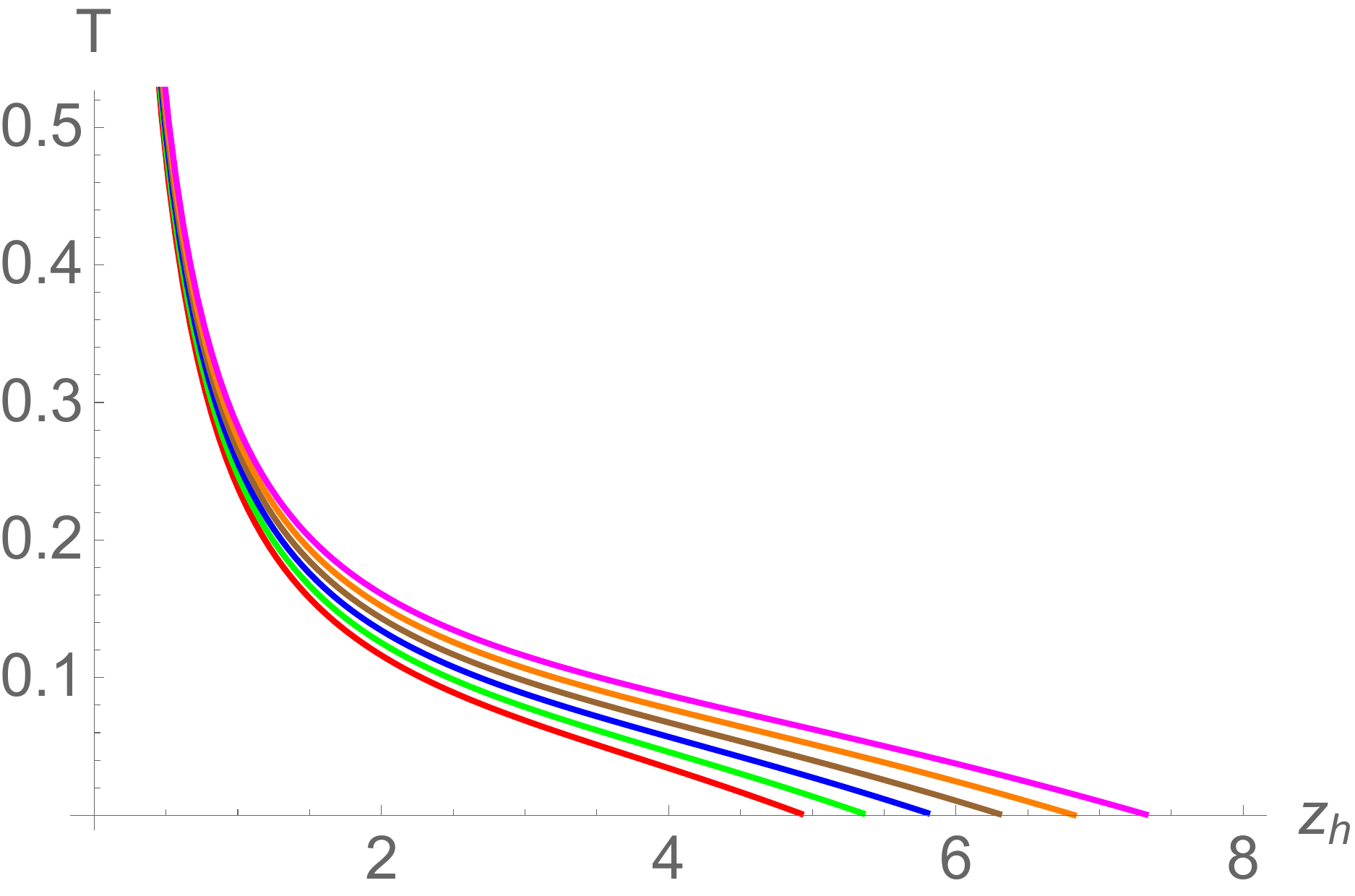}
\caption{ \small Hawking temperature $T$ as a function of horizon radius $z_h$ for various values of $a$. Here $q_e=0.1$ and $q_M=0.1$ are used. Red, green, blue, brown, orange, and magenta curves correspond to $a=0$, $0.05$, $0.10$, $0.15$, $0.20$, and $0.25$ respectively.}
\label{zhvsTvsaqEPt1qMPt1planarcase1}
\end{minipage}
\hspace{0.4cm}
\begin{minipage}[b]{0.5\linewidth}
\centering
\includegraphics[width=2.8in,height=2.3in]{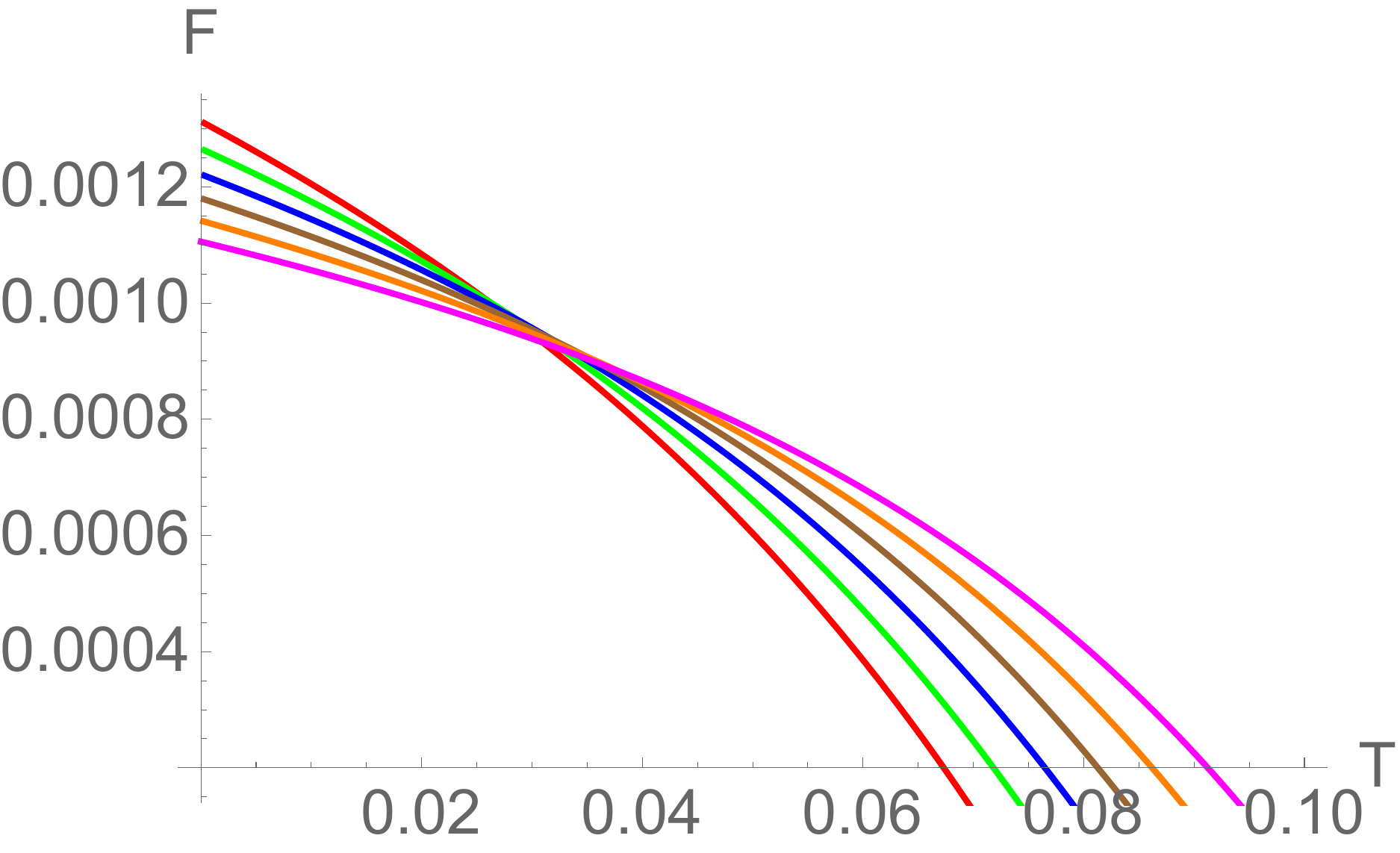}
\caption{\small Helmholtz free energy $F$ as a function of Hawking temperature $T$ for various values of $a$. Here $q_e=0.1$ and $q_M=0.1$ are used. Red, green, blue, brown, orange, and magenta curves correspond to $a=0$, $0.05$, $0.10$, $0.15$, $0.20$, and $0.25$ respectively.}
\label{TvsFvsaqEPt1qMPt1planarcase1}
\end{minipage}
\end{figure}

In Fig.~\ref{zhvsTvsaqEPt1qMPt1planarcase1}, the variation of Hawking temperature with horizon radius is shown. Here, we have kept $q_e=0.1$ and $q_M=0.1$ fixed, but similar results occur for other values of $q_e$  and $q_M$ as well. In the canonical ensemble too, the radius of the extremal horizon $z_h^{ext}$ increases with $a$. The negative slope and one to one relation between the horizon radius and Hawking
temperature further indicate that the hairy black hole does not only exists at all temperatures but also are thermodynamically stable. In particular, the specific heat at constant charge $C_q=T(\partial S_{BH}/\partial T)$ is always positive, implying the local stability of these black holes in the canonical ensemble as well.

\begin{figure}[h!]
\centering
\includegraphics[width=2.8in,height=2.3in]{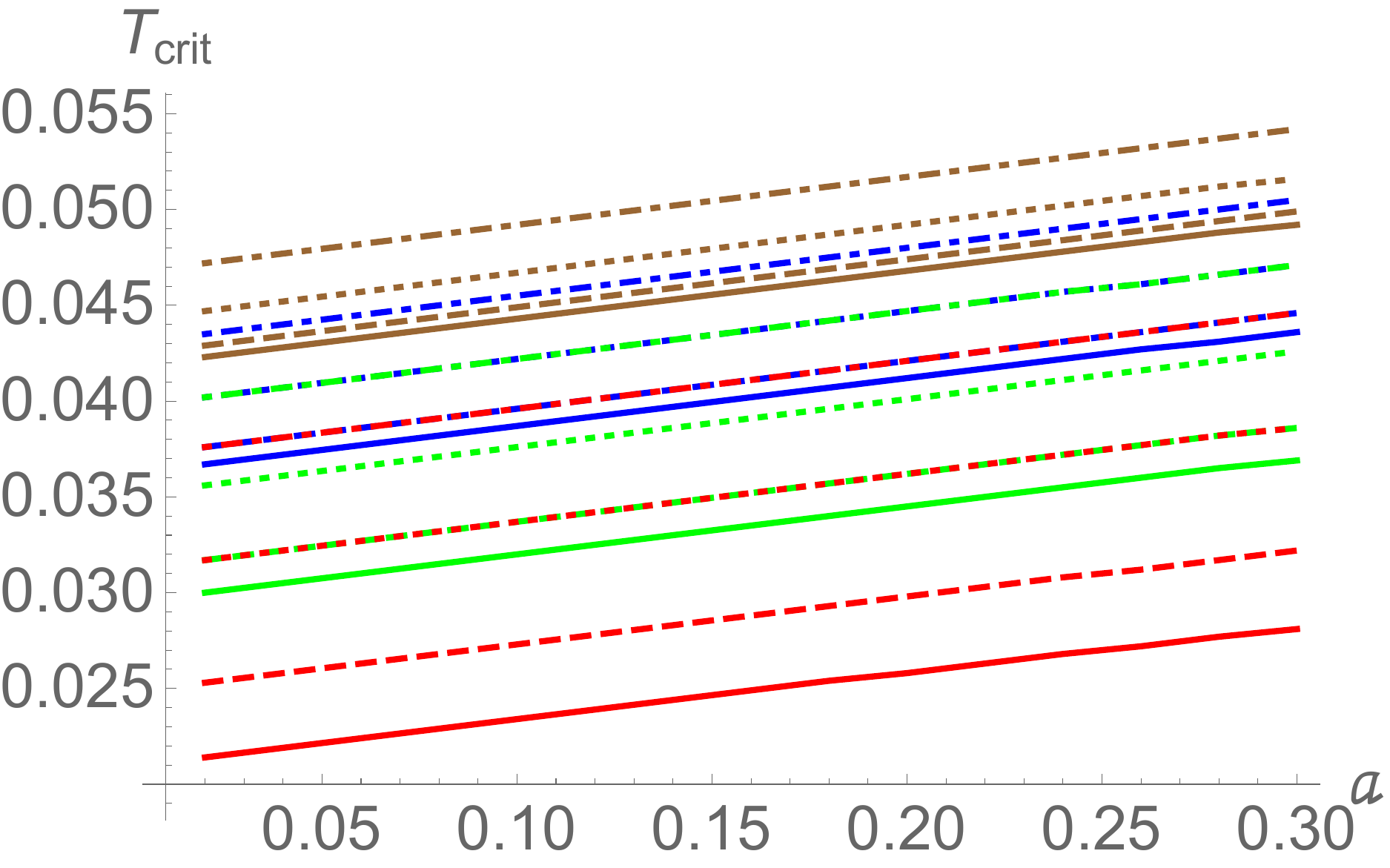}
\caption{\small The variation of $T_{crit}$ as function of $a$. Red, green, blue, and brown curves correspond to $q_e=0.1$, $0.2$, $0.3$, and $0.4$ respectively. Solid, dashed, dotted, and dot-dashed curves correspond to $q_M=0$, $0.1$, $0.2$, and $0.3$ respectively.}
\label{avsTcritvsqEvsqMplanatcase1}
\end{figure}

In Fig.~\ref{TvsFvsaqEPt1qMPt1planarcase1}, the thermal variation of Helmholtz free energy as function $a$ is shown. We again see that the free energy of the hairy black hole ($a\neq0$) is smaller than the non-hairy black hole ($a=0$) at lower temperatures, whereas it is greater than the non-hairy black hole at higher temperatures. This implies that, like in the grand canonical ensemble, the hairy black hole is thermodynamically more preferable at low temperatures compared to the non-hairy black hole in the canonical ensemble as well. We further calculated the temperature $T_{crit}$, at which the Helmholtz free energy of hairy black hole becomes lower than the non-hairy black hole, and find it to be an increasing function of $a$, $q_e$, and $q_M$. This implies that the temperature window where the hairy black hole is more preferable  increases with $a$, $q_e$, and $q_M$. The overall dependence of $T_{crit}$ on $a$, $q_e$, and $q_M$ is shown in Fig.~\ref{avsTcritvsqEvsqMplanatcase1}. Therefore, in the canonical ensemble as well, the dyonic parameter $q_M$ plays a constructive role in the thermodynamic stability of the hairy black hole. Moreover, since the case $q_e=0$ is identical to the $\mu_e=0$ case, which we have already discussed in the last subsection, the planar hairy uncharged black hole can become thermodynamically favourable in the canonical ensemble as well.

\subsection{Spherical horizon: $\kappa=1$}
We now turn our attention to dyonic hairy black hole solution and thermodynamics with the spherical horizon, corresponding to $\kappa=1$. The thermodynamic phase structure of the usual spherical non-hairy dyonic black hole has an incredibly rich structure \cite{Dutta:2013dca}. In particular, there appear interesting Hawking/page and small/large black hole phase transitions. Therefore, it will be interesting to see how the presence of scalar hair affects this phase structure.

\begin{figure}[ht]
	\subfigure[]{
		\includegraphics[scale=0.4]{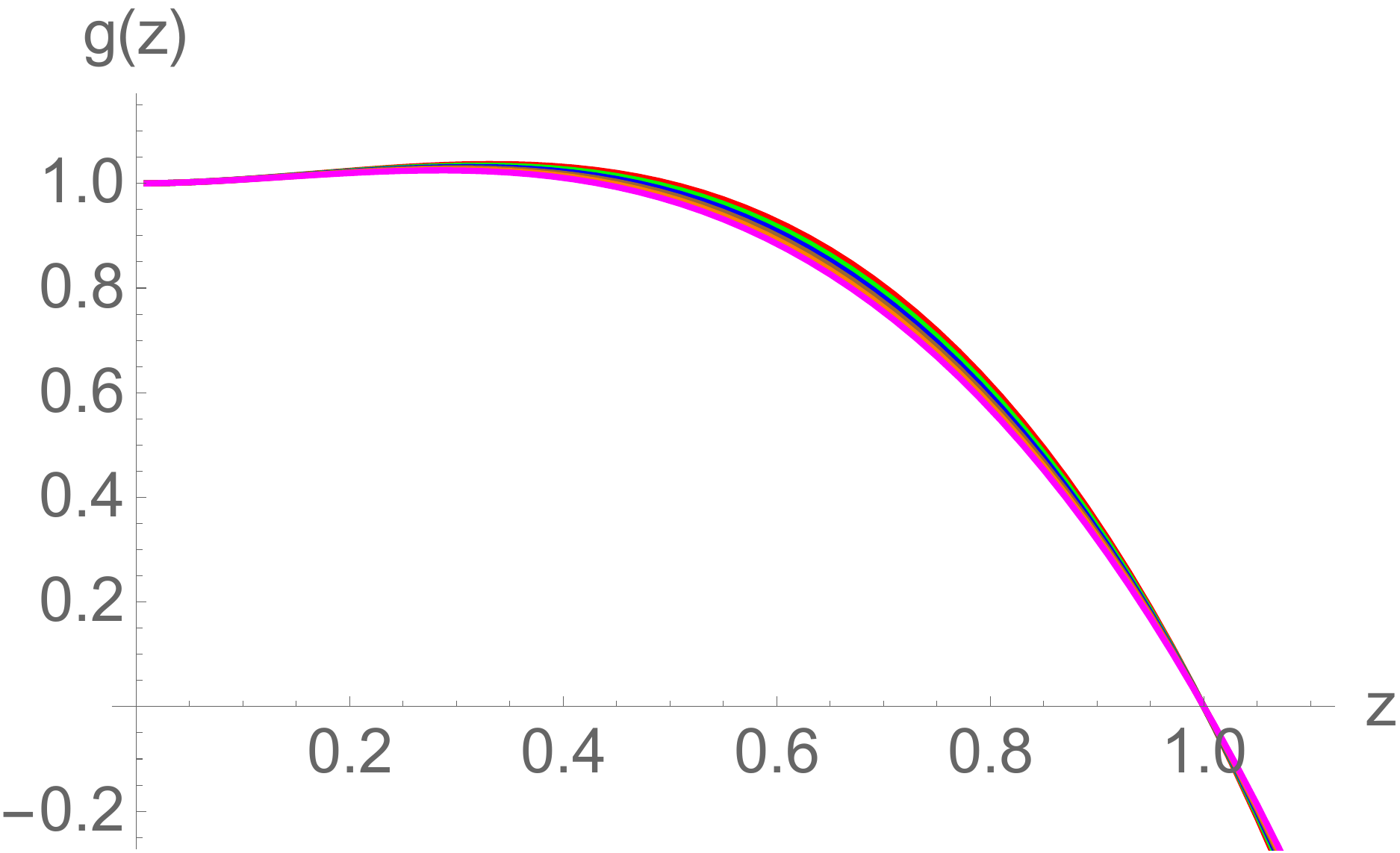}
	}
	\subfigure[]{
		\includegraphics[scale=0.4]{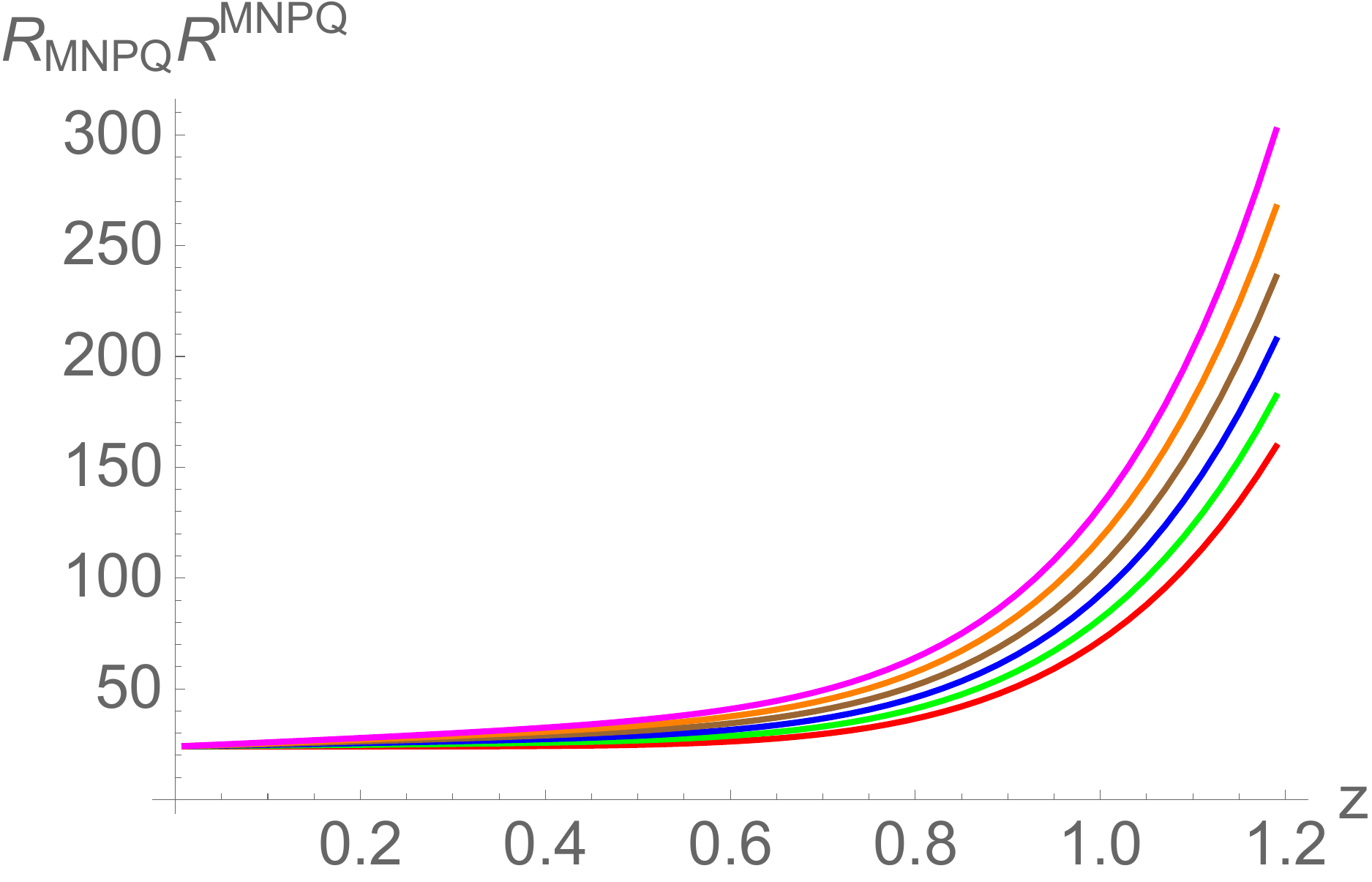}
	}
	\caption{\small The behavior of $g(z)$ and $R_{MNPQ}R^{MNPQ}$ for different values of hair parameter $a$. Here $z_h=1$, $q_e=0.1$, and $q_M=0.1$ are used. Red, green, blue, brown, orange, and magenta curves correspond to $a=0$, $0.05$, $0.10$, $0.15$, $0.20$, and $0.25$ respectively.}
	\label{zvsgvsaqEPt11qMPt1zh1f1sphcase1}
\end{figure}

Similar to the planar case, we can again obtain the analytic, though slightly complicated, expression of $g(z)$
\begin{align*}
 g(z) = 1 + \frac{a z (a z-2)+2 \log (a z+1)}{a^2 \left(a z_h \left(a z_h-2\right)+2 \log \left(a
   z_h+1\right)\right)} \Bigl\{ \log \left(a z_h+1\right) \left(\log \left(a z_h+1\right)-2 \log \left(z_h\right)-1\right)  \Bigr.\\
  \Bigl. -a^2 + a z_h \left(2 \log \left(\frac{z_h}{a z_h+1}\right)+a z_h \log \left(\frac{1+a z_h}{z_h}\right)-1\right)-2
   \text{Li}_2\left(-a z_h\right) \Bigr\}  \nonumber \\
 -  \frac{\left(q_e^2 +q_M^2  \right) \left(a z (a z-2)+2 \log (a z+1) \right)}{4 a^4 \left(a z_h \left(a z_h-2\right)+2 \log \left(a
   z_h+1\right)\right)} \Bigl\{ 2 \left(\log \left(a z_h+1\right)-3\right) \log \left(a z_h+1\right) \Bigr.\\
   \Bigl. + a z_h \left(-4 \log \left(a z_h+1\right)+a z_h \left(2 \log \left(a z_h+1\right)-1\right)+6\right) \Bigr\} \nonumber \\
 + \left(q_e^2 +q_M^2  \right) \Bigl[ \frac{2 \left(a^2 z^2-2 a z-3\right) \log (a z+1)+a z (6-a z)+2 \log ^2(a z+1)}{4 a^4} \Bigl]  \nonumber \\
  + \frac{4 \text{Li}_2(-a z)+a z (-a z+2 (a z-2) \log (z)+4)+4 \log (z) \log (a z+1)}{2 a^2} \nonumber \\
   + \frac{a z (a z-2)-2 \log (a z+1) (a z (a z-2)+\log (a z+1)-1)}{2 a^2}  \,.
  \label{metspheicalcase1f1}
\end{align*}
Where $\text{Li}_2$ is the Polylogarithm function. This expression again reduces to the standard dyonic RN-AdS expression in the limit $a\rightarrow0$. Since the expressions for $\phi(z)$ and $B_t(z)$ remain the same as in the planar case, the scalar field continues to be regular and well-behaved everywhere outside the horizon. The profile of $g(z)$ and Kretschmann scalar, shown in Fig.~\ref{zvsgvsaqEPt11qMPt1zh1f1sphcase1}, further illustrates the smooth and well-behaved nature of the spacetime.

To investigate the local and thermodynamic stability of this spherical hairy dyonic black hole, we again need to compute various thermodynamic quantities. The expressions of charge and chemical potential remain the same as in the planar horizon case. However, because of the presence of $z^3 \log z$ term in the near boundary expansion of $g(z)$, the AMD prescription does not yield a sensible result for the black hole mass. Nonetheless, we can still use the holographic renormalization method to compute the black hole mass.

\subsubsection{Mass from holographic renormalization}
The procedure for obtaining the black hole mass and other thermodynamic quantities from the holographic renormalization method is exactly the same as discussed earlier for the planar horizon case. The difference arises only in the scalar boundary counterterms in the regularize action,
\begin{eqnarray}
S_{ren}^{G} = S_{EMS}^{on-shell} + \frac{1}{8 \pi G_4} \int_{\partial \mathcal{M}} \mathrm{d^3}x \ \sqrt{-\gamma} \Theta -\frac{1}{16 \pi G_4} \int_{\partial \mathcal{M}} \mathrm{d^3}x \ \sqrt{-\gamma} \left(4 + R^{(3)}\right) \nonumber \\
+ \frac{2}{16 \pi G_4} \int_{\partial \mathcal{M}} \mathrm{d^3}x \ \sqrt{-\gamma} \left( b_1 \phi^2 + b_2 \phi^4 + b_3 \phi^6 \log \phi \right) \,.
\label{actionregGibbssph}
\end{eqnarray}
The $\phi^6 \log{\phi}$ counterterm is needed to make sure that the additional logarithmic divergences in $S_{EMS}^{on-shell}$, arising due to the structure of $g(z)$, cancel out in $S_{ren}^{G}$. This renormalized action is suitable to study fixed potential ensemble. The corresponding stress energy tensor is then given by
\begin{eqnarray}
T_{\mu\nu}^{G} =\frac{1}{8 \pi G_4} \left[ \Theta \gamma_{\mu\nu} - \Theta_{\mu\nu}- 2 \gamma_{\mu\nu} + G^{(3)}_{\mu\nu} + \gamma_{\mu\nu} \left( b_1 \phi^2 + b_2 \phi^4 + b_3 \phi^6 \log\phi \right) \right]  \,,
\label{stresstensorgibbsshp}
\end{eqnarray}
whose $tt$ component will give us the following desired mass expression \footnote{This is actually the mass per unit area $\Omega_{2,1}$ of the boundary spatial section.},
\begin{eqnarray}
& & M_{HR}^G = \frac{a^2 \left(-a \left(4 a^2+5\right) z_h^2+8 \left(a^2+5\right) z_h-60 z_h \left(a z_h-2\right)
   \left(\coth ^{-1}\left(2 a z_h+1\right)-\log (4)\right)+30 a\right)}{360 \pi  G_4 \left(a z_h \left(a
   z_h-2\right)+2 \log \left(a z_h+1\right)\right)} \nonumber \\
& &  + \frac{4 a \left(-2 a^2+15 \log \left(a z_h\right)+5+60 \log (2)\right) \log \left(a z_h+1\right)+60 a
   \text{Li}_2\left(-a z_h\right)-30 a \log ^2\left(a z_h+1\right)}{360 \pi  G_4 \left(a z_h \left(a
   z_h-2\right)+2 \log \left(a z_h+1\right)\right)} \nonumber \\
& &  + \frac{a \mu_e^2 \left(2 \left(a^2 z_h^2-2 a z_h-3\right) \log \left(a z_h+1\right)+a z_h
   \left(6-a z_h\right)+2 \log ^2\left(a z_h+1\right)\right)}{48 \pi  G_4 \log ^2\left(a z_h+1\right)
   \left(a z_h \left(a z_h-2\right)+2 \log \left(a z_h+1\right)\right)} \nonumber \\
& &  +\frac{ q_M^2 \left(2 \left(a^2 z_h^2-2 a z_h-3\right) \log \left(a z_h+1\right)+a z_h \left(6-a
   z_h\right)+2 \log ^2\left(a z_h+1\right)\right)}{48 a \pi  G_4 \left(a z_h \left(a z_h-2\right)+2 \log
   \left(a z_h+1\right)\right)} \,.
   \label{massGibbsphcase1}
\end{eqnarray}
For the fixed charge ensemble, as discussed before,  the renormalized action would need to be modified appropriately,
\begin{eqnarray}
& & S_{ren}^{F} = S_{EMS}^{on-shell} + \frac{1}{8 \pi G_4} \int_{\partial \mathcal{M}} \mathrm{d^3}x \ \sqrt{-\gamma} \Theta -\frac{1}{16 \pi G_4} \int_{\partial \mathcal{M}} \mathrm{d^3}x \ \sqrt{-\gamma} \left(4 + R^{(3)}\right) + \nonumber \\
& & \frac{2}{16 \pi G_4} \int_{\partial \mathcal{M}} \mathrm{d^3}x \ \sqrt{-\gamma} \left( b_1 \phi^2 + b_2 \phi^4 + b_3 \phi^6 \log \phi \right) +\frac{1}{16 \pi G_4}\int_{\partial \mathcal{M}} \mathrm{d^3}x \ \sqrt{-\gamma} n_r f(\phi) F^{r\mu} B_\mu \,, \nonumber \\
\label{actionregHelmholtzsph}
\end{eqnarray}
the corresponding energy momentum tensor is then
\begin{eqnarray}
T_{\mu\nu}^{F} =\frac{1}{8 \pi G_4} \left[ \Theta \gamma_{\mu\nu} - \Theta_{\mu\nu}- 2 \gamma_{\mu\nu} + G^{(3)}_{\mu\nu} + \gamma_{\mu\nu} \left( b_1 \phi^2 + b_2 \phi^4 + b_3 \phi^6 \log \phi \right) \right] \nonumber \\
 + \frac{1}{8 \pi G_4} \left[ \gamma_{\mu\nu} \frac{f(\phi)}{2}n_r F^{r\rho}B_\rho - f(\phi)n^{\rho}F_{\rho j}B_{i} \right]  \,,
\label{stresstensorHelmholtzplanar}
\end{eqnarray}
the $tt$ component of which gives us the mass
\begin{eqnarray}
& & M_{HR}^F = M_{HR}^G - \frac{q_e^2 \log \left(a z_h+1\right)}{16 \pi  a G_4} \,.
   \label{massHelmholtzsphcase1}
\end{eqnarray}
Again, this mass $M_{HR}^F$ is not exactly the mass that appears in the Helmholtz free energy but differs by a factor of $Q_e \mu_e$.

\subsubsection{Black hole thermodynamics}
Let us first discuss the black hole thermodynamics in the grand canonical ensemble. From the renormalized on-shell action, we get the Gibbs free energy per unit area as
\begin{eqnarray*}
& &G = \frac{1}{360 \pi  G_4 z_h \left(a z_h \left(a z_h-2\right)+2 \log \left(a z_h+1\right)\right)} \Bigl[
  -30 a z_h \text{Li}_2\left(-a z_h\right)-75 a z_h \log ^2\left(a z_h+1\right) \Bigl.\\
& &    \Bigr. + \log \left(a z_h+1\right) \left(a \left(-8 a^2-25+240 \log 2\right) z_h+60 a z_h \log \left(a
   z_h\right)+90\right) -5 a z_h \left(3 a^2-8 a z_h+18\right)  \Bigl.\\
 & &   \Bigr. -a^2 z_h^2 \left(a \left(4 a \left(a z_h-2\right)+5 z_h\right)-60 \log (4) \left(a z_h-2\right)+60 \left(a
   z_h-2\right) \coth ^{-1}\left(2 a z_h+1\right)\right)
  \Bigr]  \nonumber \\
& &-\frac{a \mu_e^2 \left(2 \left(a^2 z_h^2-2 a z_h-3\right) \log \left(a z_h+1\right)+a z_h
   \left(6-a z_h\right)+2 \log ^2\left(a z_h+1\right)\right)}{96 \pi  G_4 \log ^2\left(a z_h+1\right)
   \left(a z_h \left(a z_h-2\right)+2 \log \left(a z_h+1\right)\right)} \nonumber \\
& & + \frac{q_M^2 \left(\left(4 a^2 z_h^2-8 a z_h+6\right) \log \left(a z_h+1\right)+a z_h \left(a
   z_h-6\right)+10 \log ^2\left(a z_h+1\right)\right)}{96 \pi  a G_4 \left(a z_h \left(a z_h-2\right)+2
   \log \left(a z_h+1\right)\right)} 
\end{eqnarray*}
The free energy satisfies the expected thermodynamic relation $G=M_{HR}^{G} - T S_{BH} - Q_e \mu_e$, and reduces to the standard dyonic expression in the limit $a\rightarrow0$. This is a consistency check for the thermodynamic formulae found here for the spherical dyonic hairy black holes.

\begin{figure}[h!]
\begin{minipage}[b]{0.5\linewidth}
\centering
\includegraphics[width=2.8in,height=2.3in]{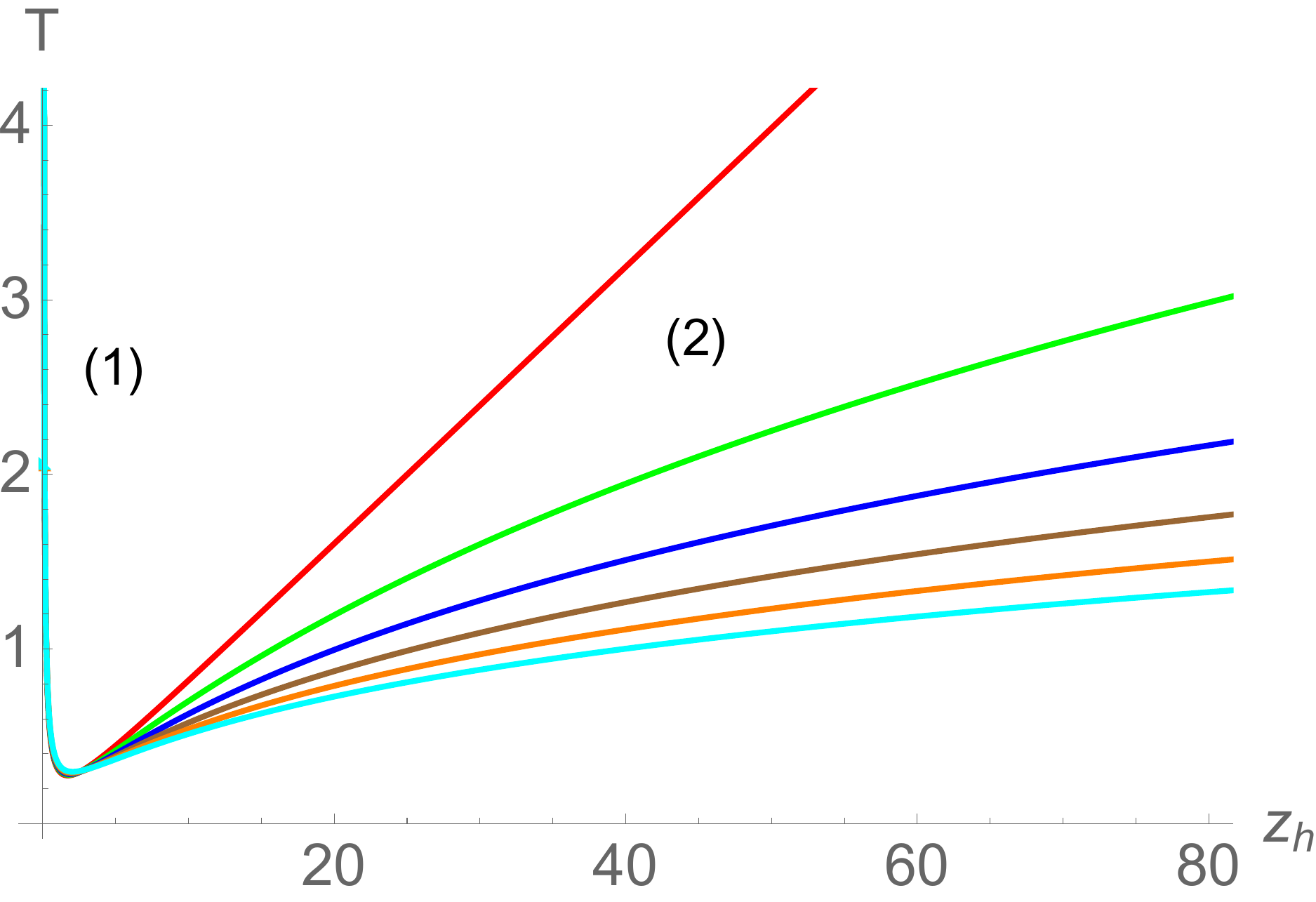}
\caption{ \small Hawking temperature $T$ as a function of horizon radius $z_h$ for various values of $a$. Here $\mu_e=0$ and $q_M=0$ are used. Red, green, blue, brown, orange, and magenta curves correspond to $a=0$, $0.05$, $0.10$, $0.15$, $0.20$, and $0.25$ respectively.}
\label{zhvsTvsaMu0qM0sphcase1}
\end{minipage}
\hspace{0.4cm}
\begin{minipage}[b]{0.5\linewidth}
\centering
\includegraphics[width=2.8in,height=2.3in]{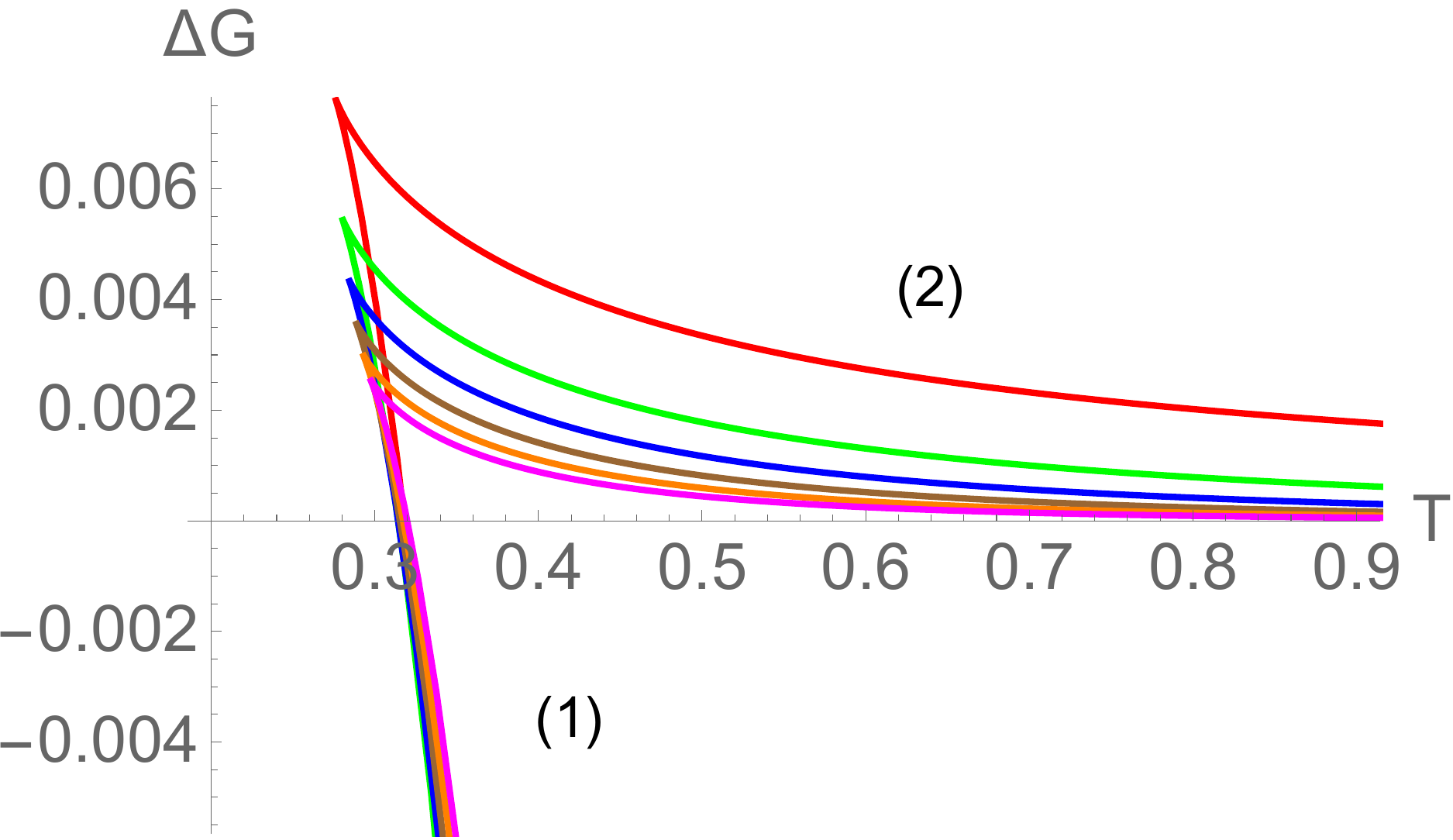}
\caption{\small Gibbs free energy difference $\Delta G$ as a function of Hawking temperature $T$ for various values of $a$. Here $\mu_e=0$ and $q_M=0$ are used. Red, green, blue, brown, orange, and cyan curves correspond to $a=0$, $0.05$, $0.10$, $0.15$, $0.20$, and $0.25$ respectively.}
\label{TvsdeltaGvsaMu0qM0sphcase1}
\end{minipage}
\end{figure}

In Fig.~\ref{zhvsTvsaMu0qM0sphcase1}, we have shown the behaviour of Hawking temperature for $\mu_e=0$ and $q_M=0$. As is well known, the usual non-hairy Schwarzschild AdS black hole exists only above a certain minimum temperature $T_{min}$ and below this minimum temperature the black hole ceases to exist, thereby exhibiting an interesting Hawking/Page transition between the Schwarzschild AdS black hole and thermal-AdS. We find that this interesting behaviour and phase transition continue to hold in the presence of scalar hair as well. In particular, there again appear two black hole branches, large and small, at all temperatures $T\geq T_{min}$. The large black hole branch [indicated by (1) in Fig.~\ref{zhvsTvsaMu0qM0sphcase1}] has a positive specific heat and is stable whereas the small black hole branch [indicated by (2) in Fig.~\ref{zhvsTvsaMu0qM0sphcase1}] has a negative specific heat and is unstable. Moreover, the free energy of the large black hole branch is always smaller than the small black hole branch. However, the free energy of the large black hole can become higher than the thermal-AdS at lower temperatures, implying the Hawking/page phase transition between them. This is shown in Fig.~\ref{TvsdeltaGvsaMu0qM0sphcase1}, where the free energy difference between hairy black hole and thermal-AdS is plotted for various values of $a$.
\begin{figure}[h!]
\begin{minipage}[b]{0.5\linewidth}
\centering
\includegraphics[width=2.8in,height=2.3in]{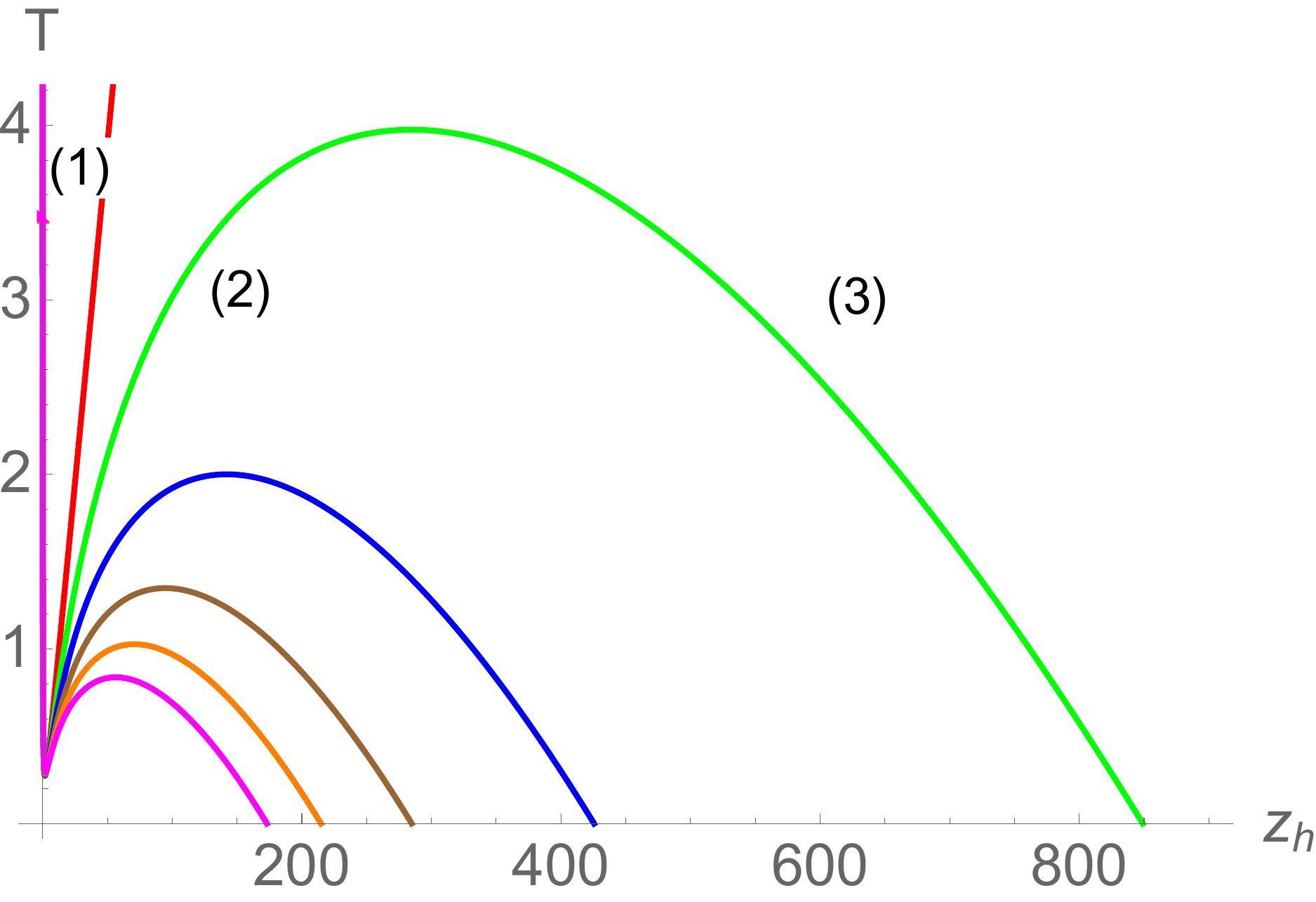}
\caption{ \small Hawking temperature $T$ as a function of horizon radius $z_h$ for various values of $a$. Here $\mu_e=0.3$ and $q_M=0$ are used. Red, green, blue, brown, orange, and magenta curves correspond to $a=0$, $0.05$, $0.10$, $0.15$, $0.20$, and $0.25$ respectively.}
\label{zhvsTvsaMuPt3qM0sphcase1}
\end{minipage}
\hspace{0.4cm}
\begin{minipage}[b]{0.5\linewidth}
\centering
\includegraphics[width=2.8in,height=2.3in]{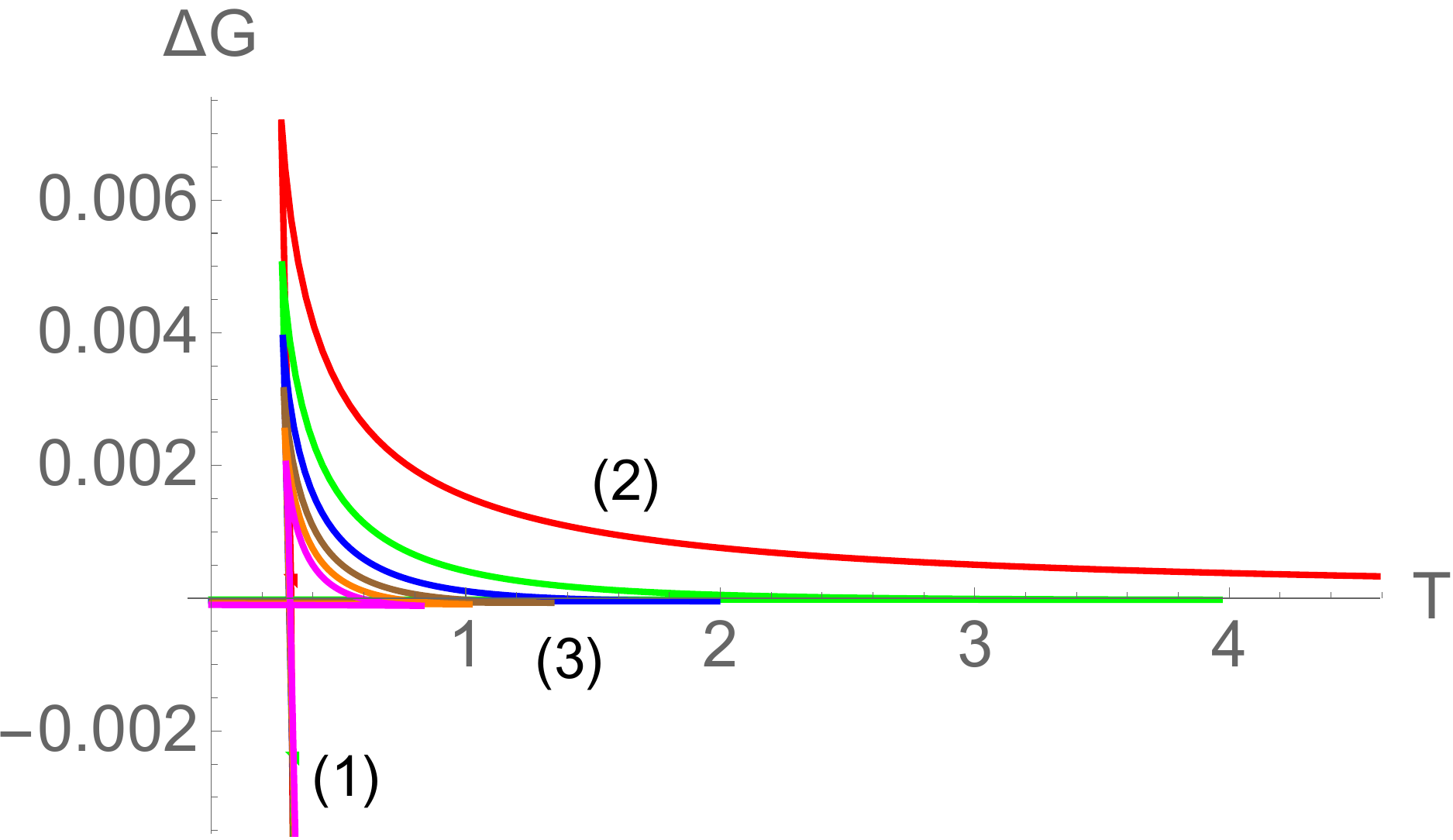}
\caption{\small Gibbs free energy $\Delta G$ as a function of Hawking temperature $T$ for various values of $a$. Here $\mu_e=0.3$ and $q_M=0$ are used. Red, green, blue, brown, orange, and magenta curves correspond to $a=0$, $0.05$, $0.10$, $0.15$, $0.20$, and $0.25$ respectively.}
\label{TvsdeltaGvsaMuPt3qM0sphcase1}
\end{minipage}
\end{figure}

The thermodynamic structure becomes even more interesting for the small but finite chemical potential $\mu_e$. For a small chemical potential, still in the case $q_M=0$, the Hawking/page phase transition continue to exists, with a large stable black hole solution dominating the thermodynamics at a higher temperature whereas the thermal-AdS solution dominating at a lower temperature.  Interestingly, with scalar hair, a new small black hole branch can appear, which remains stable at low temperature [indicated by (3) in Fig.~\ref{zhvsTvsaMuPt3qM0sphcase1}]. In particular, the Hawking temperature now has local minima and maxima and vanishes at a finite radius $z_h^{ext}$, \textit{i.e.}, at least one stable black hole branch always exists all at each temperature. Moreover, the magnitude of this $z_h^{ext}$ decreases with $a$.  The free energy behaviour, shown in Fig.~\ref{TvsdeltaGvsaMuPt3qM0sphcase1}, further suggests a first order phase transition between the large black hole branch (1) and small black hole branch (3) as the temperature is lowered. This is the famous small/large black hole phase in the context of charged AdS black hole \cite{Chamblin:1999tk,Chamblin:1999hg,Dey:2015ytd,Mahapatra:2016dae,Cvetic:1999ne}. Notice that the free energy of the unstable second branch (2) is always higher than the stable first and third branches and therefore is  always thermodynamically disfavoured.

At this point, it is important to emphasize that the usual spherical RN-AdS black hole exhibits only the Hawking/Page phase transition in the grand canonical ensemble, and the small/large black hole phase transition appears only in the canonical ensemble. Here, we see that in the presence of scalar hair the small/large black hole phase transition can take place in the grand canonical ensemble as well.

For higher values of $a$, only one stable black hole branch appears that remains thermodynamically preferred at all temperatures. In particular, the size of the second branch decreases with $a$, and then completely disappears. This leads to the merging of small and large black hole branches to form a single black hole branch that remains stable at all temperatures $T\geq 0$. Therefore, the small/large phase transition ceases to exist at higher values of $a$. A similar scenario persists for larger values of chemical potential as well. Overall, this thermodynamic behaviour is akin to the famous Van der Waals type phase transition, where a first order critical line stops at a second order critical point.

\begin{figure}[h!]
\begin{minipage}[b]{0.5\linewidth}
\centering
\includegraphics[width=2.8in,height=2.3in]{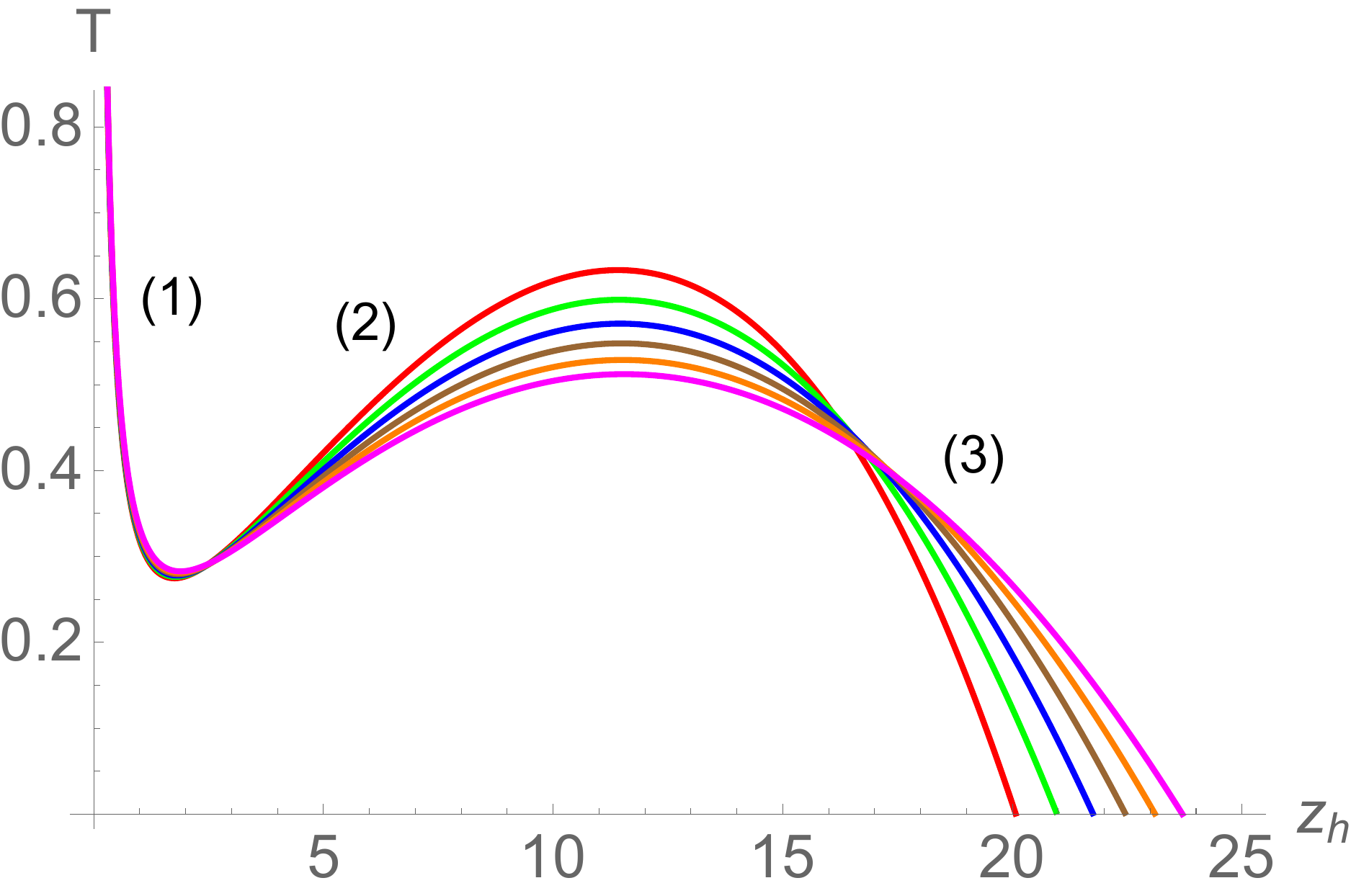}
\caption{ \small Hawking temperature $T$ as a function of horizon radius $z_h$ for various values of $a$. Here $\mu_e=0$ and $q_M=0.1$ are used. Red, green, blue, brown, orange, and magenta curves correspond to $a=0$, $0.02$, $0.04$, $0.06$, $0.08$, and $0.10$ respectively.}
\label{zhvsTvsaMu0qMPt1sphcase1}
\end{minipage}
\hspace{0.4cm}
\begin{minipage}[b]{0.5\linewidth}
\centering
\includegraphics[width=2.8in,height=2.3in]{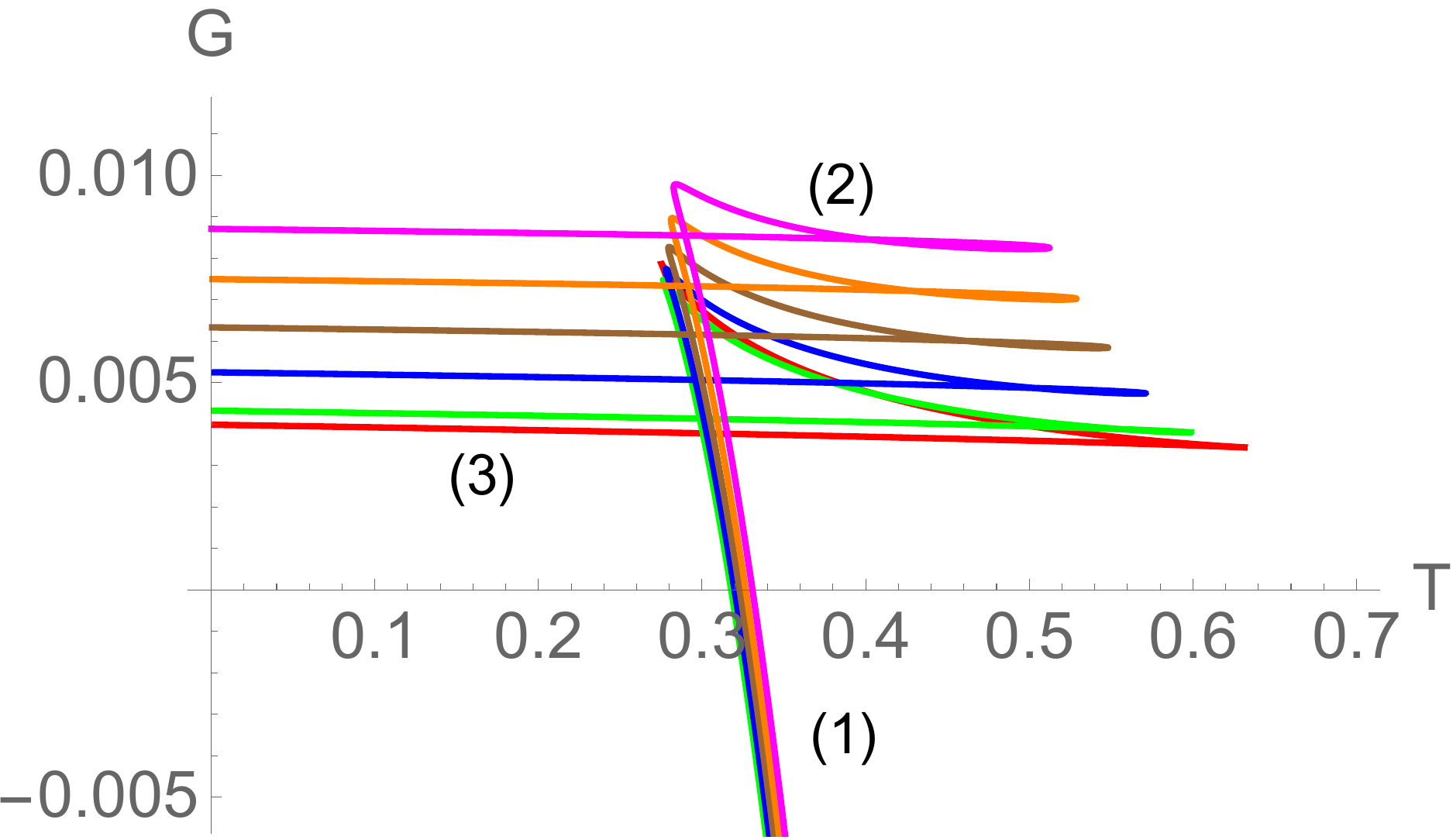}
\caption{\small Gibbs free energy $G$ as a function of Hawking temperature $T$ for various values of $a$. Here $\mu_e=0$ and $q_M=0.1$ are used. Red, green, blue, brown, orange, and magenta curves correspond to $a=0$, $0.02$, $0.04$, $0.06$, $0.08$, and $0.10$ respectively.}
\label{TvsdeltaGvsaMu0qMPt1sphcase1}
\end{minipage}
\end{figure}

\begin{figure}[h!]
\begin{minipage}[b]{0.5\linewidth}
\centering
\includegraphics[width=2.8in,height=2.3in]{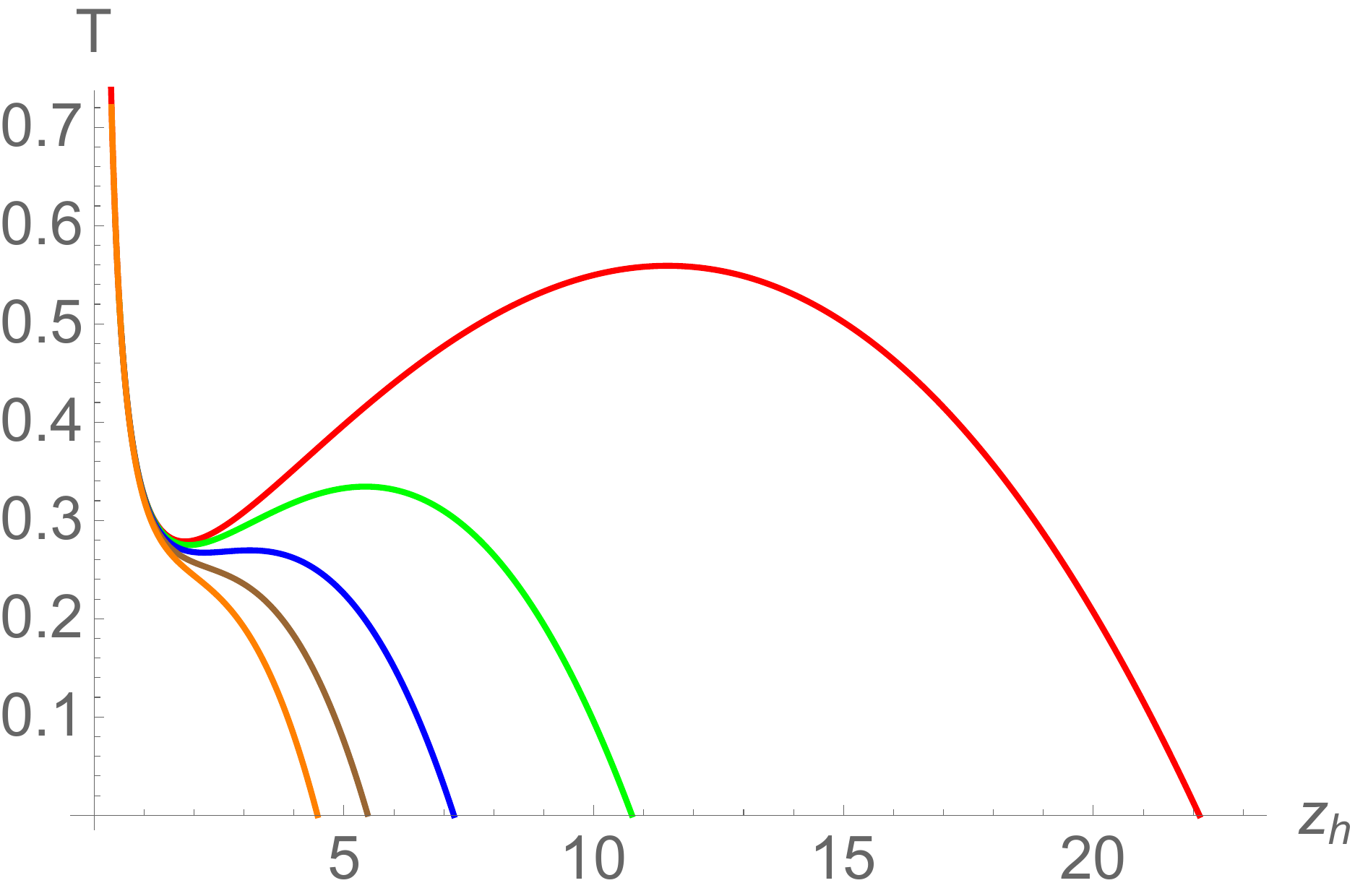}
\caption{ \small Hawking temperature $T$ as a function of horizon radius $z_h$ for various values of $a$. Here $\mu_e=0$ and $a=0.05$ are used. Red, green, blue, brown, and orange curves correspond to $q_M=0.1$, $0.2$, $0.3$, $0.4$, and $0.5$ respectively.}
\label{zhvsTvsavsqMMu0sphcase1}
\end{minipage}
\hspace{0.4cm}
\begin{minipage}[b]{0.5\linewidth}
\centering
\includegraphics[width=2.8in,height=2.3in]{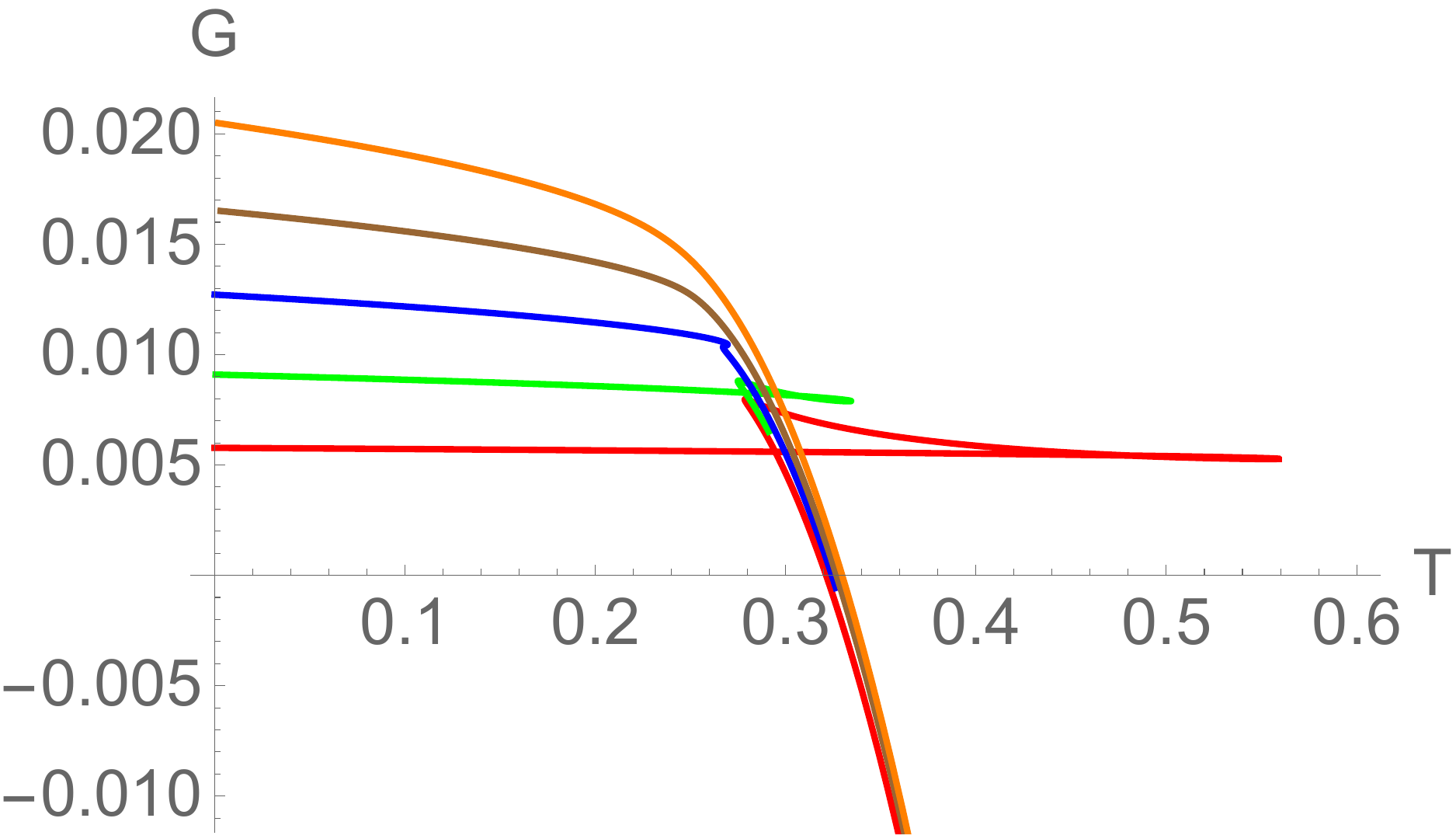}
\caption{\small Gibbs free energy $G$ as a function of Hawking temperature $T$ for various values of $a$. Here $\mu_e=0$ and $a=0.05$ are used. Red, green, blue, brown, and orange curves correspond to $q_M=0.1$, $0.2$, $0.3$, $0.4$, and $0.5$ respectively.}
\label{TvsGvsavsqMMu0sphcase1}
\end{minipage}
\end{figure}

Similar results appear for finite values of $q_M$ as well. In particular, for small $q_M$, there again exists three black hole branches: two stable and one unstable. The stable first and third branches are always  thermodynamically favoured over the unstable second branch. The stable first and third black hole branches further undergo a small/large black hole phase transition as the temperature is varied. This is shown in Figs.~\ref{zhvsTvsaMu0qMPt1sphcase1} and \ref{TvsdeltaGvsaMu0qMPt1sphcase1}.  However, for higher values of $q_M$, the size of the unstable second branch (2) start decreasing and ultimately disappear. Therefore, for large $q_M$ only one stable black hole branch exists and  there is no phase transition between black holes. This is shown in Figs.~\ref{zhvsTvsavsqMMu0sphcase1} and \ref{TvsGvsavsqMMu0sphcase1}.

We now discuss the hairy black hole thermodynamics in the canonical ensemble. Since most of the thermodynamic results are completely analogous to the above discussion, we will be very brief here. From the renormalized on-shell action [Eq.~(\ref{actionregHelmholtzsph})], the Helmholtz free energy per unit area $F$ can be obtained. We find that it is simply given by
\begin{eqnarray}
& & F =  G + Q_e \mu_e = M_{HR}^{G} - T S_{BH}  \,,
\label{Helmholtzsph}
\end{eqnarray}
thereby, again satisfying the expected thermodynamic relation.

\begin{figure}[h!]
\begin{minipage}[b]{0.5\linewidth}
\centering
\includegraphics[width=2.8in,height=2.3in]{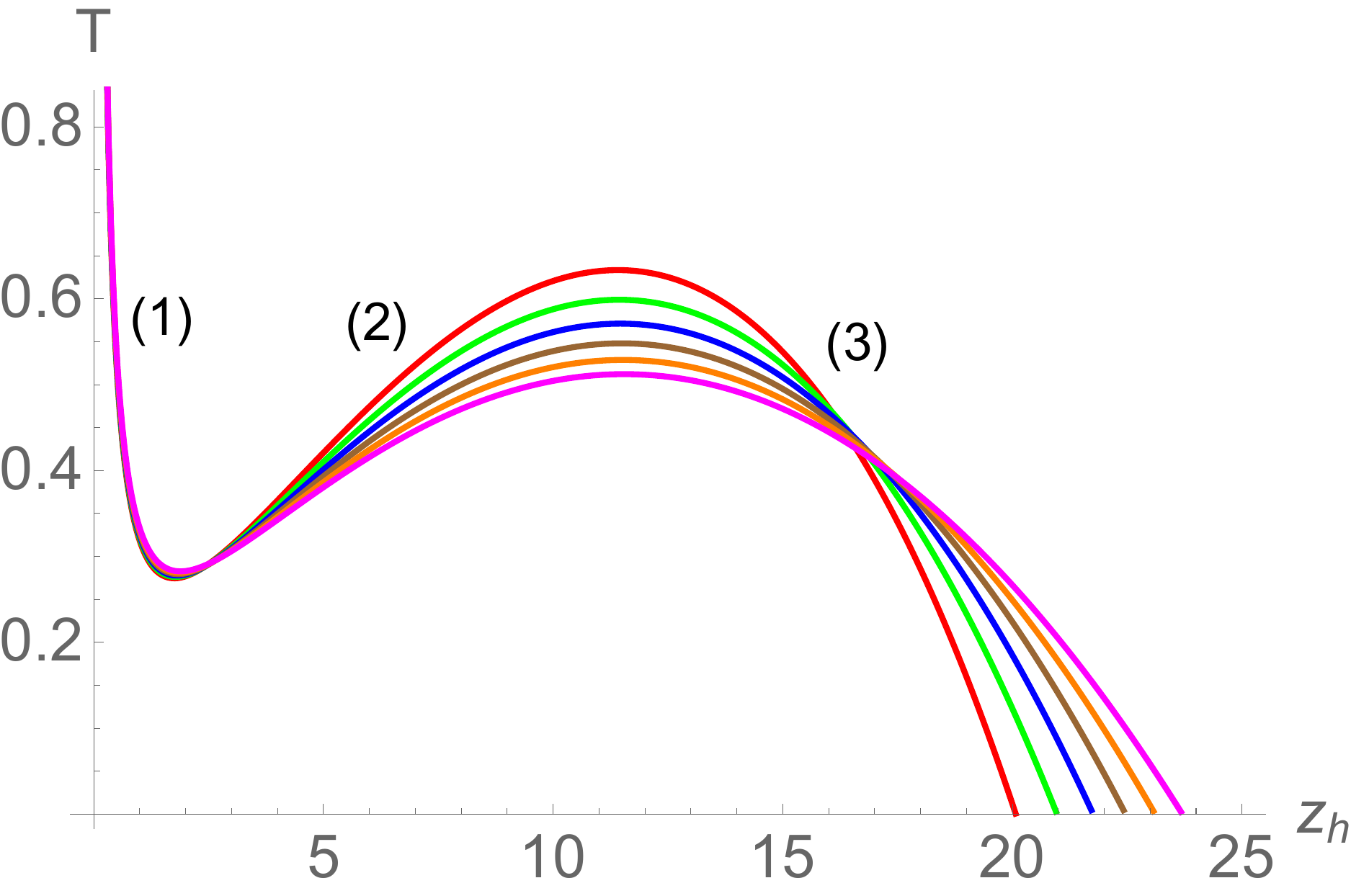}
\caption{ \small Hawking temperature $T$ as a function of horizon radius $z_h$ for various values of $a$. Here $q_e=0.1$ and $q_m=0$ are used. Red, green, blue, brown, orange, and magenta curves correspond to $a=0.0$, $0.02$, $0.04$, $0.06$, $0.08$, and $0.1$ respectively.}
\label{zhvsTvsaqEPt1qM0sphcase1}
\end{minipage}
\hspace{0.4cm}
\begin{minipage}[b]{0.5\linewidth}
\centering
\includegraphics[width=2.8in,height=2.3in]{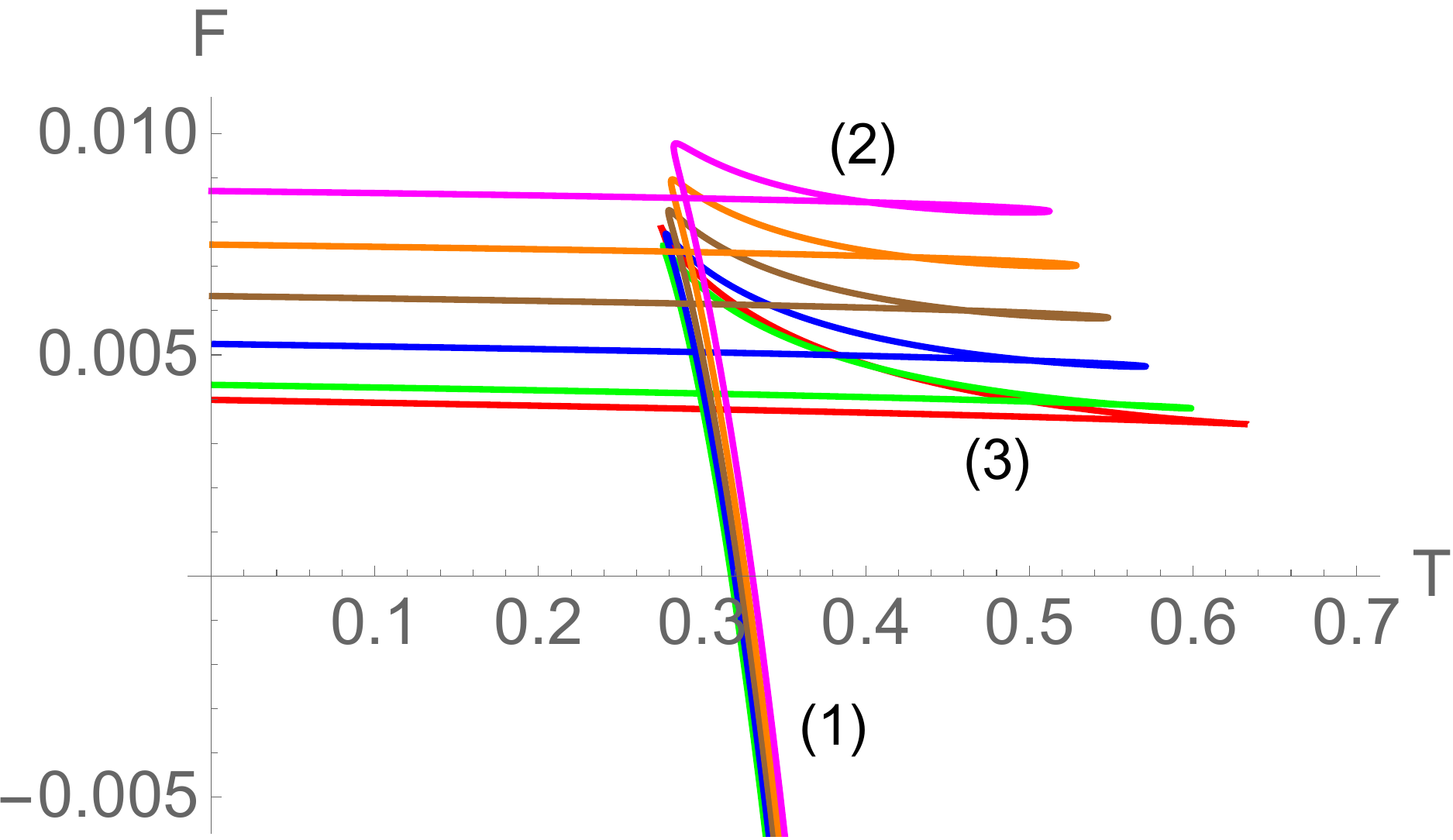}
\caption{\small Helmholtz free energy $F$ as a function of Hawking temperature $T$ for various values of $a$. Here $q_e=0.1$ and $q_m=0$ are used. Red, green, blue, brown, orange, and magenta curves correspond to $a=0.0$, $0.02$, $0.04$, $0.06$, $0.08$, and $0.1$ respectively.}
\label{TvsFvsaqEPt1qM0sphcase1}
\end{minipage}
\end{figure}

\begin{figure}[h!]
\begin{minipage}[b]{0.5\linewidth}
\centering
\includegraphics[width=2.8in,height=2.3in]{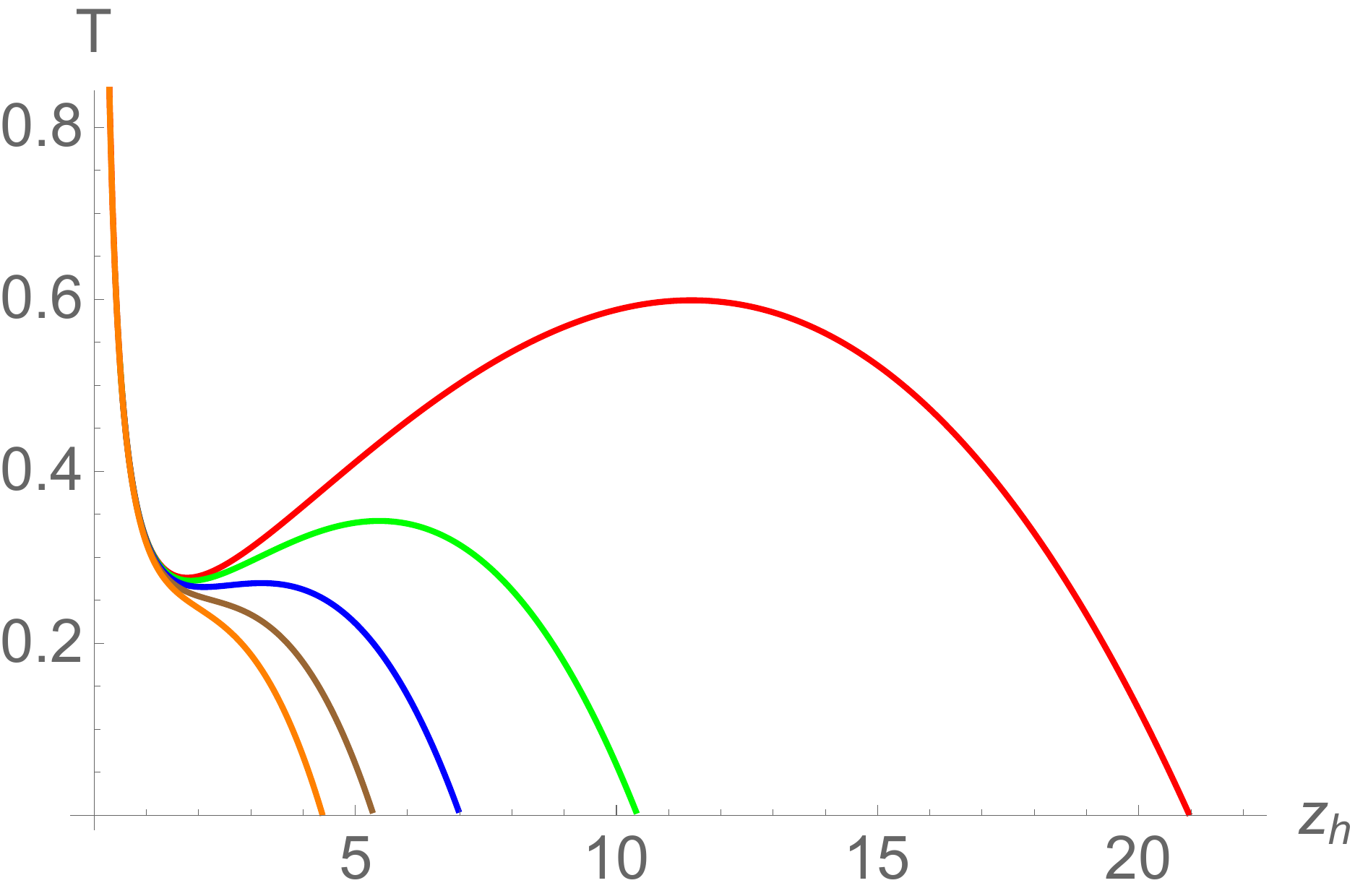}
\caption{ \small Hawking temperature $T$ as a function of horizon radius $z_h$ for various values of $a$. Here $q_M=0$ and $a=0.02$ are used. Red, green, blue, brown, and orange curves correspond to $q_e=0.1$, $0.2$, $0.3$, $0.4$, and $0.5$ respectively.}
\label{zhvsTvsqEaPt02qM0sphcase1}
\end{minipage}
\hspace{0.4cm}
\begin{minipage}[b]{0.5\linewidth}
\centering
\includegraphics[width=2.8in,height=2.3in]{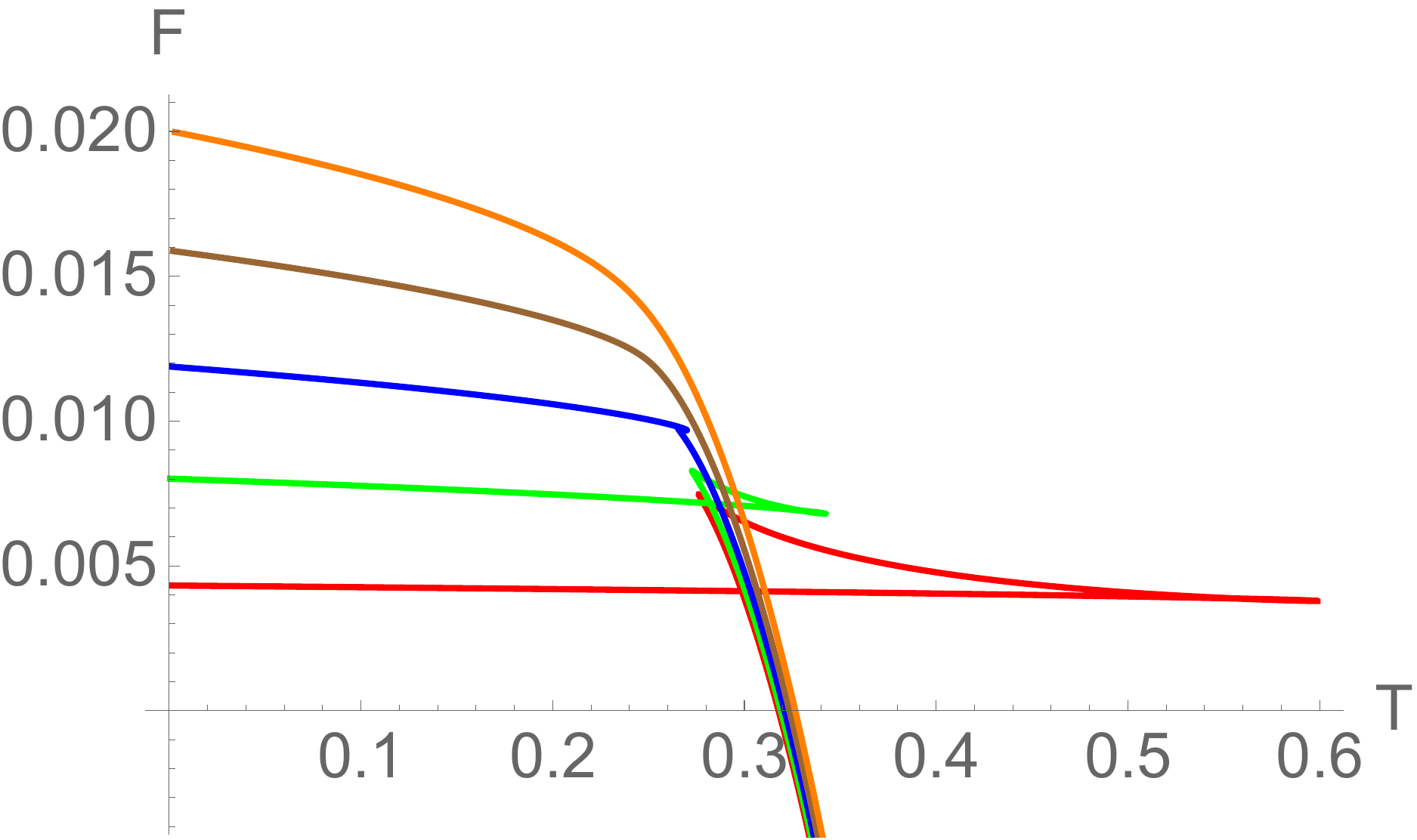}
\caption{\small Helmholtz free energy $F$ as a function of Hawking temperature $T$ for various values of $a$. Here $q_M=0$ and $a=0.02$ are used. Red, green, blue, brown, and orange curves correspond to $q_e=0.1$, $0.2$, $0.3$, $0.4$, and $0.5$ respectively.}
\label{TvsFvsqEaPt02qM0sphcase1}
\end{minipage}
\end{figure}

The thermodynamic behaviour of $q_e=0$ case is exactly similar to the $\mu_e=0$ case discussed above.  The thermodynamic behaviour for a small but finite fixed charge $q_e$ is shown in Figs.~\ref{zhvsTvsaqEPt1qM0sphcase1} and \ref{TvsFvsaqEPt1qM0sphcase1}. It is well known that the usual charged RN-AdS spherical black hole exhibits a swallow-tail like structure in the free energy and undergoes a small/large black hole phase transition as the temperature is altered in the canonical ensemble \cite{Chamblin:1999tk}. We find that similar results persist for the charged hairy cases as well. The difference arises in the magnitude of  the small/large black hole transition temperature, which increases as the parameter $a$ increases \footnote{Due to numerical artifacts, it is difficult to exactly pin point the small/large black hole phase transition temperature for large $a$ values.}. Similarly, most of the results for higher $q_e$ values persist as well. In particular, there appears a critical charge $q_e^{crit}$ above which the unstable branch [indicated by (2) in Figs.~\ref{zhvsTvsaqEPt1qM0sphcase1} and \ref{TvsFvsaqEPt1qM0sphcase1}] disappears and we have a single black hole branch, which is stable at all temperatures, \textit{i.e.}, the small/large black hole phase transition ceases to exist above $q_e^{crit}$. This is shown in Figs.~\ref{zhvsTvsqEaPt02qM0sphcase1} and \ref{TvsFvsqEaPt02qM0sphcase1}. The $q_e^{crit}$ therefore defines a second order critical point on which the first order small/large black hole phase transition line stops. As usual, the magnitude of $q_e^{crit}$ can be found by analysing the inflection point of temperature.

As we vary $q_M$, the thermodynamic phase diagram remains quite similar to what we have seen for varying $q_e$. Here as well, there occurs a critical magnetic charge $q_M^{crit}$ at which the first order small/large black hole phase transition line terminates and the nucleation of other two branches appear. The remembrance of fixed $q_e$ and $q_M$ thermodynamics is expected considering that the metric is symmetric in electric and magnetic charges. Hence, it is expected that the pure constant magnetic charge system behaves analogously to the pure constant electric charge system.

This completes our discussion on the spherical hairy dyonic black holes. We have established the resemblance of hairy dyonic black holes phase transitions with the liquid-gas phase transition. One can in principle also do a thorough analysis of the thermodynamic equation of state, the dependence of small/large black hole phase transition temperature and critical points $q_e^{crit}$ and $q_M^{crit}$ on various parameters, thermodynamic fluctuations and critical exponents etc in our model. However, we feel that investigation of these important issues is worthy of a separate independent study. The investigation of these issues here will not only make the whole paper a little bit bulky but also take us away from the main aim of our paper. Therefore, we postpone such study for future work.

\subsection{Hyperbolic horizon: $\kappa=-1$}
\begin{figure}[ht]
	\subfigure[]{
		\includegraphics[scale=0.4]{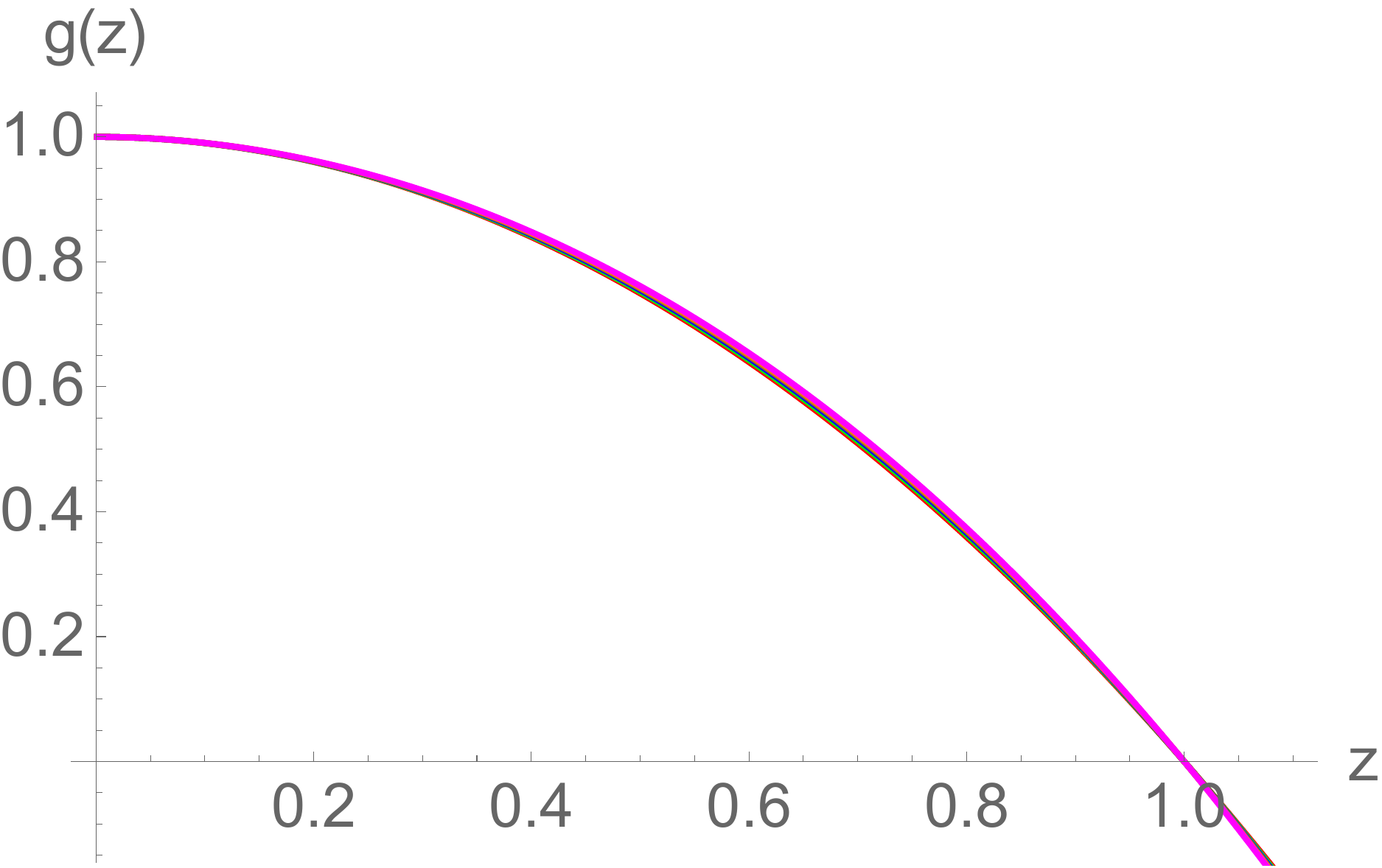}
	}
	\subfigure[]{
		\includegraphics[scale=0.4]{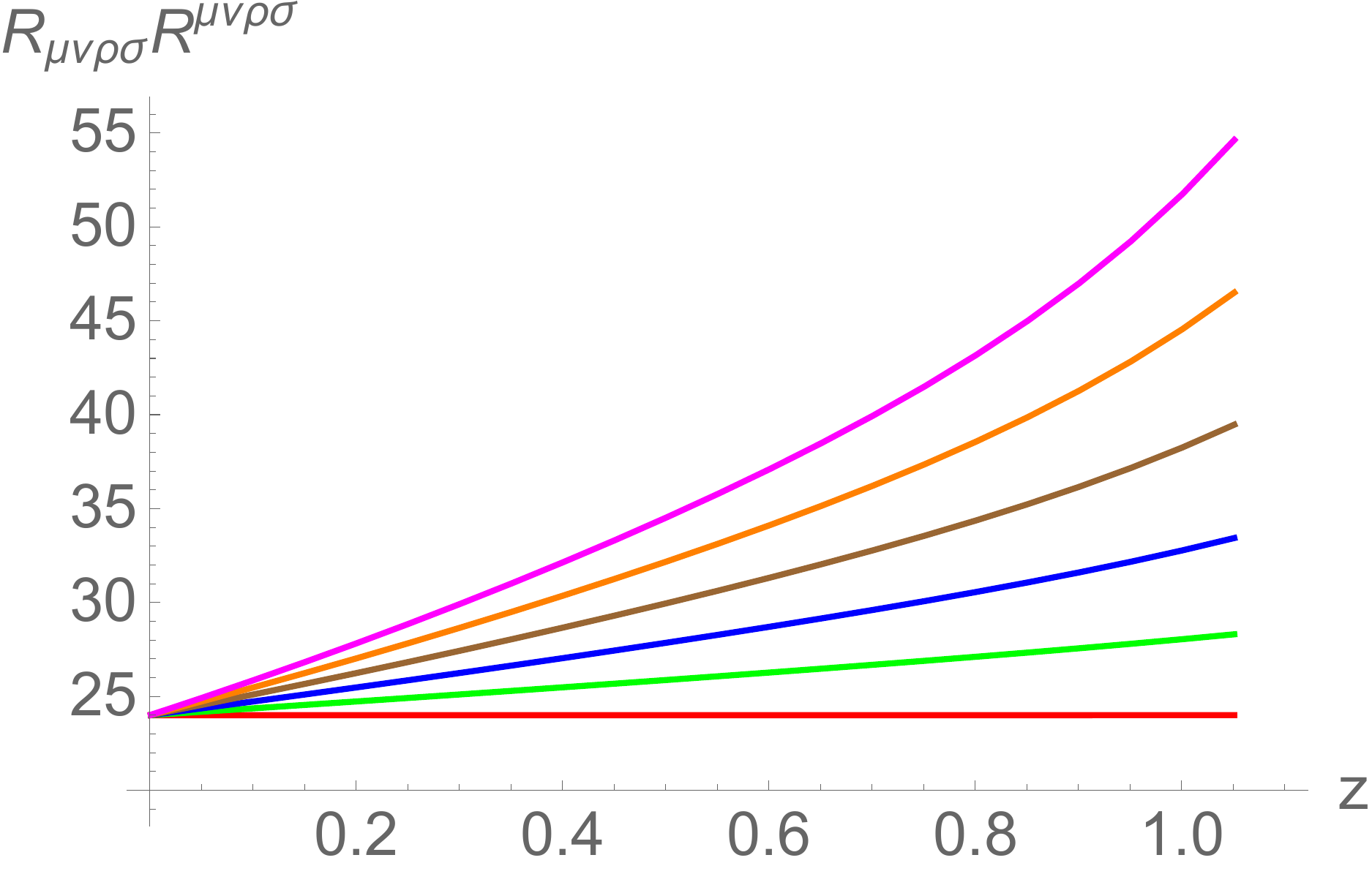}
	}
	\caption{\small The behavior of $g(z)$ and $R_{MNPQ}R^{MNPQ}$ for different values of hair parameter $a$. Here $z_h=1$, $\mu_e=0.1$, and $q_M=0.1$ are used. Red, green, blue, brown, orange, and cyan curves correspond to $a=0$, $0.05$, $0.10$, $0.15$, $0.20$, and $0.25$ respectively.}
	\label{zvskmannvsamu1qM1zh1f1hypercase1}
\end{figure}

We now briefly discuss the dyonic hairy black hole solution and thermodynamics with the hyperbolic horizon. Analytic expression of $g(z)$ can be found for $\kappa=-1$ as well. In Fig.~\ref{zvskmannvsamu1qM1zh1f1hypercase1}, the variation of $g(z)$ and Kretschmann scalar is plotted. Once again we have a well-behaved and smooth hairy black hole geometry having no divergences outside the horizon. Similarly, the scalar field remains finite and regular everywhere as well.

\begin{figure}[h!]
\begin{minipage}[b]{0.5\linewidth}
\centering
\includegraphics[width=2.8in,height=2.3in]{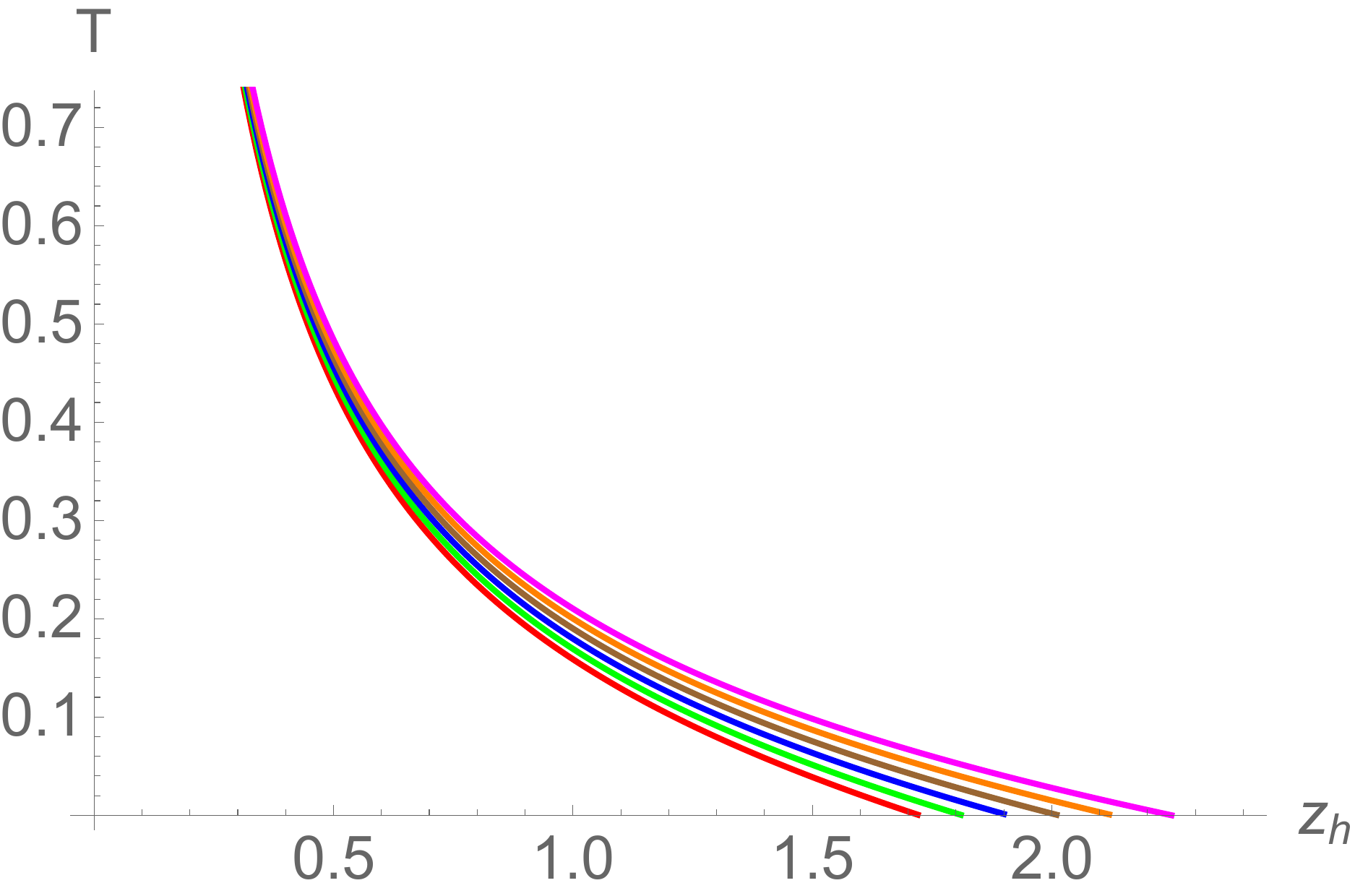}
\caption{ \small Hawking temperature $T$ as a function of horizon radius $z_h$ for various values of $a$. Here $\mu_e=0.1$ and $q_M=0.1$ are used. Red, green, blue, brown, orange and magenta curves correspond to $a=0$, $0.05$, $0.10$, $0.15$, $0.20$ and $0.25$ respectively.}
\label{zhvsTvsaMuPt1qMPt1hypercase1}
\end{minipage}
\hspace{0.4cm}
\begin{minipage}[b]{0.5\linewidth}
\centering
\includegraphics[width=2.8in,height=2.3in]{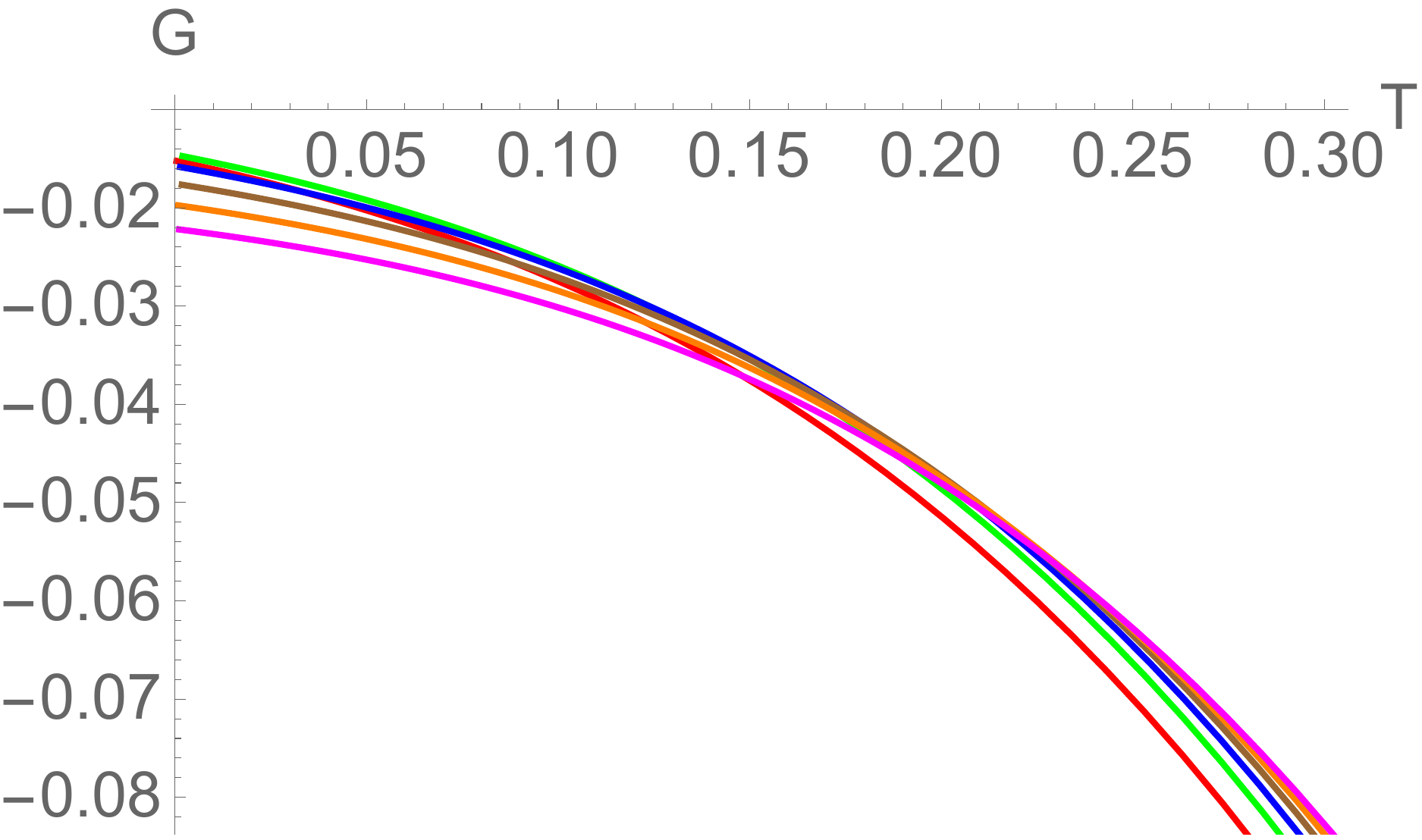}
\caption{\small Gibbs free energy $G$ as a function of Hawking temperature $T$ for various values of $a$. Here $\mu_e=0.1$ and $q_M=0.1$ are used. Red, green, blue, brown, orange, and magenta curves correspond to $a=0$, $0.05$, $0.10$, $0.15$, $0.20$, and $0.25$ respectively.}
\label{TvsGvsaMuPt1qMPt1hypercase1}
\end{minipage}
\end{figure}

Similarly, we can again obtain analytic expressions of conserved charges and free energies. No additional scalar counterterms, than those already present in the spherical case [Eq.~(\ref{actionregGibbssph})], are needed to make the on-shell action finite. In Figs.~\ref{zhvsTvsaMuPt1qMPt1hypercase1} and \ref{TvsGvsaMuPt1qMPt1hypercase1}, the thermodynamic stability and phase structure of the hairy hyperbolic black hole solution in the grand canonical ensemble is shown.  As in the planar horizon case, there is only one black hole branch. Since heat capacity at constant potential is always positive for this branch, it indicates that the hairy hyperbolic black holes are thermally stable. Moreover, there is also no Hawking/Page type phase transition and the Gibbs free energy is always negative, provided that $q_M$ is not too large. Similar
to the case of planar horizon, the free energy of hairy black hole can become smaller than the non-hairy black hole at low temperatures. This should be distinguish from \cite{Mahapatra:2020wym}, where the free energy of hairy hyperbolic black hole was found to be always higher than the non-hairy black hole. We similarly find stable and thermodynamically favoured hairy black holes for other values of $\mu_e$ and $q_M$ as well.

\begin{figure}[h!]
\begin{minipage}[b]{0.5\linewidth}
\centering
\includegraphics[width=2.8in,height=2.3in]{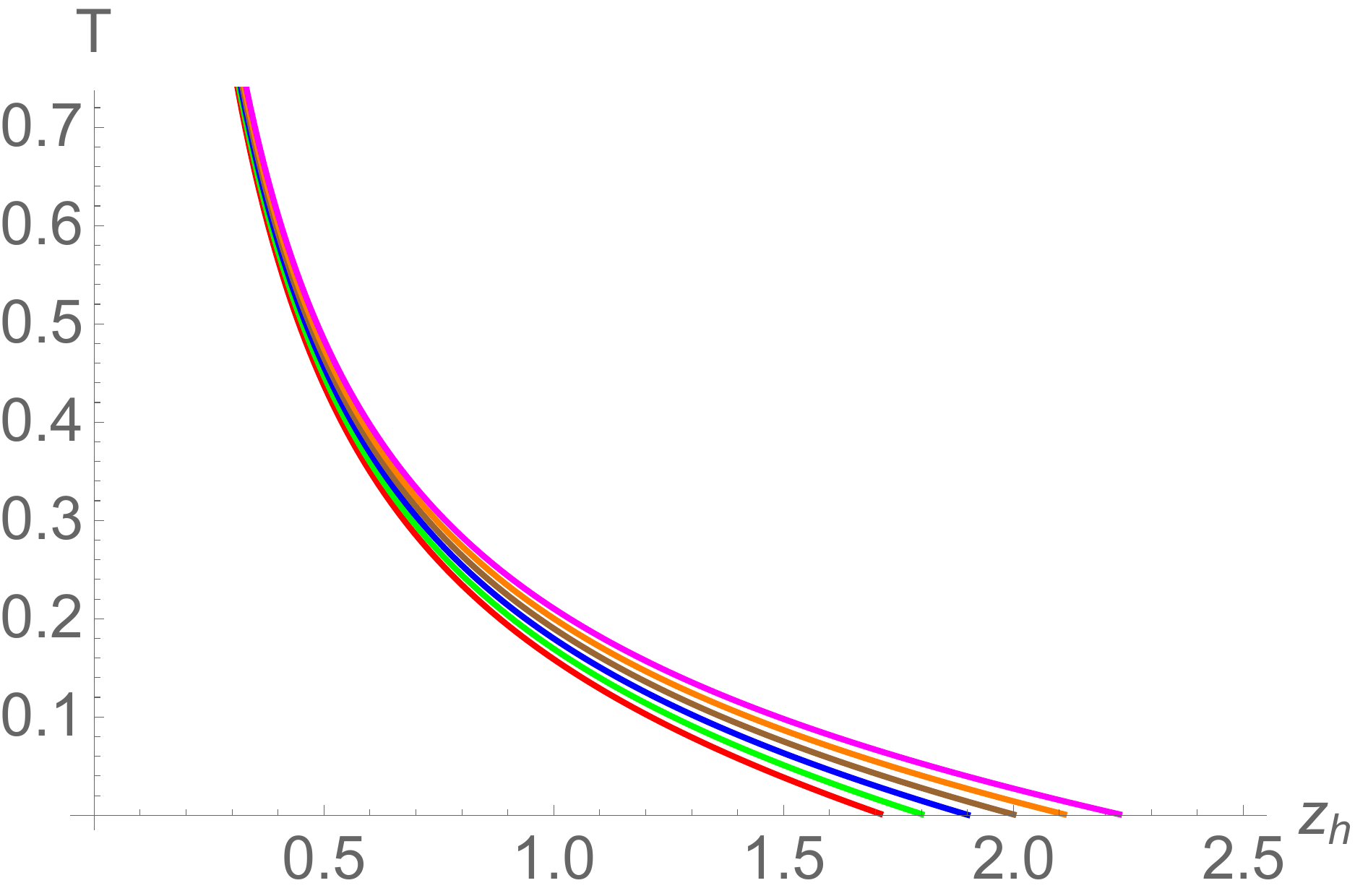}
\caption{ \small Hawking temperature $T$ as a function of horizon radius $z_h$ for various values of $a$. Here $q_e=0.1$ and $q_M=0.1$ are used. Red, green, blue, brown, orange, and magenta curves correspond to $a=0$, $0.05$, $0.10$, $0.15$, $0.20$, and $0.25$ respectively.}
\label{zhvsTvsaqEPt1qMPt1hypercase1}
\end{minipage}
\hspace{0.4cm}
\begin{minipage}[b]{0.5\linewidth}
\centering
\includegraphics[width=2.8in,height=2.3in]{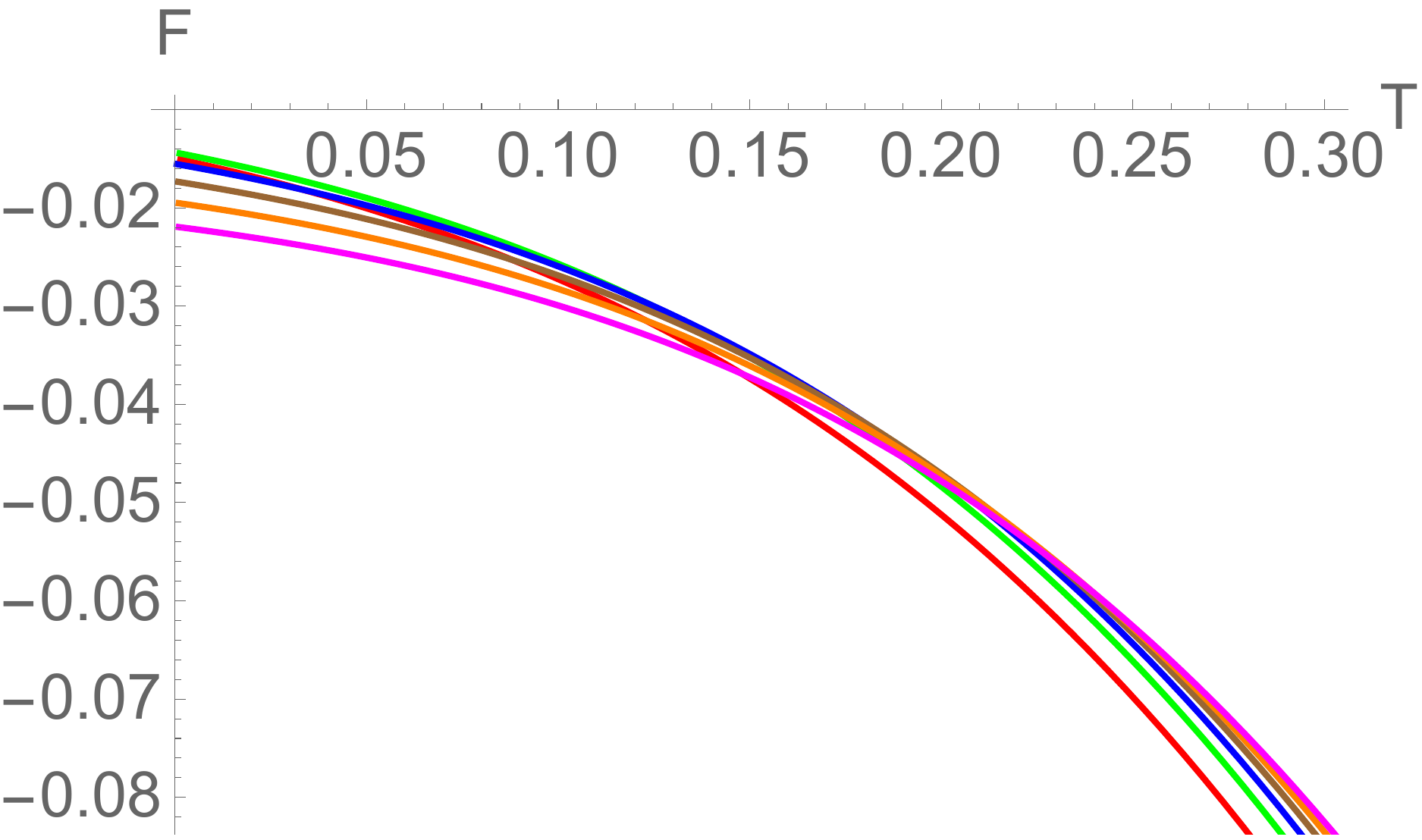}
\caption{\small The Helmholtz free energy $F$ as a function of Hawking temperature $T$ for various values of $a$. Here $q_e=0.1$ and $q_M=0.1$ are used. Red, green, blue, brown, orange, and magenta curves correspond to $a=0$, $0.05$, $0.10$, $0.15$, $0.20$, and $0.25$ respectively.}
\label{TvsFvsaqEPt1qMPt1hypercase1}
\end{minipage}
\end{figure}

Similar thermodynamic results persist in the canonical ensemble as well. This is shown in Figs.~\ref{zhvsTvsaqEPt1qMPt1hypercase1} and \ref{TvsFvsaqEPt1qMPt1hypercase1}. Again, there is a one to one relation  between horizon radius and temperature.  The specific heat of the black hole at constant charge is also always positive. The Helmholtz free energy of the hairy black hole can again become smaller than the non-hairy black hole at low temperatures. Again, for a fixed $q_e$ and $q_M$, the temperature range for which the hairy black hole free energy remains smaller increases with $a$.

\section{Black hole thermodynamics with $A(z)=-az$}
It is instructive to also investigate the hairy black hole solution and thermodynamics for a different form of $A(z)$. In particular, in order to check the universal features of the results presented above for the hairy black hole thermodynamics it is desirable to perform analogous analysis with a different form of $A(z)$. Here, we consider another simple form $A(z)=-az$. This form makes sure that constructed geometry asymptotes to AdS at the boundary.

\begin{figure}[ht]
	\subfigure[]{
		\includegraphics[scale=0.4]{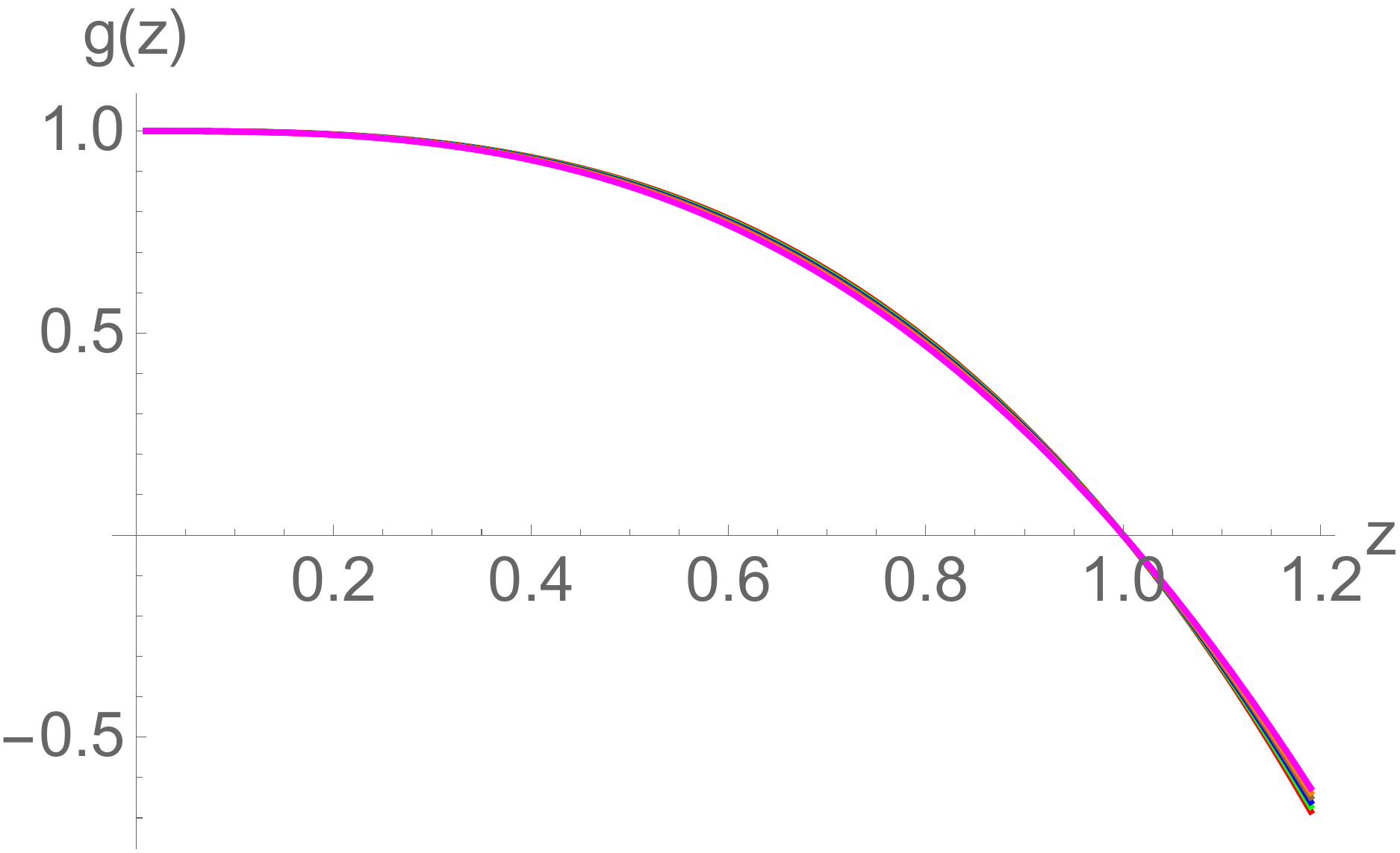}
	}
	\subfigure[]{
		\includegraphics[scale=0.4]{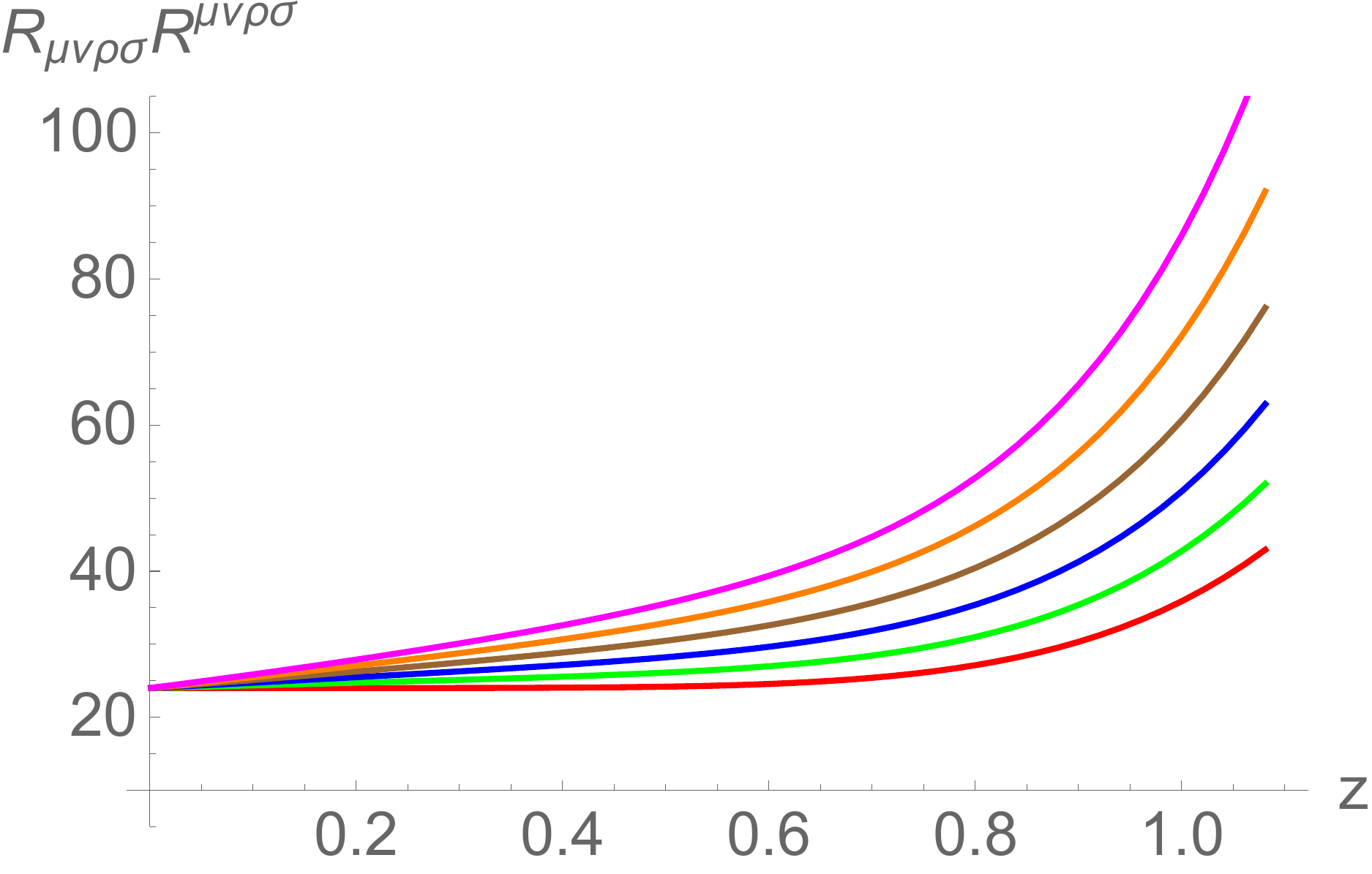}
	}
	\caption{\small The behavior of $g(z)$ and $R_{MNPQ}R^{MNPQ}$ for different values of hair parameter $a$. Here $z_h=1$, $q_e=0.1$, $q_M=0.1$, and $\kappa=0$ are used. Red, green, blue, brown, orange, and magenta curves correspond to $a=0$, $0.05$, $0.10$, $0.15$, $0.20$, and $0.25$ respectively.}
	\label{zvskmannvsaqEPt1qMPt1zh1f1planarcase2}
\end{figure}

With $A(z)=-az$, most of our results for the hairy solution remain the same as in the previous case. This is true for all horizon topology. Here we mainly concentrate on the planar black holes as analytic results are not only straightforward to obtain but also easily expressible for this case. With $A(z)=-az$, the solutions of $\phi(z)$ and $B_t(z)$ reduce to
\begin{eqnarray}
& & \phi(z) = 4\sqrt{a z}  \,,\nonumber \\
& & B_t(z) = \mu_e \left(1-\frac{1-e^{-a z}}{1-e^{-a z_h}}\right)\,,
\end{eqnarray}
together with the relation
\begin{eqnarray}
\tilde{\mu_e} = \frac{a \mu _e}{1-e^{-a z_h}} = q_e
\end{eqnarray}
Similarly, the metric function is given by
\begin{eqnarray}
 g(z) = 1-\frac{\left(-a z (a z+2)+2 e^{a z}-2\right) e^{a z_h-a z}}{-a z_h \left(a z_h+2\right)+2 e^{a z_h}-2} + \frac{e^{-2 a z}}{4 a^4} (q_e^2 + q_M^2) \times \nonumber \\
 \Bigl(1+ \frac{\left(-a z (a z+2)+2 e^{a z}-2\right) e^{a \left(z-z_h\right)} \left(2 a z_h \left(a
   z_h+1\right)-e^{2 a z_h}+1\right)}{a z_h \left(a z_h+2\right)-2 e^{a z_h}+2} \Bigr.\\
  \Bigl. +2 a z (a z+1)-e^{2 a z}\Bigr)\,. \nonumber \\
\end{eqnarray}
This is well behaved function as can be seen from Fig.~\ref{zvskmannvsaqEPt1qMPt1zh1f1planarcase2}. This, along with the fact that Kretschmann scalar is finite everywhere outside the horizon, indicating the well behaved nature of the hairy geometry. Moreover, the scalar field is finite and goes to zero only at the asymptotic boundary. It once again implies the existence of a well-behaved planar charged hairy black hole solution.

We can similarly obtain the black hole mass. The mass from the AMD prescription is given by
\begin{eqnarray}
& & \frac{M_{AMD}}{\Omega_{2,0}} = -\frac{a^3 e^{a z_h}}{24 \pi  G_4 \left(a^2 z_h^2+2 a z_h-2 e^{a z_h}+2\right)} \nonumber \\
& & -\frac{z_h^2 \left(2 a^2 \left(e^{-a z_h}-2\right)+\frac{2 a \left(e^{-a z_h}-4\right)}{z_h}+\frac{e^{-a
   z_h}+7 e^{a z_h}-8}{z_h^2}\right) \left(q_e^2+q_M^2\right)}{96 \pi  a G_4 L^2 \left(a^2 z_h^2+2 a z_h-2
   e^{a z_h}+2\right)} \,,
\end{eqnarray}
which is same as the $z^3$ coefficient of $g(z)$. Moreover, the mass from the holographic renormalization procedure also matches exactly with the AMD mass. Importantly, for $A(z)=-az$ no other counterterms are needed to renormalize the action than those already suggested for $A(z)=-\log(1+az)$. Similarly, the Gibbs free energy from the renormalized action is given by

\begin{eqnarray*}
 \frac{G}{\Omega_{2,0}} = \frac{e^{-a z_h} \left(-4 a^4 e^{2 a z_h}+8 a^4 e^{3 a z_h}-4 a^4 e^{4 a z_h}\right)}{192 \pi  a G_4
   \left(e^{a z_h}-1\right){}^2 \left(-a z_h \left(a z_h+2\right)+2 e^{a z_h}-2\right)} + \nonumber \\
  \frac{q_M^2 e^{-a z_h}}{192 \pi  a G_4 \left(e^{a z_h}-1\right){}^2} \times  \Bigl[\frac{-2 a^2 z_h^2 \left(-9 e^{a z_h}+4 e^{2 a z_h}+5\right) \left(e^{a
   z_h}-1\right)}{\left(-a z_h \left(a z_h+2\right)+2 e^{a z_h}-2\right)} \Bigr.\\
 \Bigl. + \frac{\left(17 e^{a z_h}-23\right) \left(e^{a z_h}-1\right)^3 -2 a z_h \left(8 e^{a z_h}-11\right) \left(e^{a z_h}-1\right)^2}{\left(-a z_h \left(a z_h+2\right)+2 e^{a z_h}-2\right)}  \Bigr] \nonumber\\
+ \frac{\mu_e^2 e^{-a z_h} \left(-2 a^4 z_h^2 e^{2 a z_h}-a^2 e^{2 a z_h} \left(e^{a z_h}-1\right) \left(7 e^{a
   z_h}-1\right)\right)}{192 \pi  a G_4 \left(e^{a z_h}-1\right){}^2 \left(-a z_h \left(a z_h+2\right)+2
   e^{a z_h}-2\right)} \nonumber \\
 + \frac{\mu_e^2 e^{-a z_h} \left(4 a^4 z_h^2 e^{3 a z_h}+2 a^3 z_h e^{2 a z_h}
   \left(4 e^{a z_h}-1\right) \right)}{192 \pi  a G_4 \left(e^{a z_h}-1\right){}^2 \left(-a z_h \left(a z_h+2\right)+2
   e^{a z_h}-2\right)} \,.
\end{eqnarray*}
This once again satisfies the thermodynamic relations $G=M_{HR}^G - T S_{BH} - Q_e \mu_e$ and $G=-P^{G}$.

\begin{figure}[h!]
\begin{minipage}[b]{0.5\linewidth}
\centering
\includegraphics[width=2.8in,height=2.3in]{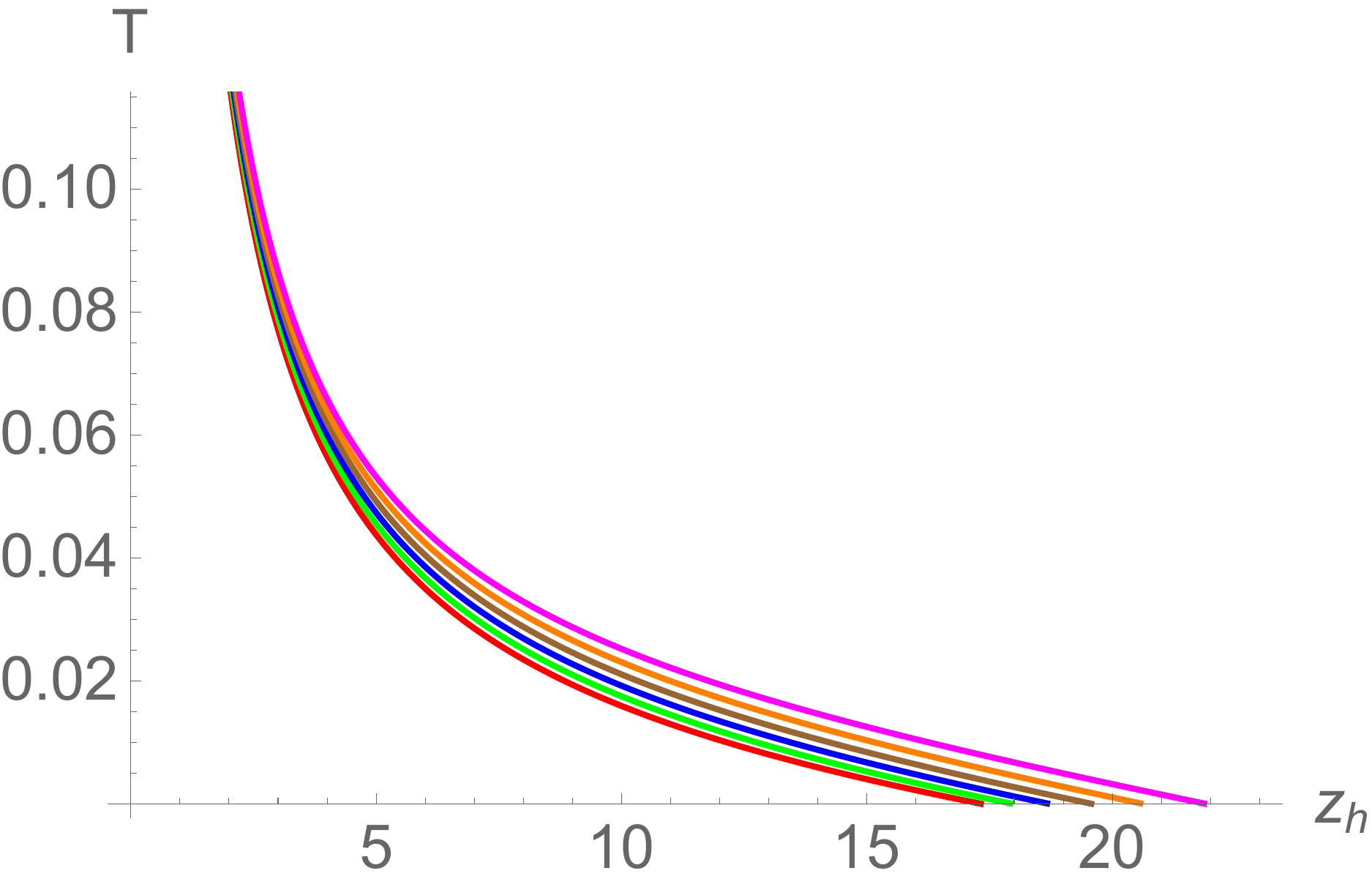}
\caption{ \small Hawking temperature $T$ as a function of horizon radius $z_h$ for various values of $a$. Here $\mu_e=0.2$ and $q_M=0$ are used. Red, green, blue, brown, orange, and magenta curves correspond to $a=0$, $0.01$, $0.02$, $0.03$, $0.04$, and $0.05$ respectively.}
\label{zhvsTvsaMuPt2qM0planarcase2}
\end{minipage}
\hspace{0.4cm}
\begin{minipage}[b]{0.5\linewidth}
\centering
\includegraphics[width=2.8in,height=2.3in]{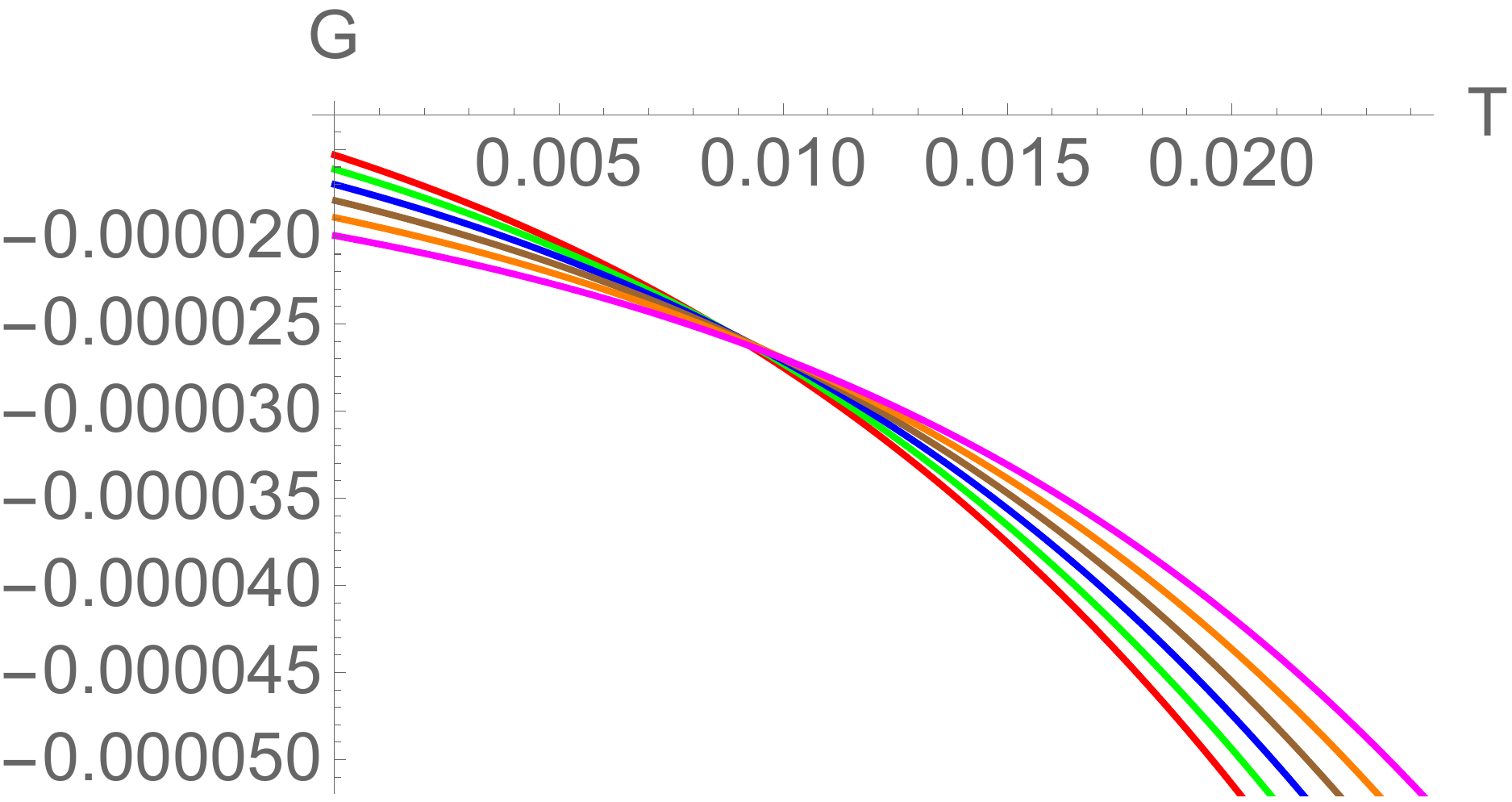}
\caption{\small The Gibbs free energy $G$ as a function of Hawking temperature $T$ for various values of $a$. Here $\mu_e=0.2$ and $q_M=0$ are used. Red, green, blue, brown, orange, and magenta curves correspond to $a=0$, $0.01$, $0.02$, $0.03$, $0.04$, and $0.05$ respectively.}
\label{TvsGvsaMuPt1qM0planarcase2}
\end{minipage}
\end{figure}

In Figs.~\ref{zhvsTvsaMuPt2qM0planarcase2} and \ref{TvsGvsaMuPt1qM0planarcase2}, the thermodynamic behaviour in the grand canonical ensemble is shown. There is again a one to one relation between the Hawking temperature and horizon radius. The black hole can become extremal for finite $\mu_e$ and $q_M$, and the magnitude of $z_h^{ext}$ increases with $\mu_e$ and $q_M$. Importantly, once again, the specific heat at constant potential is always positive, indicating the local stability of hairy dyonic black holes for $A(z)=-az$ case as well. Similarly, the Gibbs free energy of the hairy black hole can be smaller than the non-hairy black hole at lower temperatures. This is shown in Fig.~\ref{TvsGvsaMuPt1qM0planarcase2}. This suggests that for this form of $A(z)$ as well, the hairy black holes can be thermodynamically favoured at lower temperatures. Whereas at high temperatures, the non-hairy black holes have lower free energy. Moreover, the temperature $T_{crit}$  at which the free energy of hairy black hole becomes lower than the non-hairy black hole increases with $a$, $\mu_e$, and $q_M$. This once again implies that the temperature window for which the hairy black hole is more preferable widens with these parameters. The phase diagram illustrating the dependence of $T_{crit}$ on $a$, $\mu_e$, and $q_M$ is quite similar to what is shown in Fig.~\ref{avsTcritvsMuvsqMplanatcase1} for $A(z)=-\log(1+az)$.

\begin{figure}[h!]
\centering
\includegraphics[width=2.8in,height=2.3in]{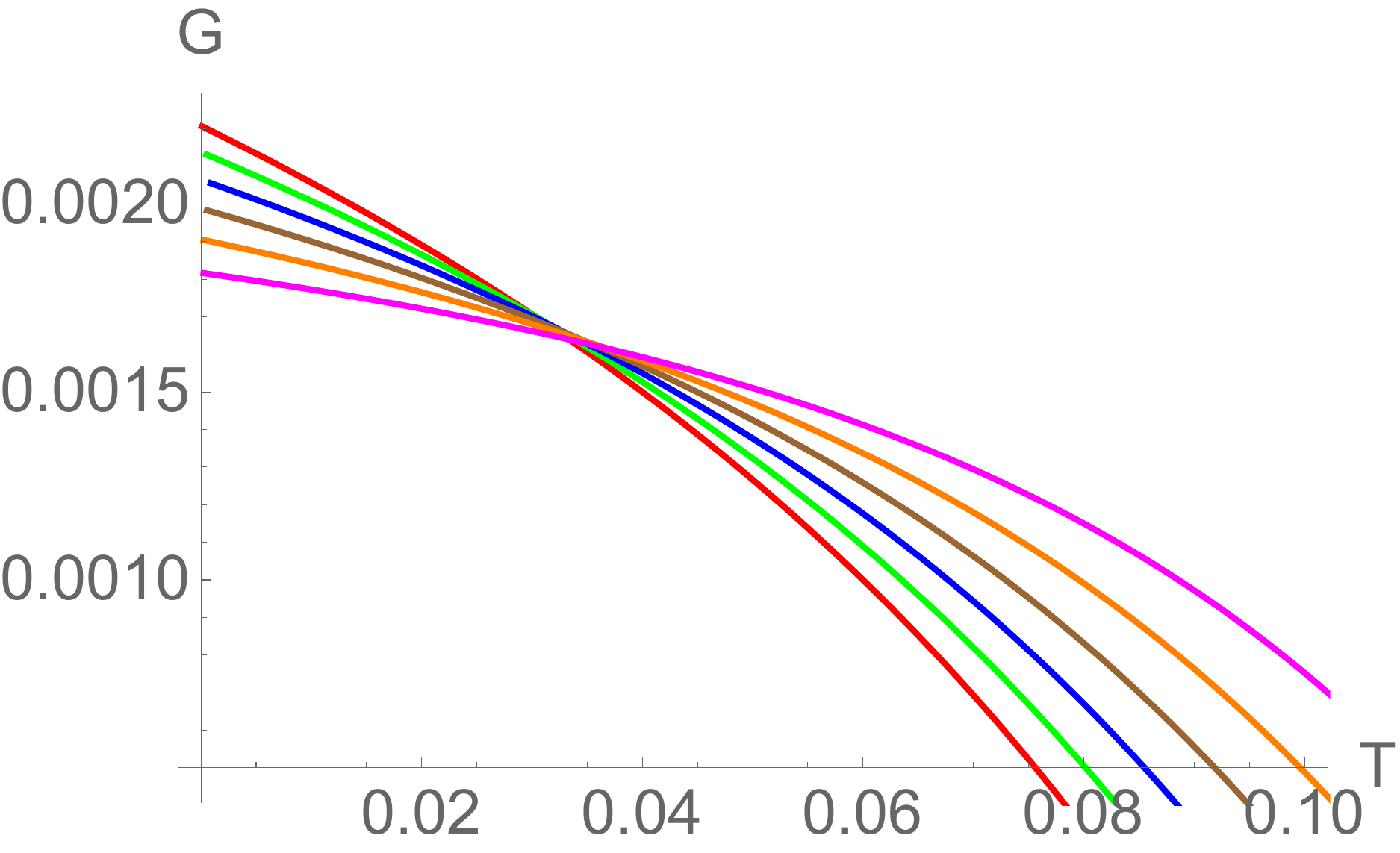}
\caption{ \small Gibbs free energy $G$ as a function of Hawking temperature $T$ for various values of $a$.  Here $\mu_e=0$ and $q_M=0.2$ are used. Red, green, blue, brown, orange, and magenta curves correspond to $a=0$, $0.05$, $0.10$, $0.15$, $0.20$, and $0.25$ respectively.}
\label{TvsGvsaMu0qMPt2planarcase2}
\end{figure}

Importantly, the parameter $q_M$ again has a constructive effect on the hairy solution. In particular, we can have a stable and thermodynamically favoured hairy solution even when $\mu_e=0$. This is shown in Fig.~\ref{TvsGvsaMu0qMPt2planarcase2} for a particular value of $q_M=0.2$. We indeed see that the free energy of the uncharged hairy black hole is smaller than the uncharged non-hairy black hole at low temperatures. Moreover, since $T_{crit}$ increases with $q_M$, it indicates that the stability of thermodynamically favoured hairy uncharged black hole enhances with $q_M$. Once again, the phase diagram
illustrating the dependence of $T_{crit}$ on $a$ and $q_M$ for the uncharged case $\mu_e=0$ is quite similar to what is shown in Fig.~\ref{avsTcritvsqMMu0planatcase1} for $A(z)=-\log(1+az)$.

\begin{figure}[h!]
\begin{minipage}[b]{0.5\linewidth}
\centering
\includegraphics[width=2.8in,height=2.3in]{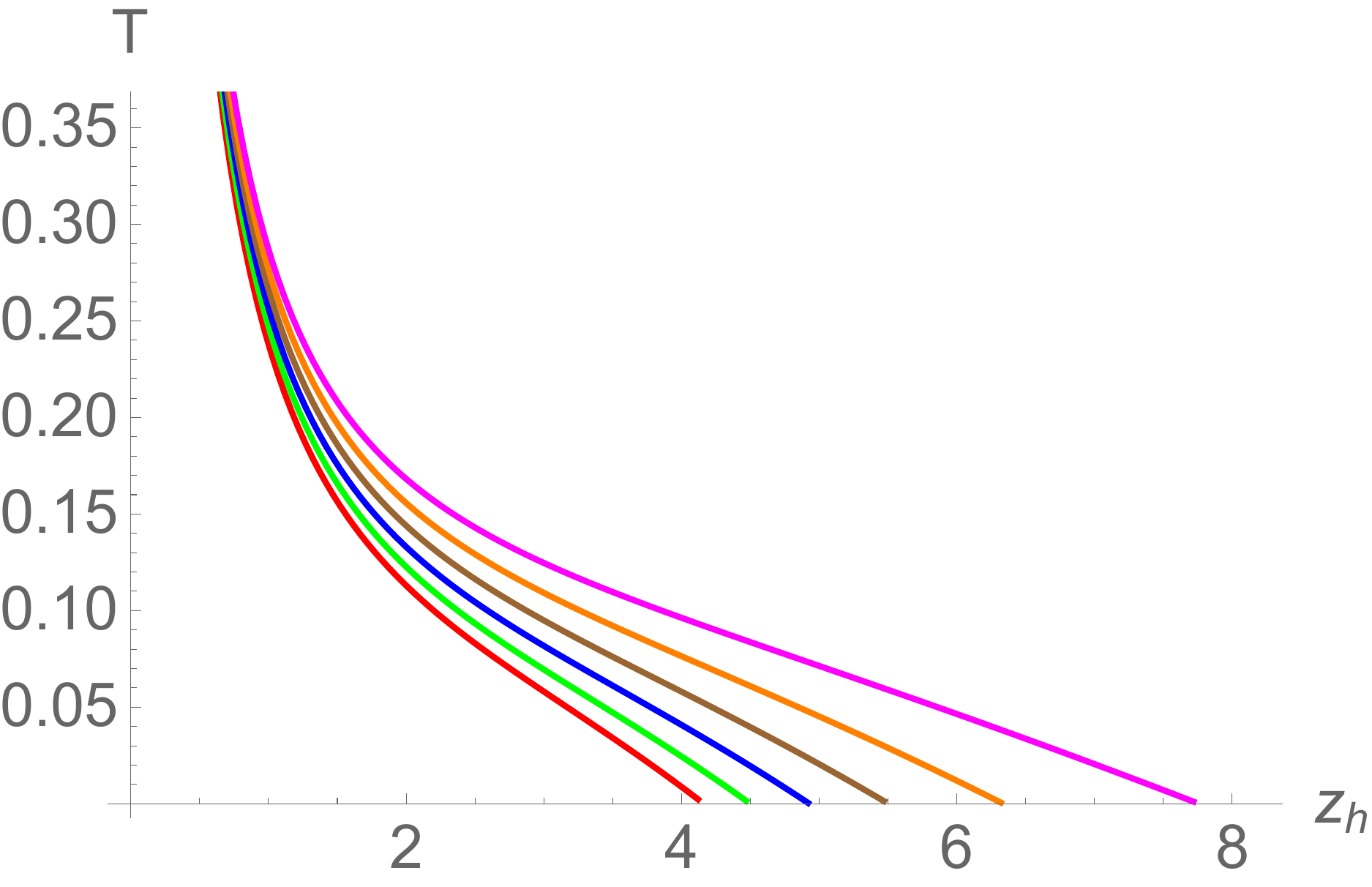}
\caption{ \small Hawking temperature $T$ as a function of horizon radius $z_h$ for various values of $a$. Here $q_e=0.2$ and $q_M=0$ are used. Red, green, blue, brown, orange, and magenta curves correspond to $a=0$, $0.05$, $0.10$, $0.15$, $0.20$, and $0.25$ respectively.}
\label{zhvsTvsaqEPt2qM0planarcase2}
\end{minipage}
\hspace{0.4cm}
\begin{minipage}[b]{0.5\linewidth}
\centering
\includegraphics[width=2.8in,height=2.3in]{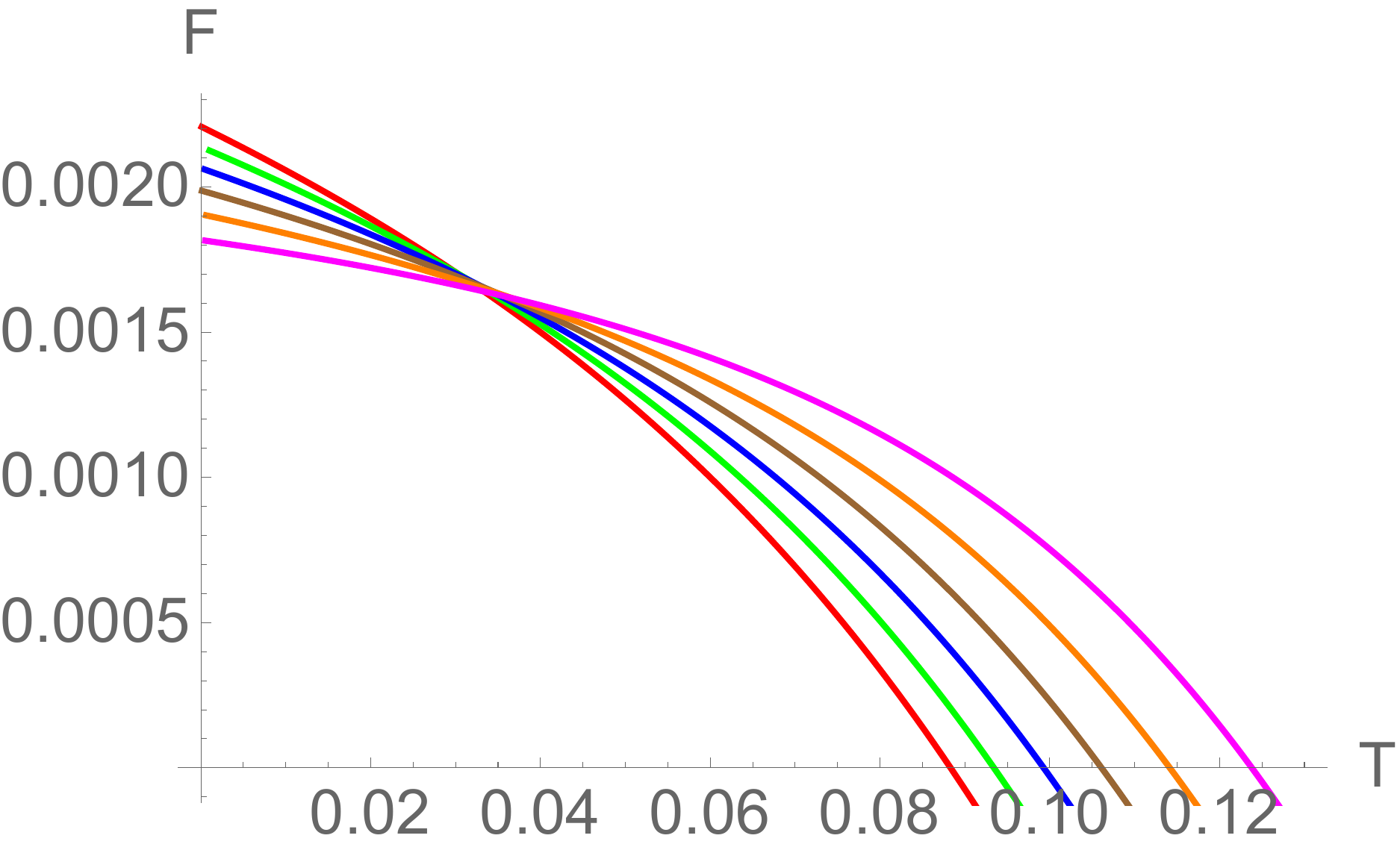}
\caption{\small The Helmholtz free energy $F$ as a function of Hawking temperature $T$ for various values of $a$. Here $q_e=0.2$ and $q_M=0$ are used. Red, green, blue, brown, orange, and magenta curves correspond to $a=0$, $0.05$, $0.10$, $0.15$, $0.20$, and $0.25$ respectively.}
\label{TvsFvsaqEPt2qM0planarcase2}
\end{minipage}
\end{figure}

The thermodynamic discussion in the canonical ensemble follows a similar trait. The relevant Helmholtz free energy again satisfies the relations $F=G+Q_e\mu_e=M_{HR}^{G}-TS_{SB}$ and $F=-P^G$. The thermodynamic behaviour is again qualitatively similar to the $A(z)=-\log(1+az)$ case. In particular, the thermodynamically stable hairy black hole exists for all temperatures and has a positive constant charge specific heat. Similar to the grand-canonical case, the free energy of the hairy black hole can be smaller than the non-hairy black hole at lower temperatures, implying the overall stability of hairy black hole over non-hairy black hole at low temperatures. This is shown in Figs.~\ref{zhvsTvsaqEPt2qM0planarcase2} and \ref{TvsFvsaqEPt2qM0planarcase2}. Moreover, the temperature $T_{crit}$ is found to be an increasing function of $a$, $q_e$, and $q_M$. Therefore, in the canonical ensemble as well, the dyonic parameter $q_M$ not only plays a constructive role on the thermodynamic stability of the hairy black hole but also make hairy uncharged black hole thermodynamically more favoured compared to the non-hairy uncharged black hole.

Above we discussed the hairy black hole solution and thermodynamic for the planar horizon and found that results remain qualitatively similar to the case of $A(z)=-\log(1+az)$. We have analyzed other simpler forms of $A(z)$ as well, and again found similar results. The results for the spherical and hyperbolic horizons, though not presented here for brevity,  again exhibit similar features for different $A(z)$ forms. In particular, for $A(z)=-az$, there are again Hawking/Page and small/large black hole type phase transitions in the spherical case whereas there are stable hairy black holes, without undergoing any phase transition, in the hyperbolic case. This shows that though the form of $A(z)$ dictates the overall hairy black hole solution and thermodynamics, however, qualitatively they exhibit similar features. Our whole analysis therefore does indicate the existence of a thermodynamically stable and well-behaved hairy dyonic black hole solution in asymptotically AdS spaces in our model.

\section{Conclusions}
In this paper, we have constructed and investigated four-dimensional hairy dyonic static black holes in AdS space in the Einstein-Maxwell-Scalar gravity theory, where the $U(1)$ gauge field carries both electric and magnetic charges.  We solved the coupled Einstein-Maxwell-Scalar equations of motion analytically and obtained the exact hairy dyonic black hole solutions with various horizon topologies. This includes planar, spherical, and hyperbolic horizon topologies. The analytic gravity solution is expressed in terms of a function $A(z)$, which allowed us to introduce scalar hair controlling parameter $a$. In the limit
$a\rightarrow 0$, our solution reduces to the standard non-hairy AdS dyonic solution. We considered two simple, and yet different, profiles for $A(z)$, each of which led to qualitatively similar features in the hairy black hole solution and thermodynamics. For these profiles, the scalar field is found to be regular everywhere outside the horizon and goes to zero at the AdS boundary. Similarly, the Kretschmann scalar
is finite everywhere outside the horizon, indicating the absence of any additional singularity in the constructed hairy black holes.

We then investigated the thermodynamic properties of the hairy dyonic black holes in the canonical and grand canonical ensembles. For this, we first obtained analytic expressions of various thermodynamic variables. To compute the black hole mass, we employed the AMD and holographic renormalization methods. We found agreement between these methods for the planar case, albeit utilising the freedom of appropriately choosing the scalar counterterms, whereas for the spherical and hyperbolic cases, due to additional logarithmic divergences in the metric function, the black hole mass expression could be obtained from the holographic renormalization method only.  We found that the specific heat is always positive for the planar and hyperbolic cases, thereby establishing these hairy black holes local stability in both these ensembles. Moreover, the hairy black holes are not only thermodynamically stable but also thermodynamically favoured. In particular, the hairy black holes have lower free energy than the non-hairy black holes at low temperatures. This interesting behaviour is quite similar to the holographic condensed matter systems, where a similar phase transition to hairy black holes, corresponding to the condensation of a dual scalar operator in the boundary theory, typically appears below a certain temperature. We further analyzed the influence of parameters $\{a, \mu_e, q_e, q_M \}$  on the temperature range for which the hairy black holes are thermodynamically favoured and on the hairy/non-hairy transition temperature. We found that all these parameters have a constructive effect on the thermodynamic stability of the hairy black hole. Importantly, thanks to finite $q_M$, the free energy of the uncharged hairy black hole can be smaller than the uncharged non-hairy black hole. This is an important improvement on the results of \cite{Mahapatra:2020wym}, where the thermodynamically favoured hairy black holes were found only for the charged case. Similarly, for the spherical horizon, like their non-hairy counterpart, there occurred Hawking/Page and small/large Van der Waals type phase transitions. Interestingly, with scalar hair, unlike their non-hairy counterpart, the small/large black hole phase transition appeared in the grand-canonical ensemble as well.

There are many directions in which our work can be extended. The first and foremost is to construct and investigate hairy black holes for non-minimal couplings, such as $f(\phi) \propto e^{-\phi}$ or $f(\phi) \propto \phi$, between the scalar and gauge field.  These types of non-minimal couplings have long been considered in the context of, e.g., supergravity and Kaluza-Klein theory \cite{Garfinkle:1990qj,Gibbons:1987ps}. A similar type of non-minimal coupling has also been used to construct spontaneous scalarized charged black hole solutions in recent years \cite{Herdeiro:2018wub,Guo:2021zed} \footnote{In spontaneous scalarised charged black hole models, generally the coupling of type $f(\phi)=e^{-\phi^2}$ are considered.}. Therefore, it is interesting to analyse the effect of this non-minimal coupling in our model. Similarly, it is also important to explicitly establish the first law of thermodynamic like relation for these hairy black holes. Since the parameter $q_M$ corresponds to a background magnetic field in the dual boundary theory, one can also use a modified version of our constructed solution to investigate important anisotropic properties in holographic QCD in the lines of \cite{Dudal:2016joz,Dudal:2014jfa,Dudal:2021jav}.  It would also be interesting to study the dynamical stability of the hairy black hole under various perturbations. Work in this direction is in progress.

\section*{Acknowledgments}
We would like to thank D. Choudhuri for a careful reading of the manuscript and pointing out the necessary corrections. The work of S.~M.~is supported by the Department of Science and Technology, Government of India under the Grant Agreement No. IFA17-PH207 (INSPIRE Faculty Award). The work of S.~P.~is supported by Grant No. 16-6(DEC.2017)/2018(NET/CSIR) of UGC, India.

\end{document}